\documentclass[a4paper,12pt]{article}

\usepackage[ left=1.0in,  
top=0.6in,
right=1.0in,
bottom=0.8in,
headsep=0.25in,
headheight=20pt,
heightrounded,
a4paper]
{geometry}

\usepackage{amsmath,amssymb,amsthm}
\usepackage{colortbl}
\usepackage{textcomp}
\usepackage{mathtools}
\usepackage{eurosym}
\usepackage{booktabs}
\usepackage{longtable}
\usepackage{rotating}
\usepackage[USenglish]{babel}

\usepackage{hyperref}

\usepackage{graphicx,subcaption}

\usepackage{pdfpages}
\usepackage{pdflscape}

\begin{document}

\title{Solar cosmic ray generation, Newtonian gravity, missing mass,
dark energy, laboratory-based nuclear astrophysics, and all that}
\author{Richard Talman\\
Laboratory for Elementary-Particle Physics\\
Cornell University, Ithaca, NY, USA\\
}

\maketitle

\tableofcontents

\clearpage

\begin{abstract}
Begun as part of a promotion of accelerator-based nuclear astrophysics, research toward this goal has shifted
to cosmic ray production within the solar system.  This has been motivated by the high quality
of data collected recently by programs such as the  International Space Station (ISS), by the fundamental importance
of the topic, and by the unsatisfactory state of our understanding of the actual source of cosmic rays.
A ``minor'' change in the Newtonian gravitational formulation converts solar production of high energy
cosmic rays from ``impossible'' to ``likely'', without much disrupting the vast existent domain of well understood
astronomical gravitational processes. This change enables the sun to ``capture'' protons of energy so high that they
would, otherwise, escape the solar system. This change in Newton's gravitational formula would disrupt current
cosmological understanding of missing mass and dark energy.
With this changed understanding of gravity, it has been quite easy to produce a semi-quantitative understanding of the
solar origin of cosmic rays up to energies at least as great as $10^6$\,GeV/nucleon that have by now been detected
and measured so persuasively. A ``Double Slingshot'' mechanism is proposed according to which at least a substantial
fraction of all cosmic rays could have begun their life within the solar system.
Accelerator-based nuclear astrophysics is promoted, beginning with the functioning of an
``{\rm E}\&{\rm M}'' storage ring for laboratory-based study of processes such as
${\rm 6LI} + {\rm 7Li} \rightarrow {\rm 13C} + \gamma$.
Especially to be emphasized are experiments made possible by
''rear-end collisions'' between nuclear isotopes of two different types traveling simultaneously at
different velocities in the same direction in the same storage ring. This amounts to ``studying nuclear physics
in a moving frame of reference''.
\end{abstract}

\clearpage

\section{Introduction}
Much of big bang nucleosynthesis (BBN)\cite{Alain-Coc-BBN}\cite{Merchant-Rowley}
concerns the production of carbon,
which is impeded by the nuclear isotope mass gap at A=8.  As one example of nuclear transmutation,
this paper analyses a storage ring process that, in a single reaction, produces carbon-13 (13C) from
lithium-6 (6Li) and lithium-7 (7Li), both available in early days of solar nucleosynthesis. No intervention
of protons, deuterons, $\alpha$-particles or, especially problematical, beryllium, is required.

  In an ${\rm E}\&{\rm M}$ storage ring with superimposed electric and magnetic bending, beams of different nuclear particle type, such as lithium-6 (6Li) and lithium-7 (7Li), can circulate simultaneously, in the same direction, at the same time, with different velocities. Intentionally choosing heavier 7Li as the faster beam, the 7Li bunches will pass through the 6Li bunches millions of times, at the same ring location, at a constant rate, occasionally, in this case, producing  $$ \rm{7Li+6Li} \rightarrow {\rm 13C}+\gamma .$$ These rear-end collisions are detectable in a tracking detector at the IP. This amounts to studying nuclear physics in a moving frame of reference, much like the same nuclear process  occurring in the sun. Essentially every two body collision in the sun can be studied in this way.  Weak interaction effects can, eventually, also be studied.

This experiment is made possible by the superimposed electric and magnetic bending in
an ``{\rm E}\&{\rm M}'' storage ring, which makes it possible, for example, for a 100 MeV 6Li beam and
a 106 MeV 7Li beam to circulate simultaneously, with different velocities. Because of their different velocities,
rear-end bunch collisions occur at a regular rate.

At solar temperatures, with Maxwell-Boltzmann (M-B) particle
distributions,  the collision rate for this process in the sun would be too small to influence solar models.  But, as a result of
(poorly understood) mechanisms involving solar magnetic fields, such particle energies are, in fact, observed to be present in
the solar system. Weak as the solar system magnetic fields are, they hold most of the (otherwise free) nuclear
isotopes captive in our solar system.

For simplicity weak interaction effects are ignored and isotopes with equal numbers of neutrons and protons,
(referred to as ``alpha material'') with
 N=P=A/2, are mainly emphasized. Their nearly identical orbits in magnetic fields make them ideally suited
for injection from the solar winds of Parker, into temporary orbits around planets or around the sun itself,
where they are accelerated by the Parker electric field component parallel to the particle velocity.

This acceleration mechanism competes favorably with the similar acceleration mechanism proposed by
Fermi, especially because nuclear collision rates between nuclear isotopes following one another along
nearly identical orbits are greatly enhanced, relative to the isotropic collisions in a Boltzmann-distributed
gas.   All alpha material'' is subsequently subject to acceleration or deceleration in ``solar system
accelerators''.  A similar superposition of magnetic and gravitational bending in the exterior fields
of actual stars and planets is contemplated.

Numerous laboratory-based storage ring experiments are described, most involving nuclear collisions between
particles of different type following almost identical orbits in the same direction. But inverse beta decay
interactions, with electrons and nuclear isotopes co-rotating can also be produced and measured.

Common to these threads is the ``alpha-deuteron cluster model'' theoretical framework\cite{Uesaka-Itagaki}.
Isotopic abundance ratios are discussed in some detail, because of the evidence they provide concerning the processes under
discussion.

\section{Nuclear $\alpha$-material}
\subsection{Novel ``$\alpha$-material'' nuclear terminology}
This paper introduces the term ``nuclear $\alpha$-material'' to refer to any assemblage of
nucleons consisting of multiples of two protons and two neutrons (that can be treated as $\alpha$
particles in effective field theory (EFT) plus, as required for odd-Z, a solitary Z=1, A=2
deuteron. \footnote{The nuclear spin is zero for even-Z, even-N nuclei, integer for all even-A nuclei,
and odd half-integer for all odd-A nuclei. Oddness of both Z and N tends to lower the nuclear binding
energy, making odd nuclei generally less stable.}

It is useful to include the deuteron as the sole $\alpha$-nuclear material with N=1. Note though,
that the neutron (because Z=0) and the proton (because N=0) are excluded from the class.
Excluding the N=1 proton allows every member of the class to have equal numbers of protons and 
neutrons: a single pair in the deuteron case. For every element there is a single entry in the
class but, especially for large Z, the N=Z isotope is not necessarily stable. Apart from $p$
and $n$ the only unstable low-Z exception is beryllium ${}^8_4$Be.  But this exception is spectacular,
in that the (extrapolated, but not measured directly) lifetime of ${}^8_4$Be is $0.819\times10^{-16}$\,s.

It is useful to introduce a small parameter, $\delta_{sol-win}$, (with subscript ``sol-win'' to associate
it with the solar wind) derived from deuteron and $\alpha$-particle masses, that symbolizes the fractional
deviation from A/Z of their mass values, allowing for their factor of two different Z factors.  Also,
to quantify deuteron parameters, the anomalous magnetic moment $G_d$ and its spin tune, $Q_s$, is
mentioned: $Q_s$, is the fraction (of 2$\pi$) angular advance of the in-plane component of the deuteron spin
vector, when the deuteron direction, itself, advances by $2\pi$:
\begin{align}
     m_{\alpha}c^2 &= 3.72737940\ {\rm GeV} \notag \\
     m_dc^2 &= 1.87561294\ {\rm GeV}      \notag  \\
     \delta_{sol-win} = (m_{\alpha}- 2 m_d)/m_{\alpha} &= -0.00639765  \\
     \hbox{deuteron anomalous\ magnetic\ moment\ }     &\equiv G_d =  -0.1432204\\
     \hbox{deuteron spin\ tune}                        &\equiv G\gamma \approx G_d. 
\label{eq:delta_solar_wind}
\end{align}
where the relativistic factor $\gamma$ can often be treated adequately as equal to 1.
These parameters are essential for the treatment of $\alpha$-particles or deuterons carried in the
solar wind or appearing there suddenly when particles in the solar wind transmutate into states
with free $\alpha$'s or deuterons.

The $\alpha$-particle itself is spinless. But in a nucleus that can be interpreted as a deuteron
bound to one or more $\alpha$-particles following a helical orbit, an internal deuteron spin
vector can be visualized as processing through $2\pi$ every six cyclotron turns. This presumably
averages out the spin dependence of all processes to be discussed. Anomalous deuteron magnetic moment
$G_d=-0.1432204$ is itself a
smallish parameter, significantly smaller (in magnitude) than most other nuclei, only about
half of which are positive.  

For purposes of this paper, the only \emph{simple particles} to enter discussion are electrons, protons,
and particles of $\alpha$-material. Remember, especially, that the proton is not a member of the
class of $\alpha$-media particles, since it does not satisfy Z=N.  Nuclei isotopes heavier
than nickel cannot be included because  N exceeds Z for every stable isotope.. 
  
Certainly there are no neutrinos, nor anything resembling $\beta$-decay, nor weak interaction
mentioned; except in reference to future experimental possibilities.  The
purpose for imposing this vocabulary is to avoid the need for endless repetition of the assumption
that weak interaction physics is being ignored. The virtue of this filtering is that all alpha material
nuclei have the same mass-to-charge ratio (except for the deviation from zero of the small
parameter $\delta_{sol-win}$).
  
The vocabulary also suggests that there is a ``favored'' isotope for every integer $Z$, even in
the common cases that the favored isotope is not, in fact, stable. In every such case, however,
like the neutron, most ``favored'' isotopes have significantly long lifetimes.
The proton must be handled separately, as an $N=0$ special case. One  point of the present paper is to
suggest that all Z=N isotopes, are susceptible to injection ``into'' and acceleration by planetary
environments or by the sun itself.  Because of their fragility, the cases of ${}^6$Li$_3$,
(especially) ${}^9$Be$_4$, and ${}^{10}$B$_5$ require special discussion. Otherwise $\alpha$-nuclear
material may be usefully employed in the cosmological production  of (otherwise hard to explain) high
energy cosmic ray particles.
  
In a certain sense
``the proton can take care of itself'' since it is the only particle that is always stable. Protons
are always present in the initial and final states of nuclear transmutation. This would be grounds, while
reporting abundance ratios measured experimentally, to use the proton as the denominator in isotopic
abundance ratios.  This is not practical however, since protons present in other nuclear isotopes
can often not be reliably counted.  

Special attention needs to be paid to the three elements Li, Be, and B.
The abundance ratios of these ``light elements'', relative to all others in various media, have been subject
to much study for the better part of a century.  This began with terrestrial ratios, meteorites, and
cosmic rays, and, more recently, with contents of the solar wind.  This paper pays special attention to the
possible influence of solar magnetic fields on the isotopic ratios resulting from the different evolution
histories of these light elements.
  
\subsection{Magnetic storage ring ``home'' for $\alpha$-material}
The universal atomic mass is defined as the mass of an unbound neutral atom of carbon-12, divided by 12.
To express the mass of the matching nuclear isotope, in the form $A m_U$ the mass of Z electrons
must have been subtracted.  The definition of ``magnetic rigidity''
of a nuclear isotope (Z,A) species, on a circular arc of radius $\rho$ in magnetic field $B$ is
\footnote{In many papers describing results from the International space station (ISS)
and elsewhere, a parameter referred to as ``rigidity'' is defined as $R = (pc)/(Ze) = B\,r_L$, which is
the same as our $B\rho$.  This means that their $R$ \emph{is not radius}; rather their $r_L$ is our $\rho$.}
        $$B\rho=\hbox{\ magnetic\ rigidity} = (A/Z)m_U\gamma\beta c/e$$
For $\alpha$-nuclear-material, with Z + N = A, the corresponding equation is
        $$B\rho=\hbox{\ magnetic\ rigidity} = Am_U\gamma\beta c/e$$

This terminology may seem to imply that any $\alpha$-material nucleus
can, for some purposes, be treated the same as any other, independent of their particular
Z-values.  This is intentionally and precisely what is being stated.
\emph{The  $Z=N$ constraint is specific to magnetic bending; all such isotopes have the same curvature while
in the same magnetic field.} Within the Parker solar wind model, all such particle orbits have
the same curvature while in the same magnetic field.

As an accelerator physicist I am enthusiastic about accelerators that work very well. To design an accelerator
injection scheme for which a large class of particles can be captured with high injection
efficiency, without violating Liouville's theorem, would be impressive indeed.  Fermi admitted, in his
conjectured astrophysical acceleration scheme, that he could conceive of no efficient injection scheme.

The first two paragraphs of Fermi's paper read as follows:
``
In recent discussions on the origin of the cosmic
radiation E. Teller has advocated the view
that cosmic rays are of solar origin and are kept
relatively near the sun by the action of magnetic
fields. These views are amplified by Alfv\'en, Richtmyer,
and Teller.  The argument against the conventional view
that cosmic radiation may extend at least to all the
galactic space is the very large amount of energy that
should be present in form of
cosmic radiation if it were to extend to such a huge
space. Indeed, if this were the case, the mechanism
of acceleration of the cosmic radiation should be
extremely efficient.\\

I [Fermi, that is] propose in the present note to discuss a hypothesis
on the origin of cosmic rays which attempts
to meet in part this objection, and according to
which cosmic rays originate and are accelerated
primarily in the interstellar space, although they
are assumed to be prevented by magnetic fields
from leaving the boundaries of the galaxy. The
main process of acceleration is due to the interaction
of cosmic particles with wandering magnetic fields
which, according to Alfven, occupy the interstellar
spaces.
``\\

With the words ``the mechanism of acceleration of the cosmic radiation should be
extremely efficient'' Fermi implies the impracticality of acceleration mechanisms
that account insufficiently for the huge range of scales spanned by Astrophysics.
Even among great scientists Fermi, because of his special brilliance concerning practical
scientific matters, was especially qualified to make such a cautionary pronouncement.
But, even Fermi, was not infallible.

One consequence of this has been, though never fleshed out in detail, that the
general perception for three quarters of a century has been that the Fermi explanation
of cosmic rays will eventually prove to be correct. It is my belief, though
Fermi's starting principles were correct, that his suggested cosmic ray acceleration
mechanism is unlikely to be correct.

Along with Alfv\'en, Richtmyer,
and Teller, my paper suggests that cosmic rays are understandable on a smaller scale, namely,
just the solar system.  It is possible though, that the final stage of the production of
ultra-energetic protons, accepts already high energy protons injected from the galaxy.

As Fermi admits, the injection efficiency requirement is
critically important. Any serious acceleration mechanism must include at least an
approximate injection design and efficiency calculation.
In his paper, Fermi arbitrarily takes a value of 200\,MeV for nominal injection energy.
This paper chooses a lower value, 10\,MeV, as arbitrary injection energy, in order to
be certain that the ``injected beam'' can have significantly large current.

Fermi proceeds, in a later section, to calculate the rate of acceleration in
a galactic space containing localized volumes of significant magnetic field that
Alfv\'en had predicted.  Other than the ``guided-wave'' feature, which is
equally essential to this paper, none of this is relevant to the present paper.
But, naturally, Fermi's discussion is concise and instructive, though not at all
promising, in my opinion.  In any case, a few of Fermi's paragraphs follow.\\
``
The path of a fast proton in an irregular magnetic
field of the type that we [Fermi, that is] have assumed will11 be
represented very closely by a spiraling motion
around a line of force. Since the radius of this
spiral may be of the order of $10^{12}$ cm, and the
irregularities in the field have dimensions of the
order of $10^{18}$ cm, the cosmic ray will perform many
turns on its spiraling path before encountering an
appreciably different field intensity. One finds by
discussion that as the particle
an elementary
approaches a region where the intensity
increases, the pitch of the spiral will decrease. One
finds precisely that
\begin{equation}
\frac{\sin^2\theta}{H} = {\rm constant},
\label{eq:Fermi-helix}
\end{equation}
where $\theta$ is the angle between the direction of the
line of force and the direction of the velocity of the
particle, and $H$ is the local field intensity. As the
particle approaches a region where the 6field intensity
is larger, one will expect, therefore, that the angle
$\theta$ increases until $\sin\theta$ attains the maximum possible
value of one. At this point the particle is rejected
back along the same line of force and spirals backwards until
the next region of high field intensity is encountered. This
process will be called a "Type A" reaction. If the magnetic field were
static, such a reaction would not produce any change in the
kinetic energy of the particle. This is not so, however, if the magnetic
field is slowly variable.  It may happen that a region of high field
intensity moves toward the cosmic-ray particle which collides
against it. In this case, the particle will gain energy
in the collision. Conversely, it may happen that
the region of high field intensity moves away from
the particle. Since the particle is much faster, it
will overtake the irregularity of the field and be
reflected backwards, in this case with loss of energy.
The net result will be an average gain, primarily for
the reason that head-on collisions are more frequent
than overtaking collisions because the relative
velocity is larger in the former case.

The equation in the preceding Fermi paragraphs has been numbered for later reference.
Quoting from the final paragraph, Fermi acceleration boiled down to `` it may happen that
the region of high field intensity moves away from
the particle. Since the particle is much faster, it
will overtake the irregularity of the field and be
reflected backwards, in this case with loss of energy.
The net result will be an average gain, primarily for
the reason that head-on collisions are more frequent
than overtaking collisions because the relative
velocity is larger in the former case.''

This is a minuscule acceleration compared to even brief presence  of any alpha material
in the longitudinal ``Parker-electric field'' illustrated in Figure~\ref{fig:Parker-solar-wind}.
Since this paper treats our solar system as closed, any Fermi-process galactic cosmic rays are
ignored by definition.

\begin{figure}[hbt!]
\includegraphics[scale=0.5]{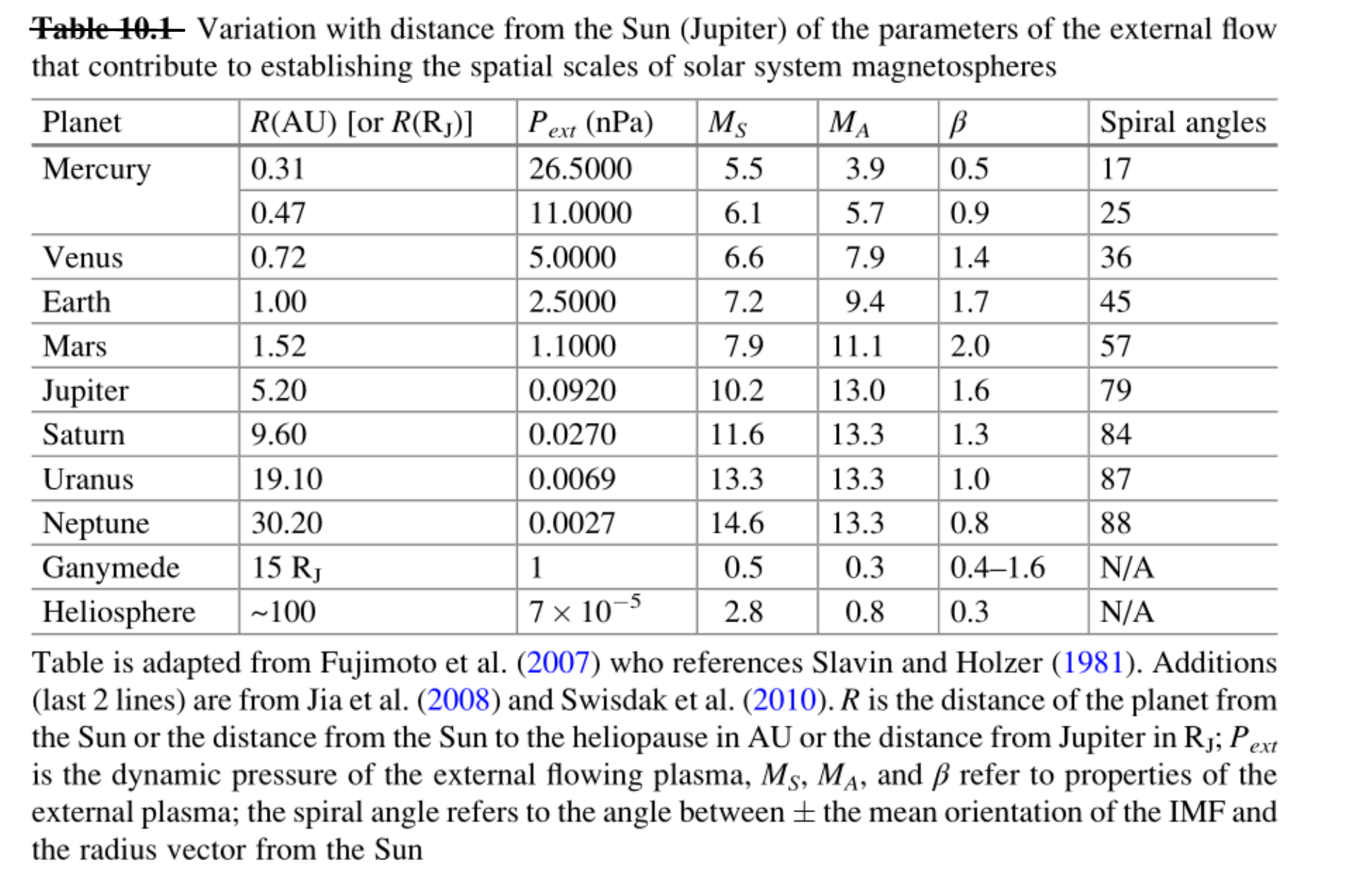}  
\caption{\label{fig:solar-wind-parameters}Parameters not defined in the original table caption
are $\beta$, which is the ratio of thermal to magnetic pressure, $M_A$ and $M_S$, which are Alfv\'en
and Mach numbers which, along with $\beta$ are parameters of of the upstream flow. For present
purposes, the most important parameter is the ``spiral angle'', call it $\theta_s$ measured in degrees
in the range $0 < \theta_s < 90^{\circ}$. For Earth, $\theta_s=45^{\circ}$, a worthwhile mnemonic.}
\end{figure}

Figure~\ref{fig:solar-wind-parameters} shows a screen-dumped table of planetary
solar wind parameters, copied from the Badman and Cowley reference~\cite{Badman-Cowley}.
Though discussed in detail later while reviewing the Parker model, this table is introduced
here in order to make available some of the important solar system parameters.  
The fact that a single theoretical calculation can form the basis for such a detailed table is
testament to both the remarkable power of the original Parker paper and to the commendable
diligence of NASA.  By luck, the 45\,${}^{\circ}$ spiral angle $\theta_P$ of our Earth's Parker
accelerator is situated where the theory is most valid, and most comprehensible.

\section{The Parker model of solar wind}
\subsection{The sun's solar wind treated as a DC arc discharge tube}
In 1926, A. A. Milne, a British writer, famously penned a puzzle in the form of a verse;\\
\noindent
``No one can tell me, Nobody knows, Where the wind comes from, Where the wind goes.``\\
Milne was referring, presumably, to the meteorological earth wind.  Though a mathematician by training,
Milne had not researched this very carefully, since, at that date, the earth's wind was fairly well
understood.  In 2025 the Milne puzzle can be applied to the solar wind.  As to ``where the wind comes from''
Milne has also become outdated; Parker has explained this.  But ``Where the solar wind goes'' remains an
active puzzle for astrophysics.

A challenge is to determine whether the solar wind can be treated as the injected beam, into an accelerator
chain formed from elements of the solar system, that produces ultra-high energy cosmic ray particles, all
within the solar system.  We begin by describing the solar wind in electric circuit terms.

The simplest possible electric circuit consists of a battery with voltage $V$ connected to a resistor with resistance
$R$, causing a current $I=V/R$ to flow in a \emph{closed} circuit.  Temporarily setting aside gravity, any increase in
the sun's central positive ions attracts electrons and repels positive ion density, produces positive
centrifugal radial ion current and symmetric centripetal electron motion. These currents cancel on large time scales,
though not instantaneously.  This is because of the different electron and ion masses, any sudden change in electric
attraction causes transient mass imbalance.  We must therefore limit discussion to slow changes.

In free space outside the sun, even if the electron charge density is small, which is guaranteed in free space, the
electrical resistance $R$ can be expected to be proportionally small.  By symmetry the resulting current is isotropic,
pointing to a distant spherical ``ground''.

In effect, space acts like a semiconductor circuit---say of $p$-type, meaning that currents consist of holes,
effectively positive, thereby causing isotropic current in the radial direction.  This represents no essential
change.  Next one could consider replacing the conducting medium by Schottky diode medium that carries current
only in one direction.  As an electric circuit, this would be problematical.  Quoting from page 99 of the 1963
Raphael Littauer book, ``Pulse Electronics'',\\
``Another rule follows from the trivial observation that diodes conduct in one direction only.
This implies that any current which flows must have a dc component, and consequently that a dc return path
must be provided around every diode. Without such a path, no current can flow at all in the long run, and
the diode may as well be omitted.''  For brevity a typically-instructive Littauer footnote attached to these
sentences is not repeated here.

Superficially one might argue that the required ``return current'' in the solar wind is eventually supplied by
the nuclear ions.  The problem with this is that the nuclear ions are traveling in the forward direction.
Because of their large mass difference, and their correspondingly large speed differential, at least temporarily,
the total charge of the sun would become negative. One concludes that electric circuit theory cannot be applied
directly to nuclear astrophysics.

A possibly related comment is that the obvious impossibility of reversing time
in a circuit application makes it impossible to apply rigorous theorems, such as CPT theorem, to electric circuits.\
In particular the time-reversal T indicated in CPT, can certainly not be applied to cosmology, since, for example,
entropy always increases.

The absence of engineering and experimenter's viewpoints is not necessarily a positive aspect of cosmological theory.
No experimentalist can be happy about Einstein's ``Cosmological Principle'', according to which the universe is uniformly
isotropic and homogeneous when viewed on a large enough scale.  In fact, there is no partial region that is at a
scale large enough to evaluate integral properties, such as charge or current conservation laws. To assume the
contrary invites the mathematical fallacy of neglecting the possibility of divergent, and hence, unphysical, behavior.
Nevertheless, the practice of describing the eventual solar wind destination as a ``ground potential'', though dubious,
can be usefully applied.

A DC arc discharge tube contains a fairly good ``vacuum'' except for ionized atomic vapor forming the
residual gas inside the bulb.  This circuit is similar to that of the  sun.  Currents are carried by the ionized plasma.
As viewed in the sky (undoubtedly with intervening transparent eye protection) the sun looks just like a bright
yellow light source; a sodium arc source for example.  Compared to laboratory light sources, a major difference for
solar wind observation is the absence of intervening light-bulb glass, with the possible exception of satellite-based
measurements in space.

Whatever particles, such as electrons, protons, deuterons, and other nuclear ions are present at their
source in the sun, except for the glass bulb, would arrive on earth in the same fractions.  It is thought that
the solar wind starts out with approximately the same nuclear isotope fractions as apply in the sun as a whole.
These ratios are labeled as ``solar'' in subsequent figures.  As measured, however, the relative abundances of
the isotopes differ appreciably.

Historically, there has been no persuasive mechanism to account for the observations shown in
Figures~\ref{fig:Simpson-Cosmic-Rays-annotated}, and \ref{fig:Wiens-solar-wind-ratios}, which exhibit
strikingly large Z-dependence of isotopic abundance ratios, especially for the ``light elements,
lithium, beryllium, and boron, by comparison  with other elements.

Because the solid angle projection of the earth is so small, only a miniscule fraction
of the solar wind points toward the earth.  But,, except for possible transmutations in flight,
and possible magnetic field effects, whatever nuclear ions leave the sun in the
portion of the solar wind aimed towards the earth should arrive here shortly later.

\subsection{The Parker solar wind model}
\begin{figure}[hbt!]
   \centering
   \includegraphics[scale=0.60]{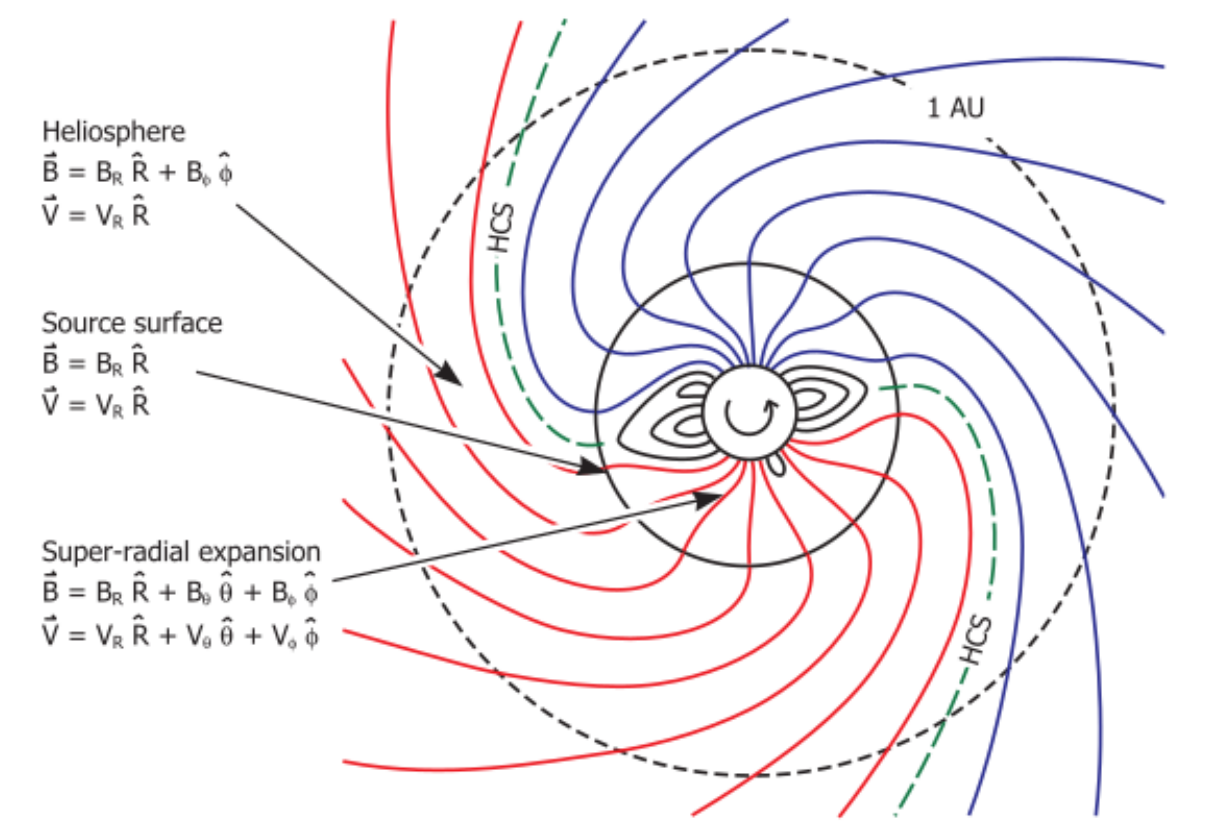}
   \includegraphics[scale=0.40]{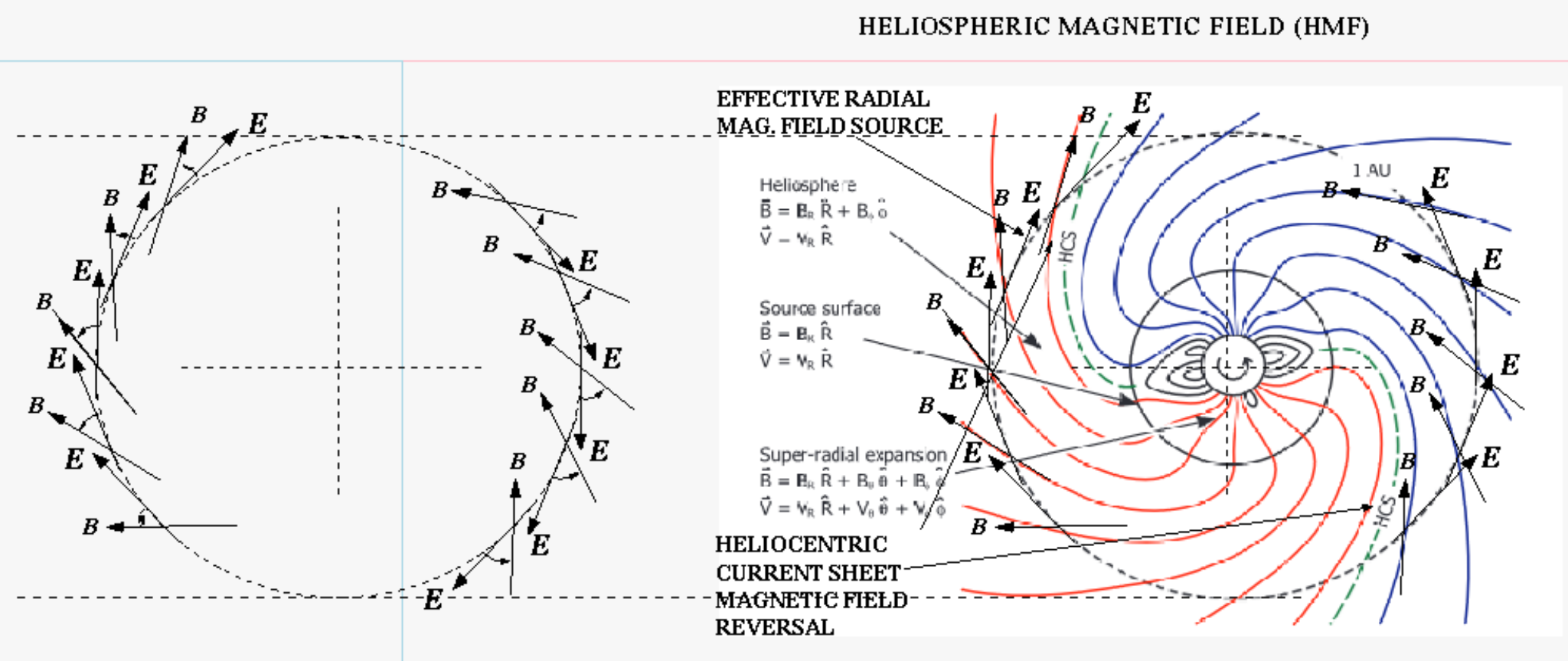}
   \caption{\label{fig:Parker-solar-wind}{\bf Top:\ }Copied from Owens and Forsythe\cite{Owens-Forsyth}, as
     follows: ``A sketch of the steady-state solar magnetic field in the ecliptic plane. Close to the sun, in a
     spatial region approximately bounding the solar corona, the magnetic field dominates the plasma flow and
     undergoes significant non-radial (or super-radial) expansion with height. At the source surface , typically
     taken to be a few solar-radii, the pressure-driven expansion of the solar wind dominates and both the field
     and flow become purely radial.  In the heliosphere, rotation of the heliospheric magnetic field (HMF) footprints
     within a radial solar wind flow generates an azimuthal component of the HMF, $B_{\phi}$, leading to a spiral
     geometry.  Regions of opposite HMF polarity, shown as red and blue, lines, are separated by the heliospheric
     current sheet (HCS), shown as the green dashed line.  Image adopted from Schatten et al. (1969).
   {\bf Bottom:\ }Embellishments of the top figure.
     The figure on the right shows the local electric and magnetic
     field directions, illustrating, for example, the Heliocentric current sheet (HCS) and its reversal.
     The figure on the left defines the Parker angle $\theta_P$ orienting the $B$-field lines relative
     to the effective radial magnetic field source and demonstrating its constancy.'' 
   }
\end{figure}

The top figure in Figure~\ref{fig:Parker-solar-wind} is copied from Owens and Forsythe\cite{Owens-Forsyth}.
The bottom figures are repetitions of the same figure, much embellished to identify the field
patterns clearly. The so-called ``Parker spiral angle'', $\theta_s$, is indicated by all the arrows in the
bottom left hand figure.  (The fact that these angles seem not quite equal is a defect of the
figure, not of the theory.)  The original version of this figure is contained (as a simple sketch) in
Parker's original paper explaining the solar wind.

One conjectures that the Parker solar wind model is even more universal.  Not many decades ago, when there was only
one solar system, one could marvel that a single theory, admittedly with multiple free parameters, could accurately
describe nine different planetary nuclear accelerators.  Today there are thousands of solar systems, and
little reason to doubt
that most of them have comparable numbers of working Parker accelerators.  In time this number will likely grow
to be vastly greater yet.

\begin{figure}[hbt!]
\includegraphics[scale=0.4]{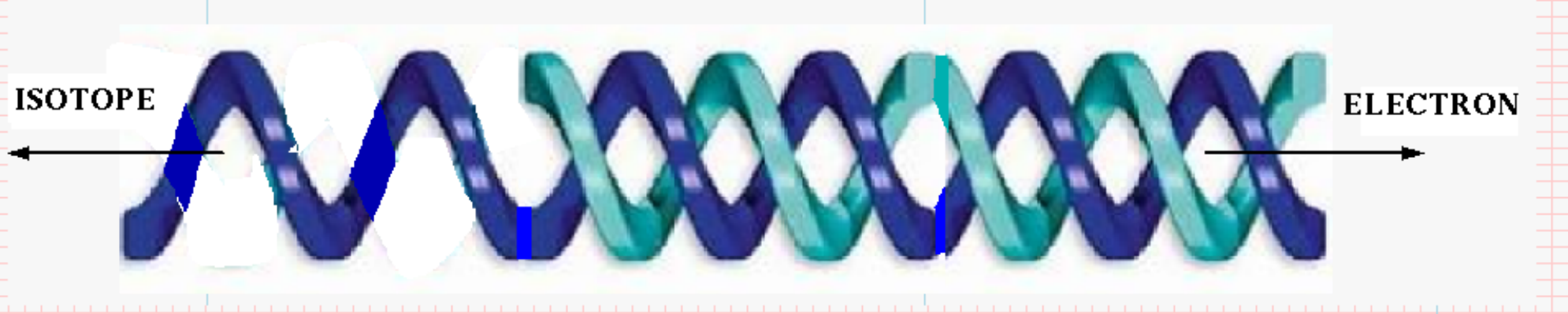}
\caption{\label{fig:Double-helix}Counter-traveling helical orbits of a single nuclear isotope and a single electron
winding around a common ``guiding center'' (not shown).  The intertwined helical coils represent stationary helices
in space. These are the orbits followed by individual electrons and ions.}
\end{figure}

Figure~\ref{fig:Double-helix} provides an artists conception of counter-traveling double nuclear isotope
and electron helical orbits. Their orbits are helical, as shown, circling around a ``guiding center'' that matches a
magnetic field line. For the plasma to be neutral requires an electron current, equal in average charge density,
though not in bunching, nor in local current density.  The figure is also misleading in suggesting electron and ion
pitch angles and radii are equal. Furthermore, electron and proton encounters are not necessarily common.

Other than causing insignificant dE/dx slowing down of the isotopes, the electrons would seem not to be
especially important.
There is, however a problem with this. With the electron to proton mass ratio being so large, the electron current
needs, at least temporarily, to be much larger than the isotope current, This discussion is continued below. 

Another confusing feature of the helical orbit concerns the pitch angle of the isotope helix. As drawn, the longitudinal
advance per turn is roughly equal to the diameter, meaning the tangent of the pitch angle is $2R/(2\pi R)$,
meaning the pitch angle is $\arctan(1/\pi)$. Both transverse and longitudinal velocities would then
be non-relativistic. With sufficiently large sustained longitudinal electric field, the longitudinal velocity
can become relativistic.  Then the ion pitch angle would become large compared to 1 and the helix would
be greatly elongated.

As complicated as it is, Figure~\ref{fig:Parker-solar-wind} requires some qualification in order to
be comprehensible. Though a nuclear isotope orbit is sometimes said to coincide with a magnetic field line,
\emph{this is categorically incorrect.}  The isotope orbits are actually helical, winding around a ``guiding center'', just
as in the earth's Van Allen Belts.  It is the guiding center that coincides with a magnetic field
line, as illustrated in Figure~\ref{fig:Double-helix}. 

\subsection{Rotational centrifugal potential energy}
It is not likely for weakly-bound charged isotopes traveling rapidly in non-vanishing magnetic fields to
break-up due to their rotational centrifugal potential energy.  Certainly, however, high velocity
neutral atoms are subject to becoming fully ionized. This section addresses this issue. 

Landau and Lifshitz, \emph{Mechanics}\cite{L-and-L-non-inertial} describe motion in a non-inertial frame of reference.
This is germane here because of the angular acceleration of nuclear ions in their helical motion relative to the
Alfv\'en guiding center in the Parker solar wind. There is a hierarchy of frames complicating this issue.
The guiding center motion on a gently curving path is mildly non-inertial, but this non-inertial deviation is
weak and can be ignored. The helical orbit, like a helical staircase favors non-Cartesian coordinates, but is inertial,
since it is not moving. However, since each nuclear ion moves along this tightly-wound helix, from the point of view
of internal nuclear dynamics, the frame of reference is non-inertial, with strength inversely dependent on helical
gyro-radius..

The Lagrangian and Hamiltonian treatment of guided-wave orbits is discussed in reasonably simple terms by
Talman\cite{Talman-Mechanics-Chapter-14} and in more authoritative and clearer terms by
J. Cary and A. Brizard,\cite{Cary-Brizard}.  For convenience a brief explanation is included as
Appendix~\ref{sec:Cary-Brizard} to the present paper.  To simplify comparison of these two sources, several important
general formulas in early pages of Cary and Brizard have been copied and specialized to match formulas in the
Talman version. Equations common to both papers are identified by matching, double, reference numbers.

Since protons are stable under all conceivable circumstances, and their radii are small, their enforced rotation around
a more or less vertical axis has little effect. But other nuclei, even as small as $\alpha$-particles, have significantly
larger radii than the proton, and their enforced rotation induces centrifugal force which has the effect of weakening their
internal mutual attractions.  This mutual repulsion becomes stronger and stronger with increasing mass number, $A$; but so
also does their mutual nuclear force attraction.

Relative to the CM of an ion, during a rotation through infinitesimal angle $d\phi$, a point's displacement
${\bf dr}$ is  $d{\bf R} + {d\phi\times {\bf r}}$ , where {\bf R} is the constant length but variable direction gyro-radius
vector.  Defining velocities ${\bf v} = d{\bf r}/dt$, and ${\bf V} = d{\bf R}/dt$, and  angular velocity
d${\pmb\phi}/dt = {\bf\Omega}$, the total velocity is 
\begin{equation}
{\bf v} = {\bf V} + {\bf \Omega} \times {\bf r}.
\label{eq:total-vel.1}
\end{equation}
In an arbitrary frame of reference the Lagrangian for a point particle is the difference between
its kinetic and potential energies. In an inertial frame, call it $K_0$,
\begin{equation}
L_0 = \frac{1}{2}m v_0^2 - U_0.
\label{eq:total-vel.3}
\end{equation}
Landau and Lifshitz proceed to show that that, for a one particle system, there is always a
frame of reference in which, other than for its instantaneous rotational state the particle is
at rest.  Its linear and angular momenta in any other inertial frame can then be obtained
by Galilean or Lorentz transformation, as appropriate.  Its energy, however, depends
on its rotational state
\begin{equation}
E = E_0 - {\bf M}\cdot {\pmb\Omega},
\label{eq:total-vel.3}
\end{equation}
where ${\bf M}$ is its angular momentum and ${\bf\Omega}=d{\pmb\phi}/dt$ is its angular velocity.
The fact that its energy is diminished (by the subtraction) reflects the fact that any net binding
force is opposite in sign to the centrifugal force; this consideration is implied by the name
``centrifugal potential energy''.

This is directly applicable to the solar wind. As each nuclear ion rotates in its helical orbit
around its guiding center, the centrifugal potential energy has the effect of \emph{reducing} the
ion's nuclear binding energy.

\begin{figure}[hbt!]
   \centering
   \includegraphics[scale=0.60]{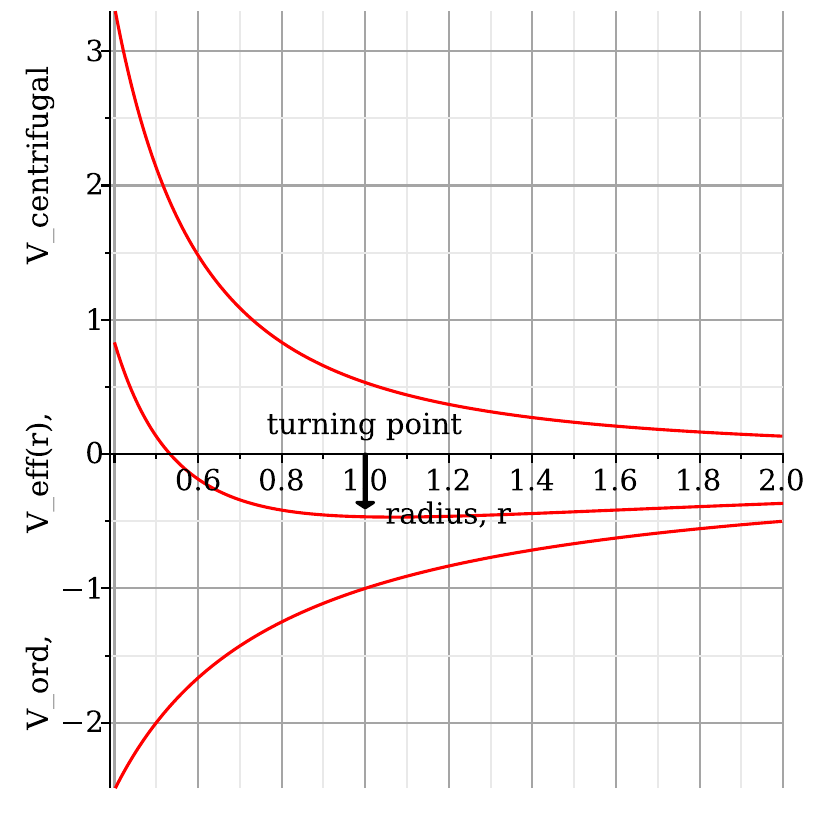}
   \caption{\label{fig:Centrifugal-potential-new} The Coulomb repulsion between positive nuclear
     isotopes is represented by positive potential energy depending inversely on separation distance $r$.
     Even in the absence of attractive nuclear force, this radial divergence is suppressed by a
     kinematic ``centrifugal repulsion'', resulting in a convergent
     effective potential,as shown in this figure. 
   }
\end{figure}

Protons are the first and only significant isotopes escaping from the solar system, accompanied, on the
average, by electrons, as required by charge neutrality.  Fractionally this loss rate is small enough to
allow our solar system (and other star systems) to be characterized as being in similar states
of ``semi-equilibrium''.  Of the nuclear isotopes it is only
the protons for which escape is likely.

Except for slowing down caused by the highly-rarefied electron
population of outer space, in equilibrium the escaping protons are largely canceled by protons impinging
from all the other stars in the galaxy.  It may not be very wrong to simply approximate the ``escaping'' protons
as being ``reflected''.  Certainly this misrepresents ``cooling'' but, in spite of their predominant number
density, the total energy of escaping protons is tiny compared to the total energy of all the
isotopes that remain captured.

Occasionally even a deuteron may escape, in spite of its doubled mass. But it will, on the average, be
replaced by an incoming deuteron.

This paper is primarily concerned with astrophysical issues;
especially as applied to nuclear transmutation of light isotopes. It is rear-end collisions of fast 6Li isotopes
as they periodically pass through slow deuterons, that enables the, previously impossible, direct laboratory
observation of, for example, the  $${}^6_3{\rm Li} + {}^2_1{\rm H} \rightarrow {}^4_2{\rm He} + {}^4_2{\rm He}$$
process.  Included is the track detection of the final state $\alpha$'s, as well as the capability of controlling
the initial state polarizations.  What makes this capability especially simple is the fact that the emitted
$\alpha$-particles are spinless.  On the other hand, as emphasized elsewhere in the papers, with both beams
highly polarized, and both incident beam spin orientations controllable, the scattering kinematics is both
well known, and accurately reproducible.

\section{Nuclear isotope abundances}

\subsection{Nuclear isotope abundance ratios}
\begin{figure}[hbt!]
  \centering
\includegraphics[scale=0.78]{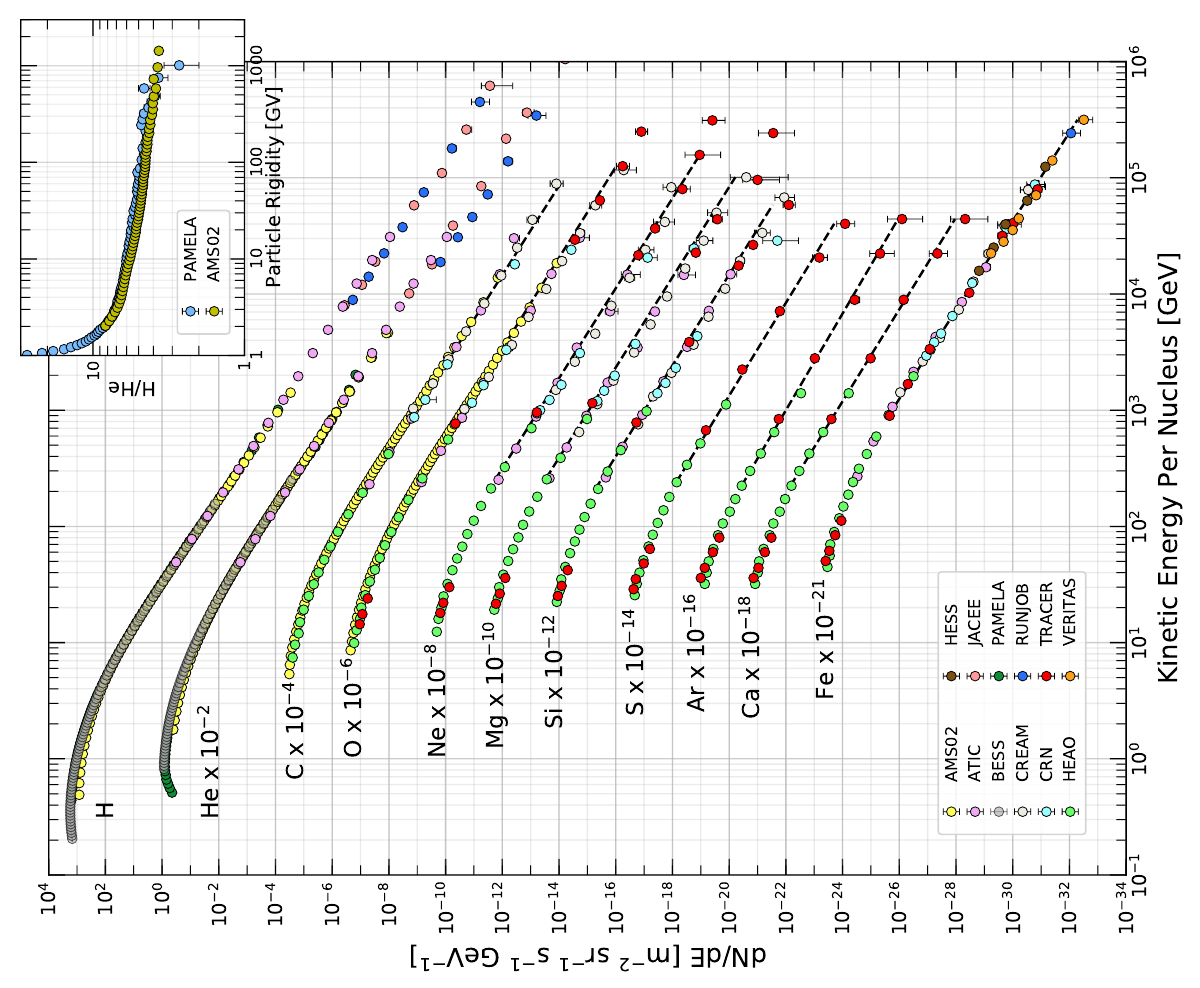}
\includegraphics[scale=0.39]{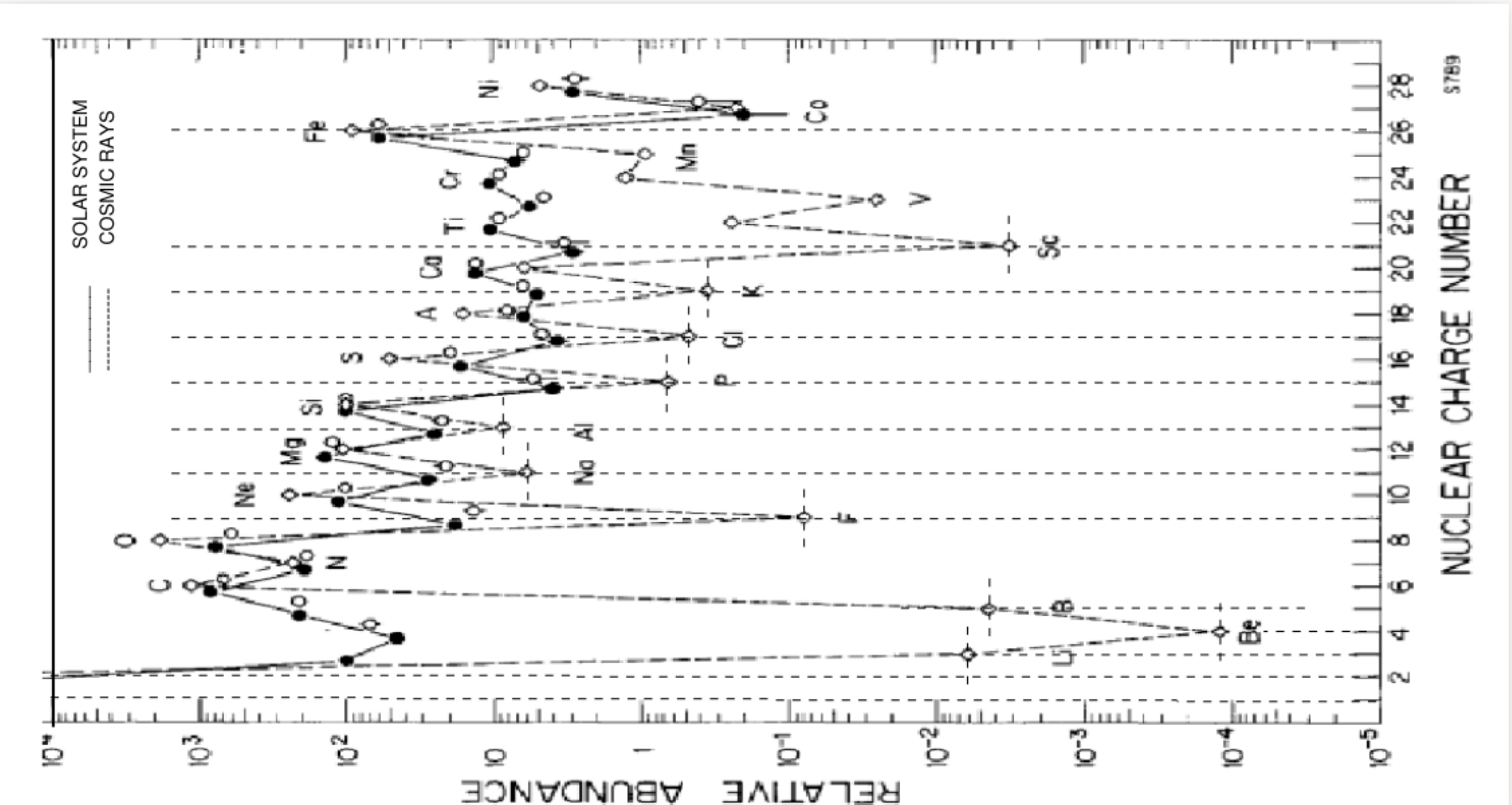}
\caption{\label{fig:Simpson-Cosmic-Rays-annotated}
The two figures on this page are explained and referenced in the text.
The lower figure is copied from a Simpson paper~\cite{Simpson}, based largely upon
data of Cameron\cite{Cameron-Fowler-1971}.
(There is a potential typo in the legend: lines joining the ``cosmic ray data points
may appear continuous at poor resolution.) Vertical broken lines in the lower figure are
drawn though nuclear abundance values that are small compared to both of their nearest
neighbors, as well as beryllium---almost the only (but the most significant) exception.
In spite of its altered format, the upper figure, copied from
reference~\cite{Particle-Data-Group-Chap-29-Cosmic-Rays}, can be regarded to be a
modernization (and at least superficial confirmation) of the data in the lower figure.  
}
\end{figure}

\begin{figure}[hbt!]
\centering
\includegraphics[scale=0.52]{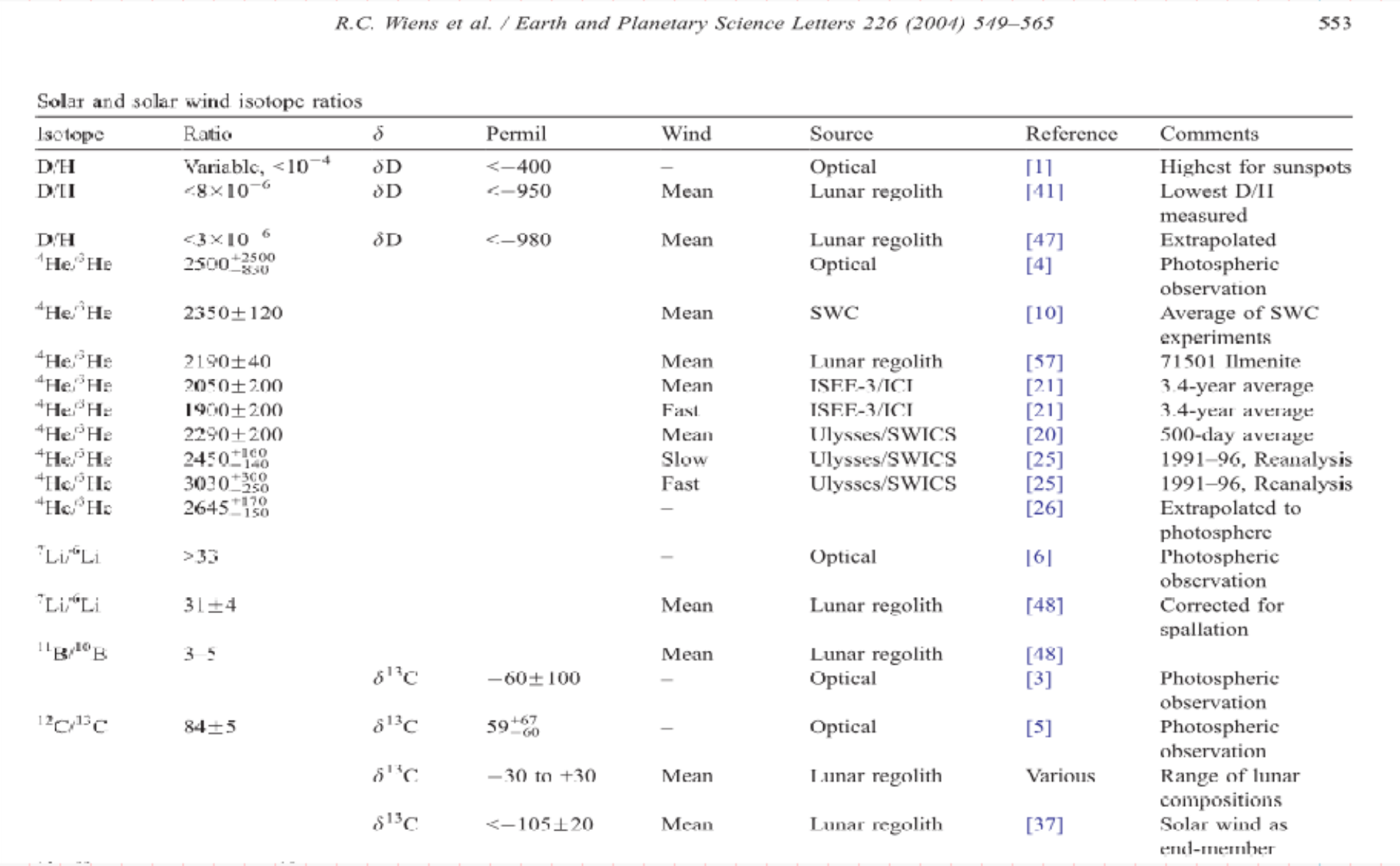}
\caption{\label{fig:Wiens-solar-wind-ratios}
  Figure (copied from Wiens)\cite{RC-Wiens}) exhibiting solar wind fractional isotope ratios.
  Especially noteworthy are vanishingly small D/H ratios, the huge ratios with ${}^4$He in the numerator, the
  large ratios with ${}^6$li or ${}^{10}$B in the denominator, or ${}^{12}$C in the numerator, and the
  complete absence of Beryllium entries.
}
\end{figure}

The lower plot in Figure~\ref{fig:Simpson-Cosmic-Rays-annotated} provides a circa 1980 measured comparison
of solar system and cosmic ray abundances of elements, Fe and below in the periodic table. It is an
annotated version of a graph copied from Simpson\cite{Simpson}.  
(The assignment of data labeled ``cosmic ray'' to galactic or extra-galactic parentage of this data
is disputed in this present paper) The criterion for plotting the broken straight lines is given later,
along with explanation of ways these lines are to be interpreted.  Figure~\ref{fig:Wiens-solar-wind-ratios}
exhibits similar, but differently obtained, NASA satellite ratios in tabular form.

\begin{figure}[hbt!]
  \centering
\includegraphics[scale=0.50]{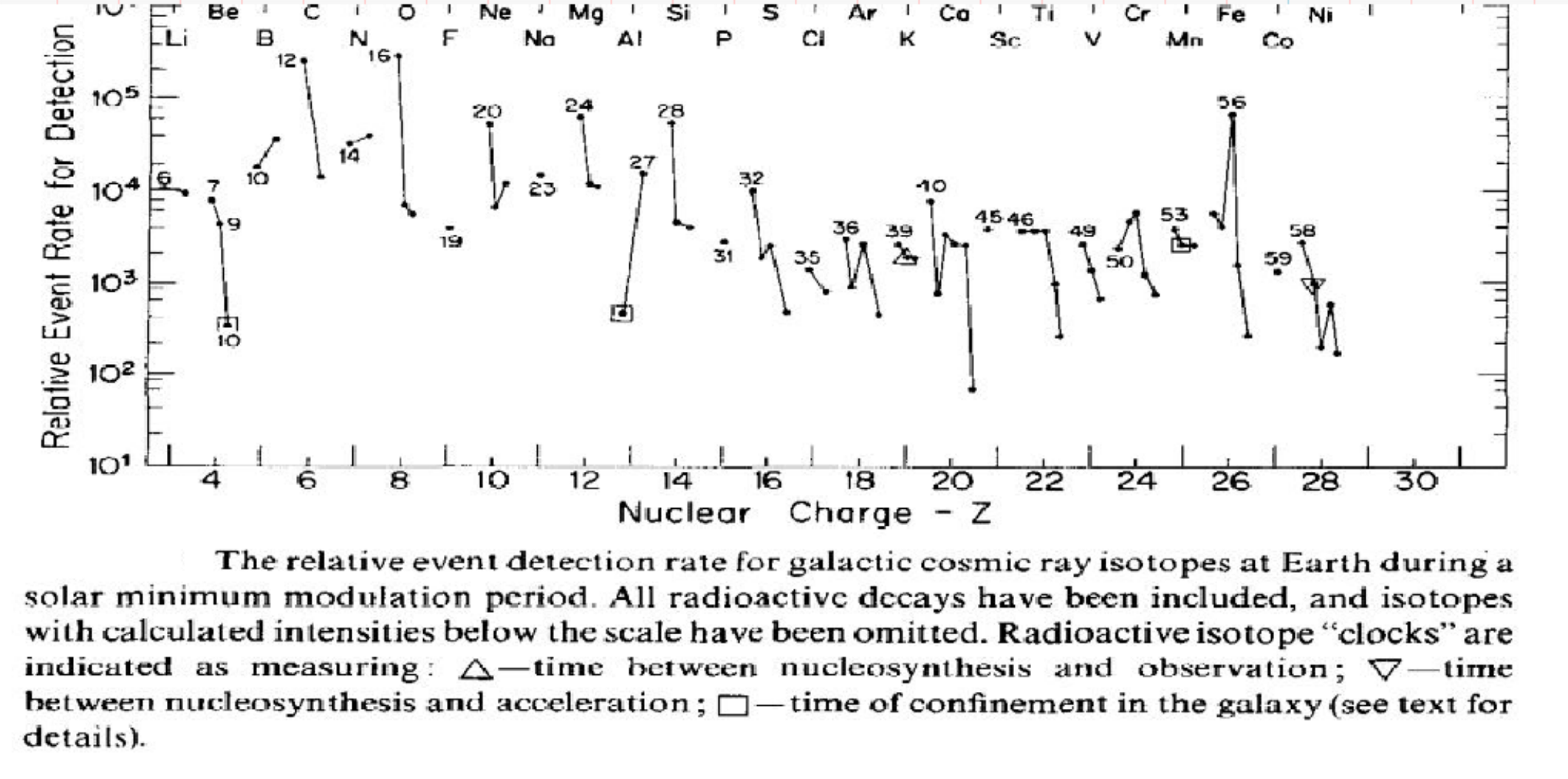}
\caption{\label{fig:Simpson-Cosmic-Rays-reactions}
Matched with the lower plot of Figure~\ref{fig:Simpson-Cosmic-Rays-annotated},
and including the original caption, radiative decays to adjacent elements are exhibited.  
}
\end{figure}
  
\clearpage

\subsection{Solar and cosmic isotopic abundances}
Cosmic ray nuclear isotopes emerging from the sun are predominantly centered in directions close to
the sun's equatorial plane.  So also are the sun's planets.  Nevertheless, because the solid angle subtended by
individual planets is fractionally so small, only a minuscule fraction of the solar wind can
be responsible for synchronous planetary effects.

In any case, the most dramatic observable manifestations of the solar wind are centered in directions toward
individual planets.  Any significant influence upon isotopic particle abundances may plausibly be associated with
the interaction between a planet and the solar wind.

Though multi-MeV cosmic ray nuclear isotopes are ultimately transported by the solar wind, it
seems unlikely for isotopes coming to the earth directly from the sun to be much accelerated
by shock waves associated with the sun. This is because the solar wind acceleration begins only
at far greater distance from the sun than the distance from the sun to the earth.

The introduction of $\alpha$-media has been motivated by the lower plot in
Figure~\ref{fig:Simpson-Cosmic-Rays-annotated} which has been copied (with added annotations)
from Simpson\cite{Fowler-70th}. This data is explained in greater detail by Cameron\cite{Fowler-70th}.  
According to the original legend shown in the upper-right corner, large solid circles joined by solid lines
represent ``SOLAR SYSTEM'' isotopic abundances, small solid circles joined by broken lines represent
``COSMIC RAYS''. As collected primarily by Cameron between 1960 and 1980, ``SOLAR SYSTEM'' meant, primarily,
isotopic abundances measured in, oceans, lakes, and in ores from mines of all kinds on earth.  ``COSMIC RAYS''
were measured ``coming out of the sky'' and \emph{were broadly interpreted as extra-galactic in origin.}.
This paper argues that \emph{the italicized extra-galactic interpretation was simply wrong.}

Though this data represented elements ``coming out of the sky'', the suggestion here is that they are
primarily produced within the solar system. At least this is the contention of this paper.

The upper plot in Figure~\ref{fig:Simpson-Cosmic-Rays-annotated} is a gloriously improved version of
the lower plot, as determined by the AMS Experiment aboard the International Space Station (ISS).
Results from various of their publications have been reproduced in
Reference\cite{Particle-Data-Group-Chap-29-Cosmic-Rays}.

Figure~\ref{fig:Wiens-solar-wind-ratios} shows isotopic abundance ratio data derived from atomic analysis of
optical spectra of light from the sun (optical) or lunar soils mineral grains from gas-rich
meteorites which were exposed to solar wind (planetary regolith) obtained with NASA satellite
from 2004, which the Wiens et al.\cite{RC-Wiens} authors refer to as ``Solar and solar-wind isotope-ratios''. 
The first two sentences of their paper read

``With only a few exceptions, the solar photo-sphere is thought to have retained the mean isotopic composition
of the original solar nebula, so that, with some corrections, the photo-sphere provides a baseline for comparison
of all other planetary materials. There are two sources of information on the photospheric isotopic composition:
optical observations, which have succeeded in determining a few isotopic ratios with large uncertainties, and
the solar wind, measured either in situ by spacecraft instruments or as implanted ions into lunar or asteroidal
soils or collection substrates.'' ``Regolith'' is the result of the impact of meteoroids as well as charged
particles from the Sun 

Copying a few entries low mass nuclear isotope ratios to isotope or atomic rates are: 
\begin{align}
  \hbox{Variable\ }  D / H\         &<10^{-6}    \quad   &\hbox{Optical}        \notag\\
             {}^4He / He\           & 2190      \quad   &\hbox{Lunar regolith} \notag\\
             {}^7Li / {}^6Li\       &>\  33     \quad   &\hbox{Optical}        \notag\\
             {}^7Li / {}^6Li\       &=\  31\pm4 \quad   &\hbox{Lunar regolith} \notag\\
             Be:                    &\ \hbox{none\ detected} &                 \notag\\
             {}^{11}B / {}^{10}B\    &=\  3-5    \quad   &\hbox{Optical}         \notag\\
             {}^{12}C / {}^{13}C\    &=\ 84\pm5  \quad   &\hbox{Optical}        
\end{align}
Noteworthy here are:

\begin{align}
\hbox{\quad (negligible\ ratio\ of\ deuteron\ atoms\ relative\ to\ neutral\ hydrogen\ atoms,})  \notag\\
\hbox{\quad huge\ ratio\ of\ charged\ helium\ nuclei\ compared\ to\ helium\ atoms,}  \notag\\
\hbox{\quad   small\ denominator of ${}^{6}Li$ nuclear isotope,}                      \notag\\
\hbox{\quad   no\ beryllium\ nuclei\ nor\ atoms\ detected,}                             \notag\\
\hbox{\quad   large numerator of ${}^{12}$C nuclear isotope.}                           
\end{align}
               
One might argue that it would have been simpler, in the present paper,
instead of introducing $\alpha$-material, to have introduced deuteron-material. This would be less effective,
however, for three reasons: $\alpha$ particles seem more robust and symmetric than deuterons, and, more
importantly, the six consecutive doubly-spaced lines, running from Z=9, fluorine, to Z=21, scandium, would
not have been noteworthy and surprising if a transition producing a single deuteron were allowed to be
emphasized by a broken-line. This is the only way to distinguish the single line spacing (on the left and right
of the plot) along with double spacing (in the middle). 

A feature of the solid line data that is not obvious at a glance, is the so-called ``odd-even effect'';
the even-Z data points are systematically \emph{above} both of their nearest odd-Z data points. In the
hundred or so ``solar system'' data points there is only one exception to this rule; it is beryllium, with Z=4,
and less abundant than both of its nearest neighbors. (In this particular plot this deviation is masked by
optical illusion resulting from the upper point-joining line and the lower annotation line
\emph{seeming to be} identical).

\section{Some features of stars}
\subsection{Temperature and energy, proportional, but not equivalent \label{sec:TempEnergy}}
It is thought that solar nuclear isotope ratios in effect $5\times10^9$ years ago have almost all been accurately
determined by modern day experiments. But, in the period from their creation  until the present, materials
containing the various isotopes have undergone vastly different histories, during which the relative abundances
of different isotope pairs (even of perfectly stable isotopes) could have changed substantially.  Such changes
can be expected to have been correlated especially with charge number $Z$, but also with
other properties such as $A/Z$. For example, massive (large A) material participating
in nuclear bombs or nuclear power production are certain to have changed.  So also has (low $A$) material
participating in a hydrogen bomb or other cosmological explosions.

Such isotope ratio changes are also likely to be correlated with ambient temperature, as measured either
in degrees Kelvin (K), or MeV energy units. (For present mnemonic purposes, the Boltzmann
equation $E = {\rm k_B}T$ equates  a temperature of $10^{10}$\,K to an energy of 1\,MeV.)
Of these parameters, only the magnetic field $B$ can be set to relevant values in laboratory experiments while
temperature can only be inferred by astrophysical measurements, expressed in either temperature T, measured
in K degree units, or as kinetic energy, measured in eV units.
\footnote{In anything resembling a competition between laboratory-based astrophysics, and Astrophysics itself, there is one main way in which they are ``evenly-matched''. It is that the nuclear energies match. In the sun the energies are achieved by heating, while storage ring energies are achieved by passage through RF cavities. In this case, in order of  magnitude terms, extensive variables are related by the ratio of the mass of the sun, to the mass of the storage ring. For producing a suitably large number of events for scientific measurement, the storage ring efficiency easily makes up for this huge difference, even producing energies orders of magnitude higher than can be produced at the center of the sun.  The E\&M storage ring gains other huge advantages in information quality, compared to isotropically produced laboratory rest-frame events, by detecting resonantly-enhanced, differential moving frame cross sections from mono-energetic, collinear incident beams.}
For the present paper the distinction between intensive and extensive variables is especially  important.
In a storage ring it is routinely
possible to raise particle kinetic energies to values unachievable in an isotropic Bolzmann distribution at the
hottest conceivable astrophysical temperature.
\footnote{For bridging low energy and high energy physics there are serious terminology and notational complications. especially concerning energy definitions and symbols. Not counting weak interactions, which are discussed only secondarily in the paper, non-relativistic kinematics, such as $KE=mv^2/2$, along with masses retained to the highest possible precision, would be adequate for the pure nuclear physics. Yet fully relativistic formulas are needed, and used, for accelerator physics, such as spin procession. That is $\mathcal{E}\equiv E_{\rm rel}=\sqrt{p^2c^2 +m^2c^4}$, along with $E_{\rm rel}=mc^2 + KE = \gamma mc^2$, where for semi-relativistic nuclear isotopes $\gamma$ is not much greater than 1, and $\beta=v/c$ is always significantly less than 1. Momentum $p$, as used in Eq.~\ref{eq:B-rho.1} is the true relativistic momentum. Its great virtue is that Eq.~(\ref{eq:B-rho.1}), which is one of the two most important formula in the paper, is valid both classically and relativistically.}

Temperature and energy have the peculiar Boltzmann property of being strictly proportional in spite of having
different physical dimensions.  This is nowhere more significant than in nuclear astrophysics.  Rare, but stable
nuclear isotope species that have come into existence only because of weak interaction beta decay processes survive
comfortably for millions of years at the center of a star where the temperature is 15 million degrees Kelvin.
Even unstable isotopes can regenerate in this environment, though not in an environment where regeneration is too
slow.  Yet the chemically equivalent neutral atom loses its electrons spontaneously at a much lower temperature,
such as 5 thousand degrees Kelvin, a temperature which is readily achievable in a terrestrial laboratory.

What discounts for this extreme sensitivity is the fact that the Boltzmann product $kT$, where $T$ is
the temperature in Kelvin and $k=1.380/10^{23}$ Joules per Kelvin, enters the Maxwell Boltzmann distribution
in the exponential form $\exp(\pm kT/E)$, which dominates the high temperature behavior.  The special power
of Boltzmann's contribution to this distribution function is that
that the energy can include kinds of energy other than $KE$.  Treating rotational
energy as negligible, kinetic energy is the only form of energy that enters storage ring behavior.

What makes the Boltzmann $KE=k_BT$ formula confusing is that $KE$ and $T$ are not commensurate quantities,
since $KT$ is an individual particle property while $T$ is a property of a medium, in the form of a parameter of
a distribution function.  One way that $KE$ and $T$ can be said to be commensurate is that $KE$ could be set, for
reference purposes, to match the energy of the peak energy in the M-B distribution.  

Typical nuclear binding energies are measured in multi-keV or MeV units; typical atomic binding
energies are measured in eV or keV units. This accounts for the hypersensitivity distinguishing between the burning
rates of nuclear isotopes and atomic material.

\subsection{Layering by element type in stars like the sun}
\begin{figure}[hbt!]
  \centering
  \includegraphics[scale=0.6]{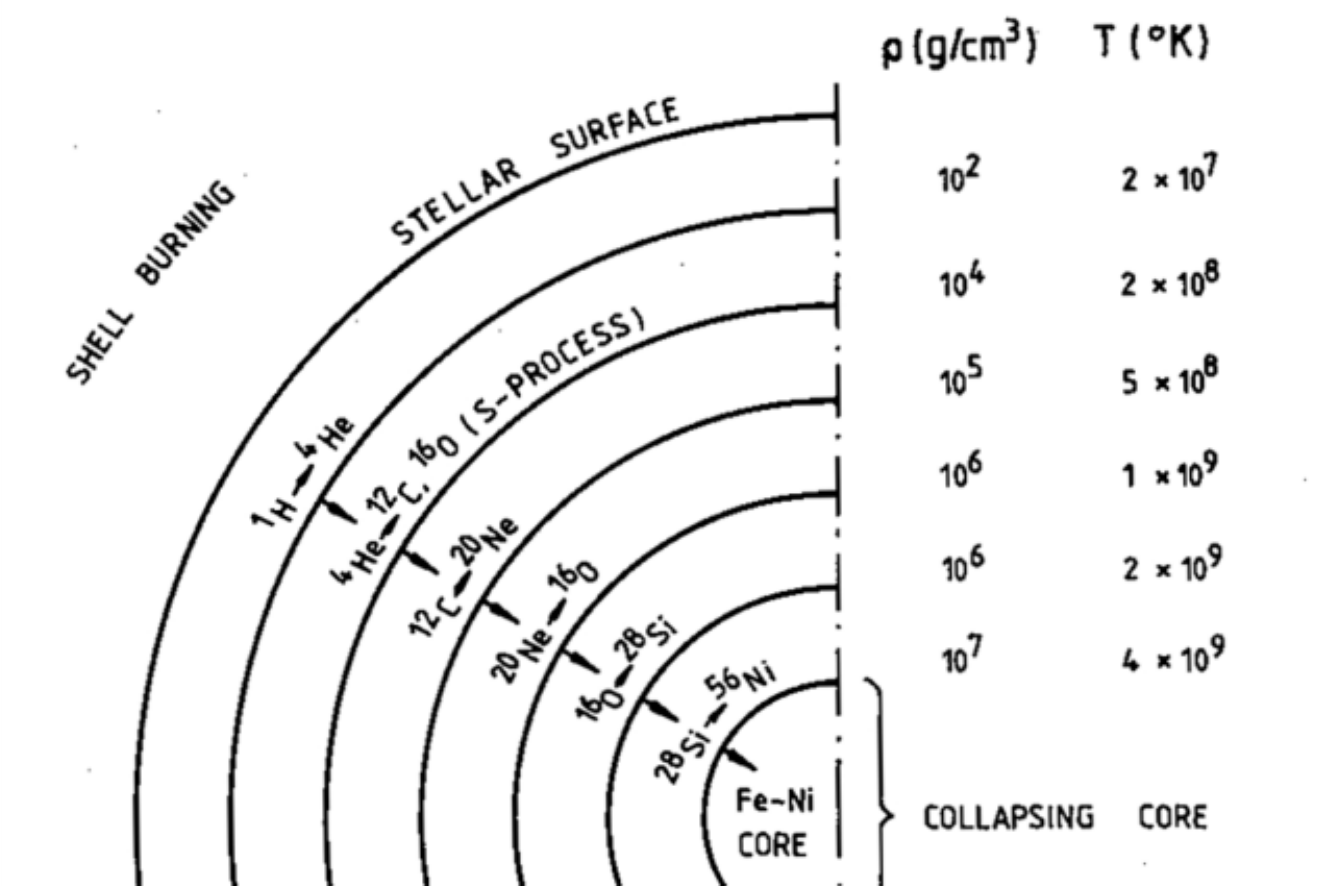}
  \caption{\label{fig:Cauldrons-nuclear-shell-quarter} Clipped from a figure
    in the Rolfs and Rodney ``Cauldrons'' book,  this figure exhibits a theoretical simulation of a so-called
    ``white-dwarf'', ``pre-supernova'' star of mass 1.4 times the mass of the sun.
    What is left of the original light isotopes is in the form of light-element gas
    near the surface of the star.
  }
\end{figure}

So far this discussion has been only semi-quantitative. Figure~\ref{fig:Cauldrons-nuclear-shell-quarter},
copied and sliced from Rolfs and Rodney\cite{Rolfs-Rodney},
shows the ``onion-shell'' structure of a star somewhat more massive (about 4/3 as massive) as the sun.
Very conveniently, this figure, copied from ``Cauldrons'', page 437, provides a pictorial representation
of the dependence of temperature vs radius, that should be similar to the corresponding layering in the sun.

Many solar astrophysical papers have been written suggesting that the planetary structure
of the solar system has been interrupted by ``Giant impacts'' of some sort.  Gabriel and
Camboni\,\cite{Gabriel-Cambioni} have surveyed some of the possibilities. These proposals are too
complicated to be discussed even cursorily here. Rather, a tentative narrative is produced,
arranged to be helpful in interpreting and understanding the physical processes responsible
for the isotopic abundances being discussed in the present paper.

Because Jupiter and Saturn are so massive and so close, it is commonly supposed
that, of individual planets, these were the most probably implicated
in the impacts. Accepting this as given, the assumption here is that some sort of giant impact
participated in the present structure of the solar system, producing the present sun, with the
Saturn-Jupiter combination as the major secondary product and the remaining planets as peripheral
debris. In the present context, Jupiter and Saturn are important because of their large masses
and strong magnetic fields.

Curiously, though having been born from the sun, the magnetic moment of Jupiter, though it
wanders a bit, it does not vary sinusoidaly with time.  On the other hand, the sun's magnetic
fields varies sinusoidally with a period of 22 years. This will be seen below to have significant
importance.

\begin{figure}[hbt!]
\centering
\includegraphics[scale=0.40]{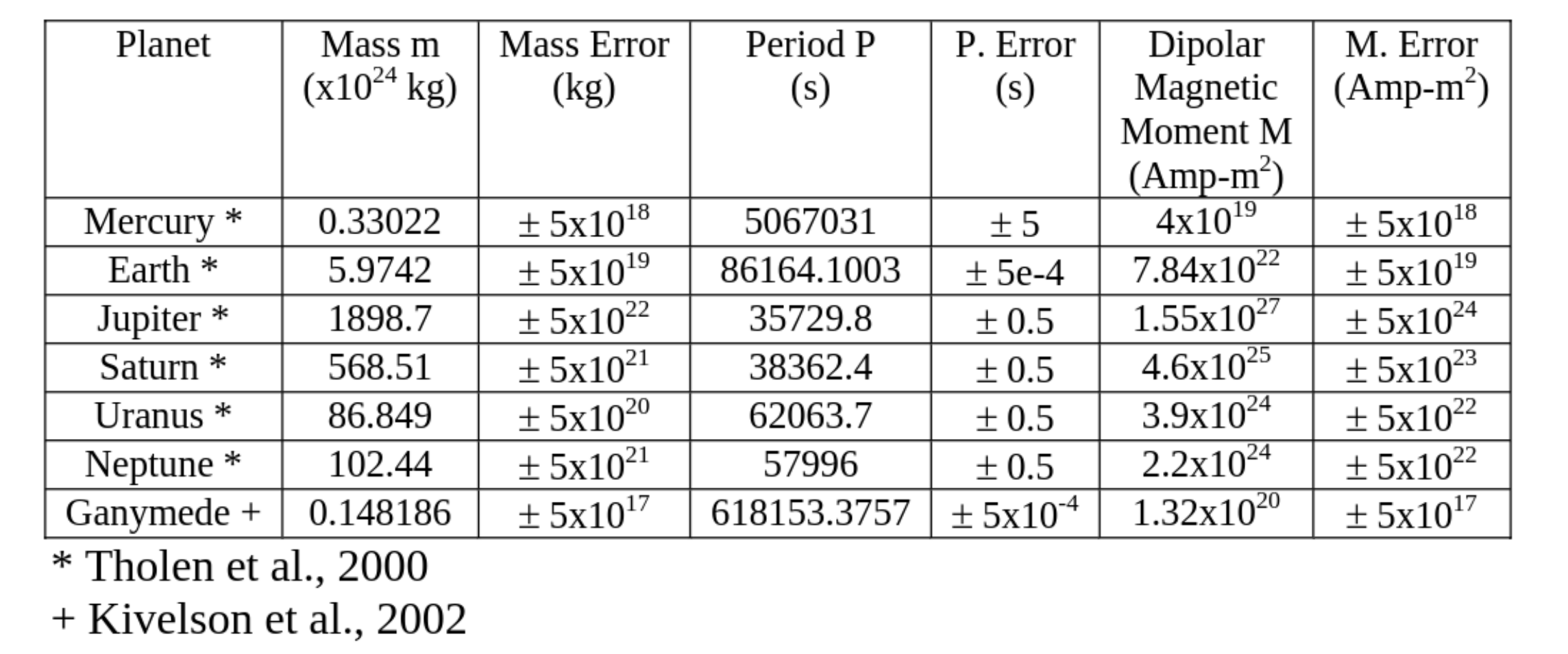}
\caption{\label{fig:Durand-Manterola-planet_MDMs}Mass, orbit periods, P, and magnetic moments, M of the planets.
  Copied from  Biemond\cite{Biemond}.  The sun's dipolar magnetic moment, though highly
  variable, has maximum value of approximately $10^{30}\,{\rm Amp-}{\rm m}^2$.}
\end{figure}

Figure~\ref{fig:Cauldrons-nuclear-shell-quarter},
shows the distribution of elements in radial shells of a white dwarf star.  Since the sun is NOT a
``white-dwarf'' this figure would be irrelevant, except for its elegant display of the radial distribution of
element types, with correlated temperatures and densities, in a star with mass not very different (4/3) from the sun
In this case hydrogen and helium fuel in the core has been exhausted and the only remaining heating is provided
by nuclear fission. By this time the lightest remaining elements, hydrogen and helium, have migrated to
the surface where they reside in ``cool'' areas near the surface of the star. Meanwhile the intermediate mass
isotopes have become arrayed in shells with radii varying inversely with their mass A numbers.

This particular star is destined to produce a supernova explosion, but a slightly less
massive star such as our sun, can be expected to have acquired similar layering.

\section{Magnetic fields in and around the sun and planets}

\subsection{Mass distribution in the sun\label{sec:Sun-mass-distribution}}
\begin{figure}[hbt!]
  \centering
  \includegraphics[scale=0.40]{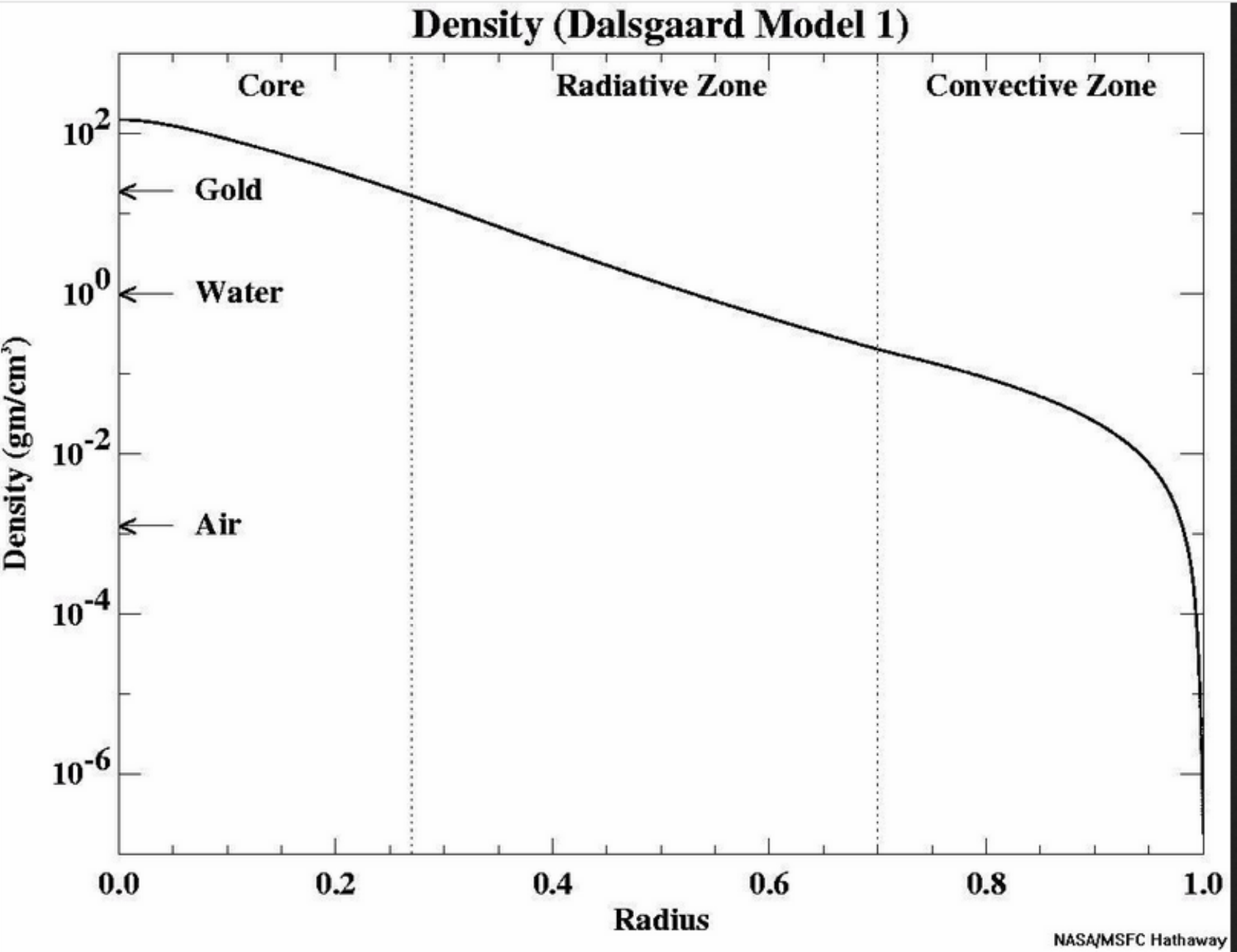}
  \caption{\label{fig:Sun-density-profile}Mass density within the sun, plotted as a function of radius,
  copied from Dalsgaard\cite{Dalsgaard}.}
\end{figure}

The mass density distribution in the sun is shown in Figure~(\ref{fig:Sun-density-profile}).

\subsection{Solar and planetary magnetic moments}
\begin{figure}[hbt!]
   \centering
   \includegraphics[scale=0.45]{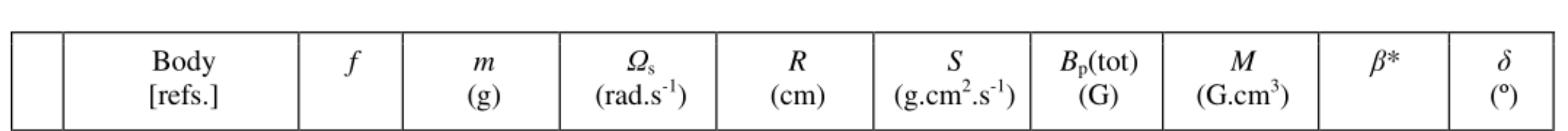}
   \includegraphics[scale=0.45]{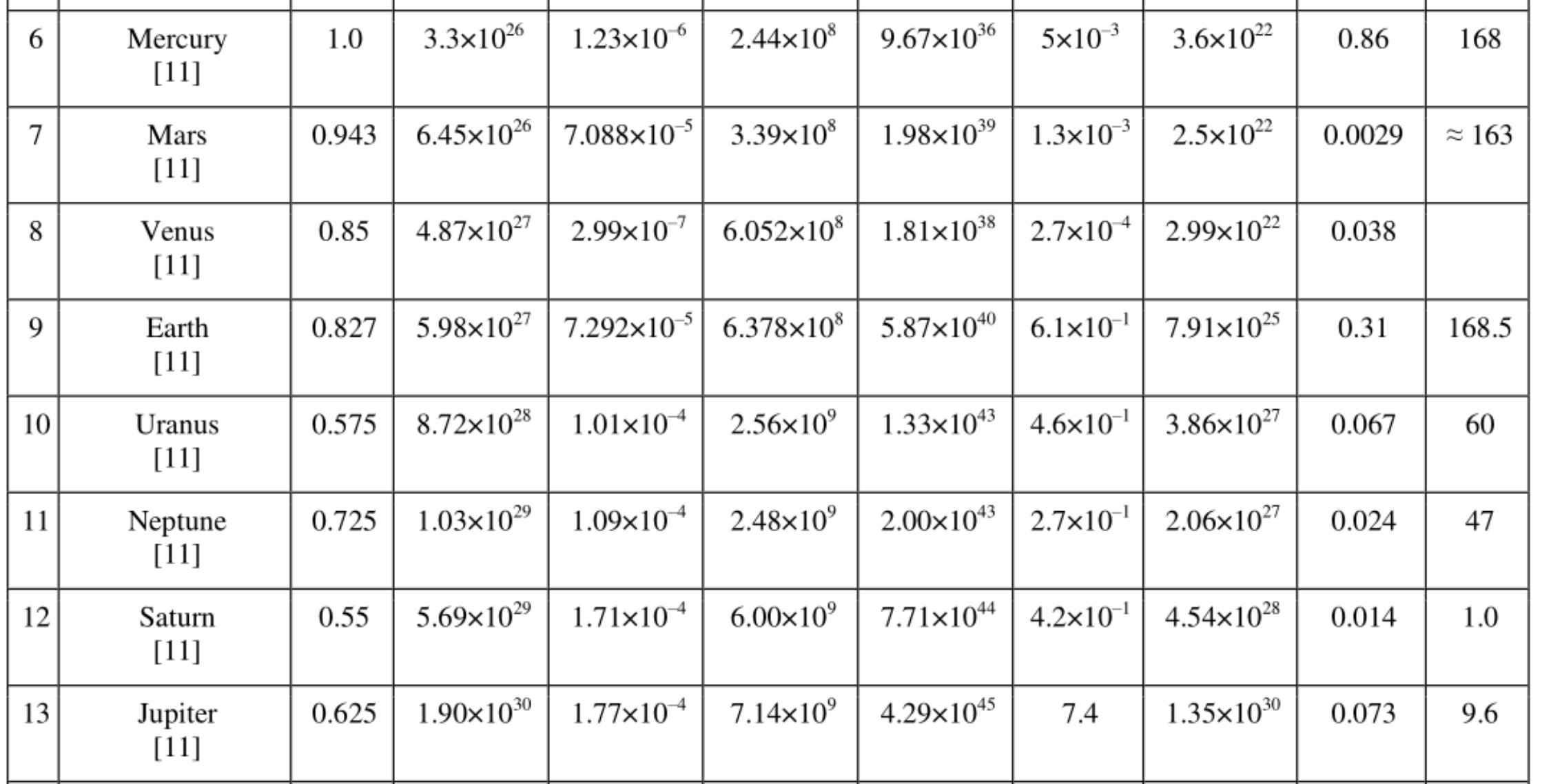}
   \includegraphics[scale=0.45]{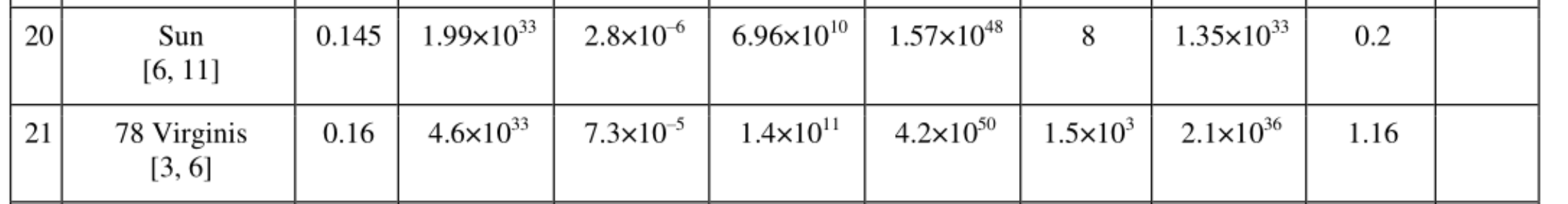}
   \caption{\label{fig:Solar-system-parameters}Planetary parameters, copied from Biemond\cite{Biemond}.
     Angular velocity $\Omega$, magnetic moment $M$, and angular momentum $S$, are especially
     important for issues concerning solar magnetic fields.}
     \end{figure}

\begin{figure}[hbt!]
   \centering
    \includegraphics[scale=0.6]{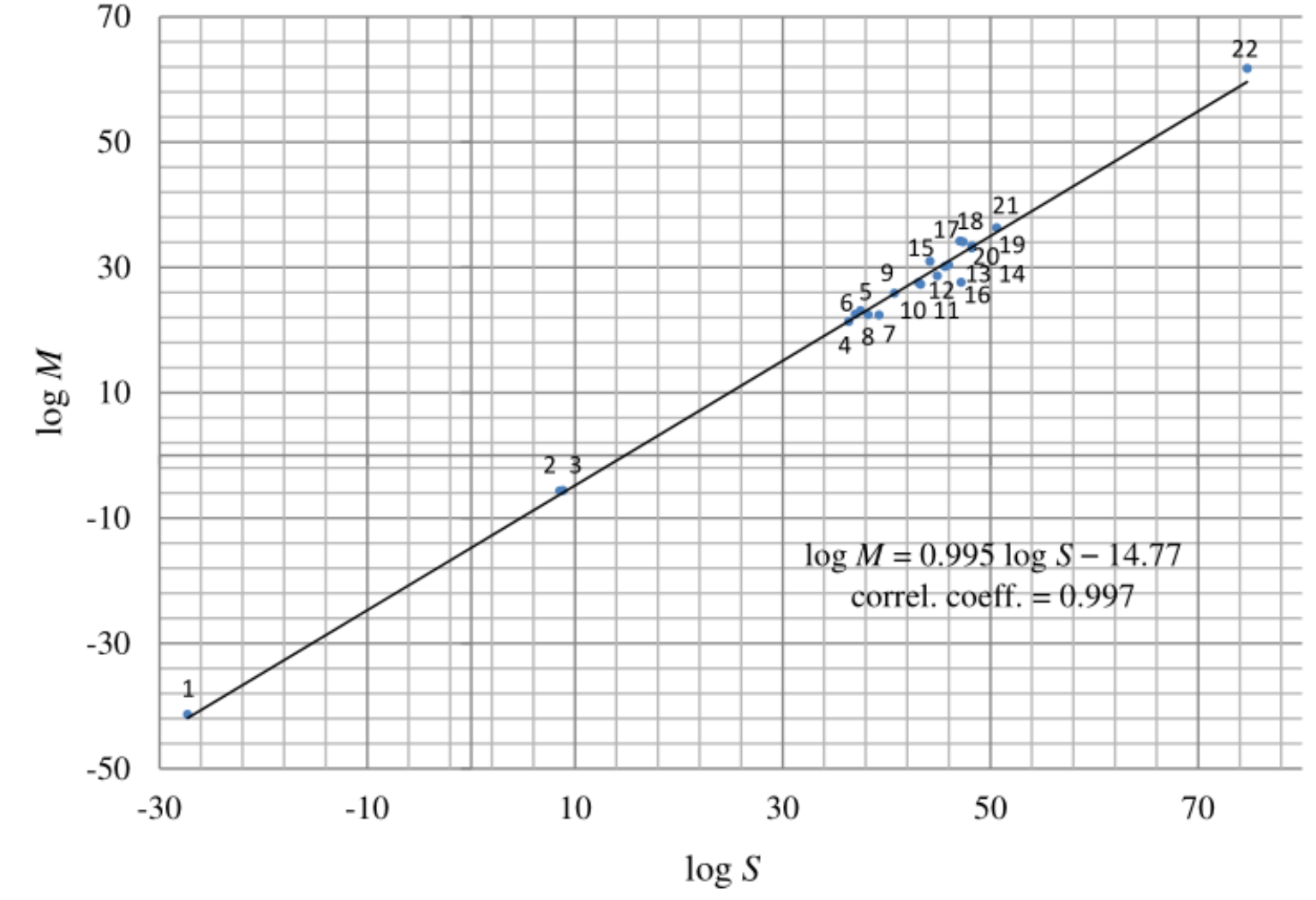}
    \caption{\label{fig:log-log-plot}Log-log-plot copied from the same
    Biemond paper as Figure~\ref{fig:Solar-system-parameters}.}
\end{figure}

The sun, itself, as well as the major planets, can serve as predominantly magnetic accelerators (i.e. with electric
longitudinal Parker acceleration or deceleration).  Relevant planetary parameters are given in
Figures~\ref{fig:Durand-Manterola-planet_MDMs}, \ref{fig:Solar-system-parameters}, and \ref{fig:log-log-plot}, 
copied from Biemond\cite{Biemond}.  These plots exhibit experimental evidence of correlation between magnetic
moment $M$, and angular momentum $S$ that is currently not understood theoretically;
\begin{equation}
{\bf M} = -\frac{1}{2}\frac{\beta^*}{c} \sqrt{G}{\bf S}, 
\label{eq:Blackett-empirical}
\end{equation}
where $\beta^*$ is a dimensionless empirical fitting factor
(not to be confused with relativistic factor $\beta$), and G is the gravitational constant.
The other factors are defined in Figure~(\ref{fig:Solar-system-parameters}).
The validity of this formula can be judged by the log-log plot in Figure~((\ref{fig:log-log-plot}).

For purposes of the present paper, the appearance of gravitational constant $G$ in
Eq.~(\ref{eq:Blackett-empirical}) is not particularly significant; it's presence is needed
only to understand the proportionality of {\bf M} and {\bf S} in Eq.~(\ref{eq:Blackett-empirical}),
without having to wrestle with the units of G. More noteworthy is that the numerical values of
$\beta^*$ range over three orders of magnitude.

The angular momentum {\bf S} in Eq.~\ref{eq:Blackett-empirical} for a spherical star of
radius $R$ can be calculated from
\begin{equation}
{\bf S} = I{\bf \Omega} = \frac{2}{3}fmR^2{\bf\Omega} , 
\label{eq:Blackett-empirical.2}
\end{equation}
where $m$ is the mass of the star, $\Omega$ is its angular velocity
and $I = 2/5 f m R^2$ is its moment of inertia'
The factor $f$ is a dimensionless factor depending on the homogeneity of the mass
density in the star (for a homogeneous mass density $f = 1$).
The magnitude of the gravitomagnetic dipole moment $M$ can be calculated from the expression

\begin{equation}
 M = \frac{1}{2}R^3B_p ;
\label{eq:Blackett-empirical.3}
\end{equation}
where $B_p$ is the magnetic field measured, say, at the north pole.

The fact that all points are close to the same straight line in Fig~(\ref{fig:log-log-plot})
is not very significant; this is guaranteed by the presence of the arbitrary fitting factor $\beta^*$.
It can be seen from the second to last column in the table,
that $\beta^*$ is far from constant.  The major content, therefore, is that for empirically adjusted
$\beta^*$, the various proportionalities among $M$, $G$, and $S$,  are satisfied by all bodies listed.

\clearpage

\subsection{Solar magnetic fields}
Magnetic fields in the interior of the sun are hopelessly chaotic. Interior magnetic field structures, referred to as 
``flux tubes'' are distributed throughout the sun's interior. These are in the form of more or less cylindrical, randomly
oriented, structures that are quite well modeled in isolation. Fortunately, for present purposes, for reasons to be
explained, it will not be necessary to understand how the internal stellar matter is organized.  Only the magnetic fields at the
surface of the sun are sufficient, as boundary conditions, to establish the external magnetic fields.  It is,
of coarse, necessary for these surface fields to be reliably known. 

\begin{figure}[hbt!]
   \centering
   \includegraphics[scale=0.33]  {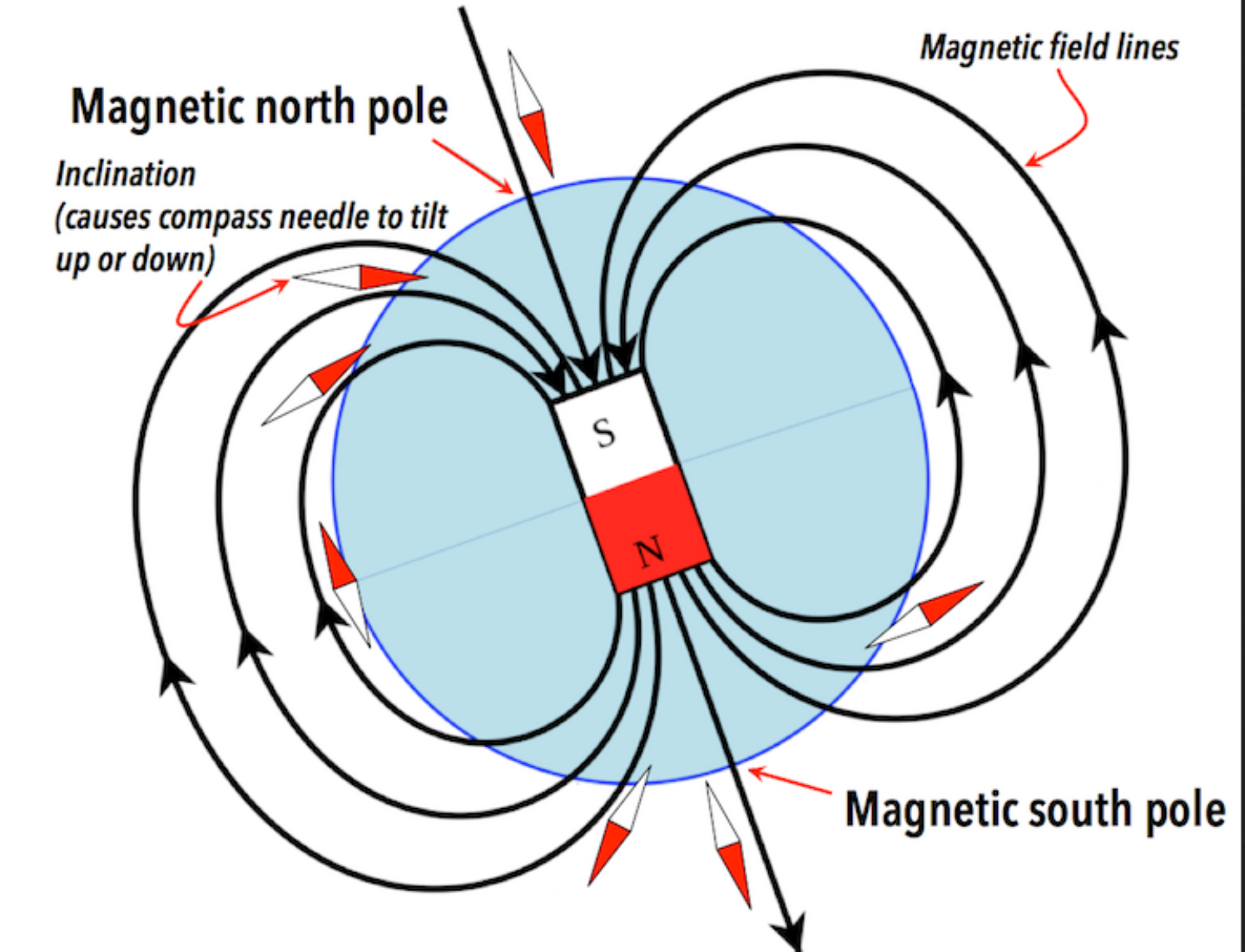}
   \includegraphics[scale=0.33]  {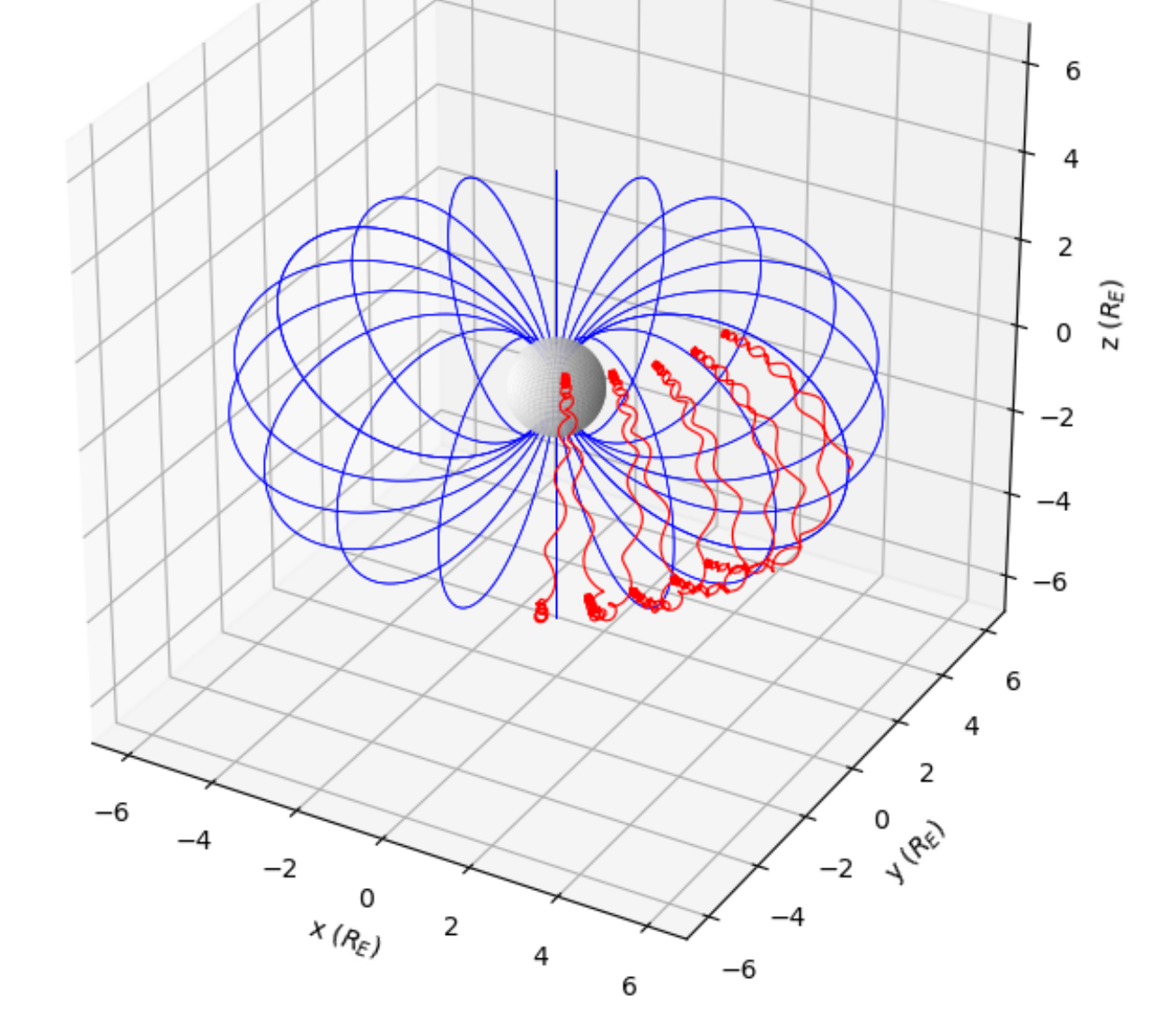}
   \includegraphics[scale=0.45]{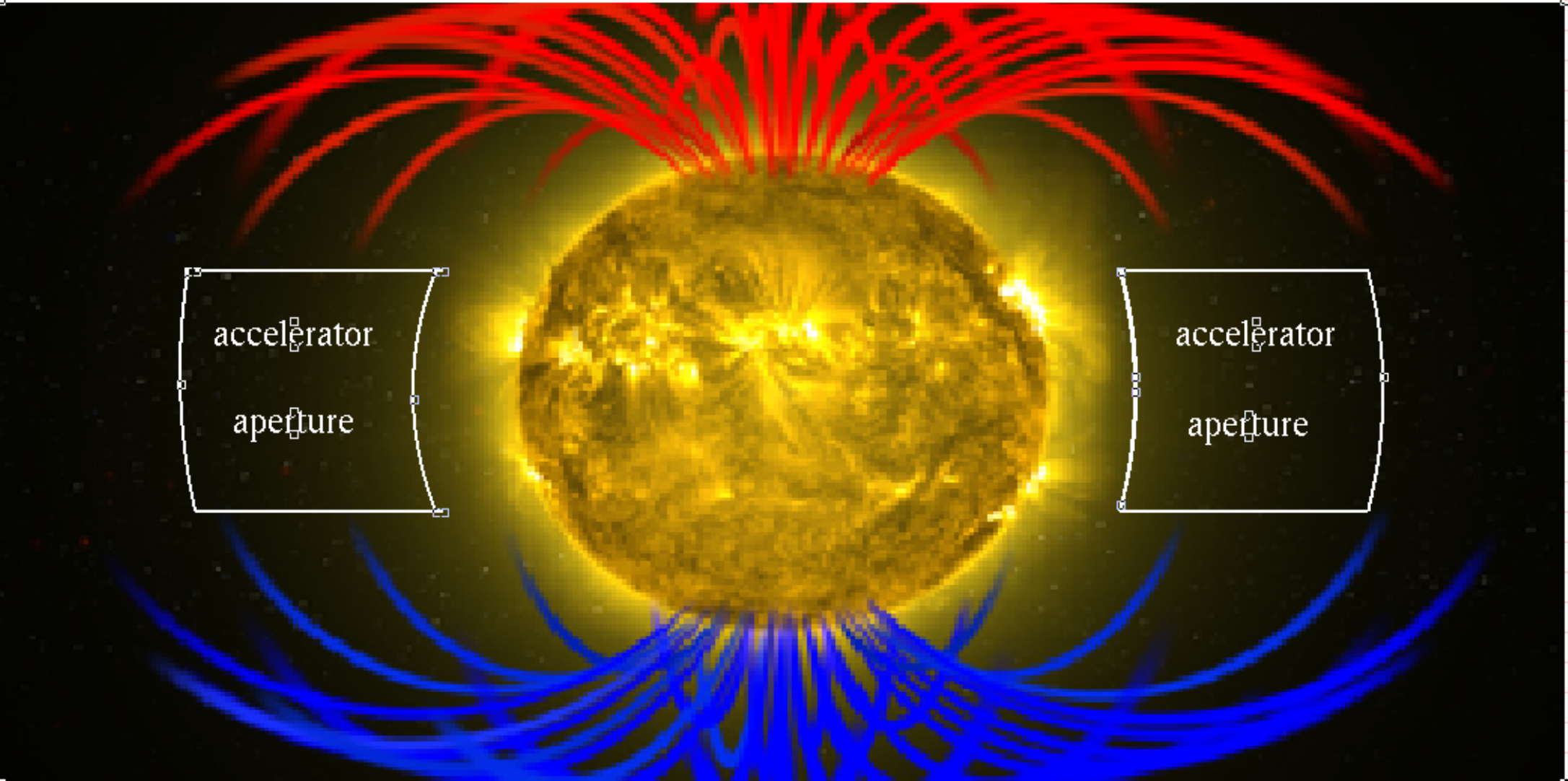}
   \caption{\label{fig:sun-interior-magnetic-field}{\bf Top left:\ }Magnetic dipole field pattern.
     {\bf Top right:\ } Perspective view of dipole field pattern. 16 field lines are shown
     (actually +1, including the straight line from observer's view point).
     {\bf Bottom:\ } 2025 image shows magnetic fields radiating from the sun's poles. 
     Courtesy of NASA's Goddard Space Flight Center.  Superimposed is the outline of the
     aperture of the sun as a particle accelerator.  Interpolated onto the photograph are
     outlines of a virtual vacuum chamber for the sun as particle accelerator}
\end{figure}

\begin{figure}[hbt!]
  \centering
  \includegraphics[scale=0.4]{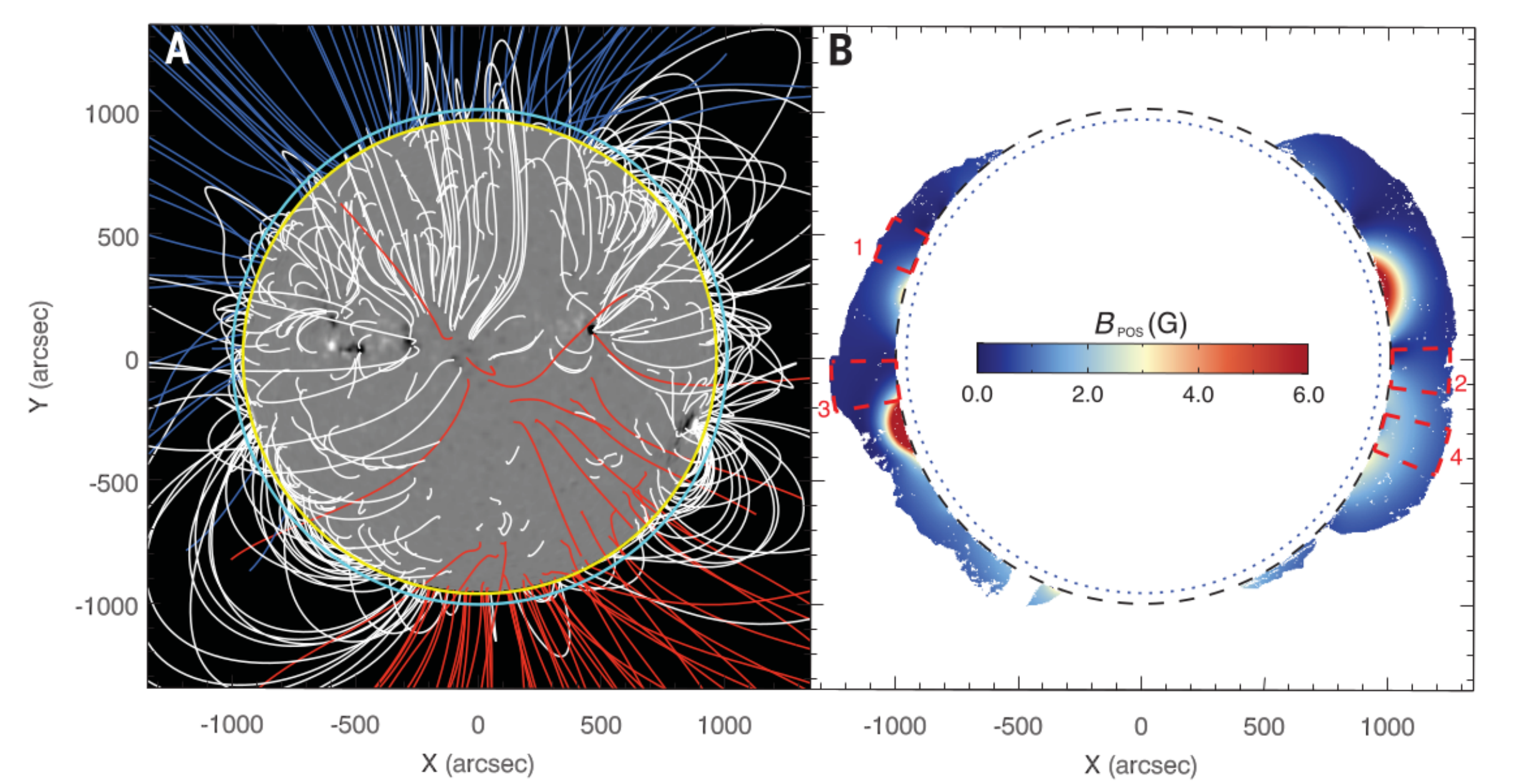}
  \caption{\label{fig:Solar-external-magnetic-field}Copied from a Yang et al.\cite{Yang-Betge} article,
    this figure shows typical magnetic fields  in the solar corona.
 }
\end{figure}

Magnetic fields external to the sun are shown in Figures~(\ref{fig:sun-interior-magnetic-field})
and (\ref{fig:Solar-external-magnetic-field}).  Since magnetic flux lines are continuous the average magnetic
fields, even in the interior of the sun, would seem to be required to match the fields shown in the upper two plots in
Figure~\ref{fig:sun-interior-magnetic-field}.  It is clear, in the bottom picture, that the surface itself
is represented by actual data, made available from the NASA's Goddard Space Flight Center.  From their
description it seems likely that the external magnetic field lines have been constrained to best fit the
available measurements to the theoretically well known magnetic dipole pattern.

As an aside, accepting all this to be true, along with the surface magnetic fields shown in Figure~(\ref{fig:Solar-external-magnetic-field}),
copied from Yang et al.\cite{Yang-Betge}, it suggests that the mean magnetic field along the axis of the core
at the center of the sun is roughly 0.01\,T, and aligned with the axis.

For present purposes it is only the magnetic field in the free space outside the sun that is needed for evaluating the
acceleration of particles during their temporary storage and acceleration while passing by the sun.

\clearpage

\subsection{Disambiguation of the term ``cosmic ray''}
Protons are essentially the only hadronic particles escaping from the solar system. Furthermore, these escaping
protons can be conjectured to be being replaced (on the average) by immigrating protons from other stars.
(To the very limited extent that other light nuclear particles, such as deuterons, escape, they can
be treated in the same way as escaping protons.)
On a hierarchy of  solar, then galactic, then cosmic, sequentially increasing time scales, there can be
balanced immigration and emigration of protons, on solar, then galactic, then cosmic time scales. 

As regards cosmic rays, my arbitrary policy in this paper is constrained to discuss only cosmic ray
acceleration within the solar system treated as a closed and isolated system.  This leaves open the
possibility that a significantly large flux of ultra-high energy protons produced in the galaxy are being
injected into the solar system.

An unfortunate consequence of this policy is that it renders the term ``cosmic ray'' \emph{hopelessly ambiguous}.
Ever since their discovery in 1912 by Victor Hess, it has been mainly assumed that all particles ``coming out of the sky''
originated outside the solar system.  This is very nearly opposite to the policy in this paper, which treats as
``cosmic-rays'' \emph{only nuclear particles created within the solar system which have acquired energies well in access of
their nuclear creation energy, say greater than 10\,GeV, while remaining within the solar system.} 

Though confusing, this is not unprecedented.  It simply means that one must admit the possible existence of two
classes of particles coming out of the sky.  One is ``solar system cosmic rays'', the other  ``non-solar system cosmic rays''.
It seems certain that both classes will consist primarily, of protons, deuterons, and $\alpha$-particles,
in that order. At least for the time being in the present paper, this relieves me from the responsibility of explaining
in more than semi-quantitative detail, a significantly large fraction of the cosmic rays being produced.

Regrettably, since cosmic ray identifications never establish their source,  it must therefore always be accepted that what
has been measured is an incoherent sum of solar and non-solar cosmic rays.

It is my prejudiced belief that most ``cosmic rays'' are ``solar''.

As an aside, it can be mentioned that golf ball sized ice balls routinely come out of the sky
during electrical storms in Texas.  Hundreds of golf ball size pockmarked automobiles guarantee that this is
not ``fake news''.  Apparently this phenomenon is caused, in the presence of electrical storms, by vertical winds, that
temporarily prevent the ice balls from falling. In the presence of water vapor, and extremely low temperature,
it is no surprise that the ice balls grow quickly.

One supposes that a similar phenomenon can occur when the solar wind holds up counter-traveling nuclear ions
long enough for partially charged atomic balls containing radioactive nuclear isotopes to join.  The real mystery
is how such spheres can subsequently retain or recover semi-relativistic velocities high enough to produce
$\mu$-mesons, or other incontrovertible evidence of their nuclear history.

\section{Astronomical bodies as multi-isotope accelerators}
\subsection{Rough parameter values \label{sec:Rough-values}}
During isolated collisions between two nuclear particles the ``strong'' nuclear force competes with
the ``medium strength''electric force.  At nuclear scale the nuclear force wins; at atomic scale
the electric force wins. At astronomical scale, because of charge neutrality (cancellation
of nuclear and electron charge densities to a first approximation) the competition
shifts to gravitation versus magnetism.

As earth observers we are poorly equipped to assess any of these forces.  With magnetic compass we can
detect, in round numbers, a half Gauss earth magnetic field, which is amply strong enough for a
light needle on a decent bearing to measure the field direction---in other words negligibly small
for doing real work.  As it happens, a typical magnetic field at the surface of the sun is
not very different---only several Gauss, with the further complication of varying more or less
sinusoidally with a period of 22 years.

But, by no means, may the sun's magnetic field be ignored,
since the sun's radius exceeds the earth's radius by five orders of magnitude. Furthermore, treated
as a magnetic dipole field, the maximum magnetic field is at the center of the sun.

Using Eq.~\ref{eq:B-rho.1}, storage rings, such as the Parker ring around the sun, or the
earth's Van Allen belts) will store particles with momentum proportional to the ring radius.
With the sun/earth radius ratio being $10^5$, and the momentum of a 2\,MeV electron in the Van
Allen belt around the earth being possible, the momentum of an electron or proton stored in the Parker ring
around the sun could be 200\,GeV. \footnote{In this calculation an electron has been chosen to avoid needing
to use an exact relativistic formula for the deuteron in an only semi-relativistic context.
Both particle energies are large enough for their momenta to be roughly 
proportional to their energy. In both cases referring to spherical bodies as accelerators is misleading.
Actual accelerator have apertures of, perhaps 10\,cm. Astrophysical accelerators require no beam
tubes, and have nearly unlimited (one-sided) apertures.}

Jupiter could also serve as a curiously powerful particle accelerator, as can be seen from the table shown
in Figure~(\ref{fig:Solar-system-parameters}).  Its surface magnetic field is (by $7.4\,{\rm G}/8.0\,{\rm G}$)
nearly the same as the sun, and its radius almost exactly 1/10 as great. Though the radial dependence follows
an inverse cube (1/r³) law for the magnetic field strength, it is effectively constant (weakly defocusing)
for elevations small compared to the Jupiter's radius.  With injection from the solar wind, and bending
by the natural equatorial longitudinal (or ``vertical'' for a horizontal bending magnetic spectrometer
in the equatorial plane).  Along with the latitudinal electric field, provided by the Parker mechanism,
Jupiter can serve as a very powerful cosmic particle accelerator magnet.  

Currently, the highest energy terrestrial accelerator (the LHC at CERN) has a maximum proton energy
of 7000\,GeV, with an average magnetic field of about 6\,Tesla.

Somewhat similar considerations apply to the relative importance of gravitational
and magnetic forces in stars.  Cosmologically natural magnetic
fields, either on the earth's surface or on the surface of the sun are measured
in Gauss units. Measured exterior solar magnetic fields were shown in
Figure~\ref{fig:Solar-external-magnetic-field}, copied from Yang et al.\cite{Yang-Betge} 
%


\subsection {Magnetic bending on star and planet equators}
Later in this paper laboratory storage rings with superimposed magnetic and electric bending will
be discussed in considerable detail.  Meanwhile, one notes that the magnetic dipole moments common to all
stars and planets provide a magnetic field (with magnetic dipole pattern).
The magnetic field to be emphasized in this section is aligned with the rotational axis and uniform
close to the equatorial plane.  By symmetry, external to the sun, this field 
parallel to the axis of rotation of the sun and sufficiently uniform for large aperture treatment.
The ``natural'' cyclotron orbits are circles of radius greater than the solar radius, and
centered on the axis in this equatorial plane. Any cosmic ray orbit discussed  in this paper follows a helix of
increasing or decreasing radius, slowly drifting ``out of plane'', or `` vertically'', in accelerator
terminology, away from, or toward the equatorial plane.  

The Parker solar wind provides another essential field component.  It is the latitudinal (or ``tangential''
in Frenet-Serret accelerator terminology) component of electric field, shown in the lower left corner of
Figure~\ref{fig:Parker-solar-wind}. Though small at the surface of the sun, this electric field, which is
caused by rotation of the sun's magnetic moment, becomes substantial at the surface labeled ``effective radial
magnetic field source'' in the figure. This CW or CCW electric field can serve the same role as the RF cavities
in terrestrial circular accelerators or storage rings.

When referring to a star or planet as an accelerator, these circles are slowly expanding or
contracting, nearly-closed orbits. Since free space is a fairly good vacuum, and only a quite
small number of turns expected, no vacuum chamber is required. In a linearized sense then, the
ring aperture is the infinite half-space external to this sphere; or, rather, to the nonlinear
dynamically limited portion of this half-space. Unlike any functioning terrestrial accelerator,
there is no significant focusing, other than the weak geometric focusing of closed circular orbits.
As a consequence nonlinear bending will surely be negligible. 

To make this kind of celestial accelerator better yet one could wish for focusing to better preserve beam emittances;
otherwise known as ``Courant invariants''. These were  named after ``Ernest'', who was the the son of ``Richard'', co-author
of ``Courant and Hilbert'', with the ``Hilbert'' of Hilbert space fame. Their colleagues in Gottingen, Germany,
include far too many to list famous physicists and mathematicians, including Riemann in 1859. 

Helical Alfv\'en guiding centers could supply
such focusing if they lay in the equatorial plane. But the only important Alfv\'en guiding centers are aligned with
magnetic fields perpendicular to the equatorial plane.  Any useful net beam acceleration must,
therefore, be accomplished in a fairly small number of turns, or possibly just a fraction of one
turn.

Though some protons are constantly being lost, they remain predominant, since they are
constantly being regenerated.  Note, though, that it is only low energy protons, not high energy
cosmic rays that are being regenerated.

\subsection{Review of superimposed ${\rm E}\&{\rm M}$ storage ring bending}
This section, considers the simultaneous storage of beams in the  circular arcs
of a predominantly electric``E\&M'' ring with superimposed magnetic bending, or a predominantltly
gravitational ``G\&M'' storage ring. 

We start by reviewing the E\&M case, which is also discussed later in the context of
laboratory-based, as contrasted with solar-based storage rings.
For simplicity the arcs are assumed to be perfect circles,
of bending radius $r_0$. Fractional bending coefficients $\eta_E$ and $\eta_m$ are defined by
\begin{equation}
\eta_E = \frac{qr_0}{pc/e}\,\frac{E_0}{\beta},\quad 
\eta_M = \frac{qr_0}{pc/e}\,cB_0,
\label{eq:BendFrac.2}
\end{equation}
neither of which is necessarily positive. In both cases, the right hand factors have been
arranged to be both dimensionally equivalent and comensurate.

In a storage ring with predominantly electric bending, to represent a small part of the required bending
force at radius $r_0$ being provided by magnetic bending while preserving the orbit curvature,
we define ``electrical and magnetic bending fractions'' $\eta_E$ and $\eta_M$ satisfying
$$\eta_E + \eta_M=1, \hbox{\ where,\ say,\ } |\eta_M/\eta_E| < 1/3. $$
The resulting magnetic force dependence on direction causes an $\eta_M>0$ (call this ``constructive'') 
or $\eta_M<0$ (``destructive'') perturbation to shift opposite direction orbit velocities (v) of the same radius, 
one up in radius and one down, resulting in two stable orbits in each direction.  For stored beams, any further 
$\Delta\eta_M \ne 0$ change causes beam velocities to ramp up in kinetic energy ($KE=\mathcal{E}-mc^2$) 
in one direction, down in the other.\footnote{In this paper there are several different types of ``energy''.
The most important of these is the total relativistic particle energy $\mathcal{E}$, with font chosen to be
calligraphic, in order to avoid clashing with $E$, the symbol for electric field.}

Depending on the sign of magnetic field $B$, 
either the lighter or the heavier particle bunches can be faster, ``lapping'' the slower bunches 
periodically, and enabling ``rear-end'' nuclear collision events. (The only longitudinal complication 
introduced by dual beam operation is that the ``second'' beam  needs to be injected with accurate velocity, 
directly into stable RF buckets.)

Only in such a storage ring can ``rear-end'' collisions occur with heavier particle bunches
passing through lighter particle bunches, or vice versa.  From a relativistic perspective, treated as 
point particles, the two configurations just mentioned would be indistinguishable.  As observed in the 
laboratory, to the extent the particles are composite, such collisions would classically be expected to be
quite different and easily distinguishable.

Pavsic\cite{Pavsic}, in a 1973 paper reproduced in 2001, developed a 
``mirror matter'' Hamiltonian formalism, distinguishing between ``external'' and ``internal'' symmetry.  
He points out, for example, that ``the existence of the anomalous proton or neutron magnetic moments indicates 
the asymmetric internal structure of two particles''; a comment that applies directly to the present paper.
Otherwise, Pavsic is agnostic, suggesting that his formalism provides only a parameterization for experiments
sensitive to internal structure, with possible implications concerning mirror matter.

\section{``Minor'' revision of Newton's gravitation law}
The relativistic forms of Newton's mass, force, acceleration, momentum and energy laws and
formulas are reviewed in Appendix~\ref{sec:RelativisticMechanics}, which is copied almost
\emph{verbatim} from reference~\cite{Talman-Mechanics-Chapter-8}.

For purposes of the present paper, in a hybrid form of classical and relativistic mechanics,
classical for the sun, relativistic for the proton, Newton's universal gravitation formula
for the force between a proton and the sun will be assumed to take the form
\begin{equation}
F_p = G\frac{\mathcal{E}_1\mathcal{E}_2}{r^2} \approx G\frac{m_1\gamma_pm_p}{r^2},
\label{eq:Newton-gravitation-law-a} 
\end{equation}
where $r$ is the radius of the sun,
$G$ is the gravitational constant, $c=1$, $\mathcal{E}_1$ and $\gamma_1m_1 \approx m_1$ are the
energy and mass of a ``heavy'' body, such as the sun, $\mathcal{E}_2$ and $m_2$ are the energy
and mass of a ``light'' (meaning ``not-heavy'') particle, such as a proton or a uranium nucleus,
with $\gamma_2=1/\sqrt{1- \beta_2^2}$ and $\beta_2$ the usual Einstein relativistic parameters for the
light particle, and $\gamma_1=1$ for the heavy particle.

Newton, himself could hardly have anticipated nuclear particles. The following alternative discussion
assumes implicitly that particle~1 is heavy and particle~2 is light.

Though the first form of the revised Newton formula in Eq.~(\ref{eq:Newton-gravitation-law-a}) is
intended to be exact in general, for simplicity, the present paper assumes more limited kinematics.
Our revised Newton formula assumes the particle mass $m_1$ is so ``heavy'' that its
center can be taken to be the center of mass (CM), and that the particle  of mass $m_2$ is a
``light'' elementary particle or nucleus.

The reader is expected to notice that a sacred formula is being tampered with.  Perhaps
the revised Newton formula is valid in General Relativity under the special conditions for which
I will use it?  I doubt that this is possible, since some physically well-defined quantities have
different values, depending on whether or not mass is replaced by energy (with $c=1$).

\emph{I am forced to revise Newton's gravitational law in this new way in order to explain how most of
the cosmic rays observed in nature, at least up to $10^6$\,GeV/nucleus for example, can have
been produced within the solar system.}

For ordinary astronomy this change in Newton's law has no effect, since all heavenly bodies
are ``heavy'', meaning $m=\mathcal{E}$. For laboratory atomic, nuclear and particle physics the
change will also have negligible effect, since all elementary particles are ``light'' and gravity
can always be neglected for individual processes. The influence on terrestrial statistical
mechanics should also not be controversial, because all neutral atoms, molecules, and other
gaseous particles are moving slowly enough to be treated as being heavy.

In other words, formula~(\ref{eq:Newton-kinetics-revised-a}) is distinguishable from the original
Newton version only for evaluating the force between a ``perceptibly weightless''
particle on a human scale (such as a uranium atom, abbreviated as ``light'' for the
time being) and an ``unimaginably heavy'' particle on a human scale (such as a star or planet,
abbreviated as ``heavy'').

As far as I know,  no laboratory measurement has ever been motivated by the possibility of
testing the sensitivity of Newton's gravitational law to any alteration resembling that
in formula~(\ref{eq:Newton-gravitation-law-a}), or demonstrating the unphysicality  of
such an alteration.

Changing Newton's law in this way would, it seems, have a major impact on cosmological issues;
for example on both ``missing mass'' and ``dark energy''.

It will be shown next that the replacement of mass $m$ by energy $\mathcal{E}$ (dimensionally
equivalent quantities with $c$=1) in the gravitation formula makes it easier to believe that
cosmic rays are being generated within the solar system.  It is simply because the gravitational
force increases linearly with $\gamma_2$ once particle~2 has become fully relativistic.

From then on, the acceleration that charged nuclear isotopes receive while circulating around
the sun, from the Parker longitudinal component of electric field, increases their centripetal
gravitational force by exactly the compensating amount needed for them to continue on
the same orbit on which their energy had been being increased.

The growth in energy proportional to $\gamma_2$ allows solar particle accelerators to be more or
less as effective as multi-turn laboratory particle accelerators for energy ramping to higher and
higher energies. Quantitative description of this acceleration follows.  To emphasize that
the force is purely gravitational, and proportional to the praticle rest mass, we will use $A$,
as in $A=Z+N$, as subscript for particle~2, meaning, roughly, that $m_A=Am_p$ or that the rest
mass of particle~A, is $A$, as measured in GeV units.  

To simplify the treatment of ultra high energy performance we specialize the discussion to
nuclear particles of mass index $A$.  Eq.~(\ref{eq:Newton-gravitation-law-a}) needs to be complemented
by the kinetic equations for acceleration, $a_A$ and kinetic energy $KE_A$;
\begin{align}
  a_A  &= \frac{v_A^2}{r_A},\\
 KE_A  &= {\mathcal E_A} - m_A \approx \frac{\gamma_Am_Av_A^2}{2} = \frac{G\,M_{\rm sun}\gamma_A\,m_A}{2\,r_A},
\label{eq:Newton-kinetics}
\end{align}
This is still not the equation we need, since it does not take proper account of center
of mass (CM) motion. In the sun as accelerator, the ``planetary object'' is not a single nucleus; it
is a belt of nuclei uniformly distributed above the sun's equator\cite{Talman-centrifugal}.
In this situation the CM would remain perfectly at rest.  (If the particl~A were ``heavy'' this equation
would be seriously incorrect.) 

To force the orbit to be circular and concentric with the sun, and to generalize Eq.~({\ref{eq:Newton-kinetics})
to be independent of nuclear particle type, while requiring the radius to be greater than the radius of
the sun, we apply the constraint $$r_A=r_{\rm A-lim},$$ to represent the the radius at which the particle
energy (otherwise known as its relativistic mass) first becomes (essentially) proportional;  
in spite of the fact that the specification, is poorly defined to the same extent that the word ``essentially'' is
poorly defined.

In Newton's $F=ma$ equation, the acceleration $a_A=v^2/r$ appears explicitly, which provides explicit
access to the radius factor $r_A$, but at the seeming cost of restoring the classical $v_A^2$ factor
in the kinetic energy.  However, in Eq.~(\ref{eq:Relint.twentythree2}), one sees in the relativistic
formulation of Newton's law, a new multiplicative factor of $\gamma_A$ appears. Subsequently, the
$v_A^2$ factor appearing in the final factor of Eq.~(\ref{eq:Newton-kinetics}) cancels out, but the force on
particle~2 will still be proportional to $\gamma_2$.

A clearer way of writing Eq.~(\ref{eq:Newton-kinetics}) is then
\begin{equation}
  r_{\rm A-lim} = \frac{G\,M_{\rm sun}}{2}\,\frac{\gamma_A\,m_A}{KE_A}
  \overset{\rm also}{=} \frac{G\,M_{\rm sun}}{2}\,\frac{\gamma_A\,m_A}{\mathcal{E}_A -m_{\rm A}},
\label{eq:Newton-kinetics-revised-a}
\end{equation}
along with the understanding that $r_{\rm A-lim}$ is the limiting radius of a ring of nuclear isotope-A,
such as the proton, around the sun and NOT the radius around the sun of a planet of mass equal to that
of nuclear isotope A. We must reject solutions for which $r_A \le r_{\rm sun}$ for which
particle~A's orbit is inside the sun, which is physically impossible.

To account for this situation implicitly we assume the particle~2 energy is large enough, compared to its mass,
for the energy to scale exactly proportional to $\gamma_2$, and revise
Eq.~(\ref{eq:Newton-kinetics-revised-a} accordingly, with the limiting result,
\begin{equation}
  r_{\rm A-lim} = \frac{G\,M_{\rm sun}}{2} = 6.6 \times 10^{9}\,{\rm m}.
\label{eq:Newton-kinetics-revised-b}
\end{equation}
Following Eq.(\ref{eq:Relint.thirteen2}) in Appendix~\ref{sec:RelativisticMechanics}, the final factor in
Eq.~({\ref{eq:Newton-kinetics-revised-a}) has been set equal to 1. This result no longer depends on the mass
of particle~2.  It can be compared numerically with the radius of the sun, which is
\begin{equation}
  r_{sun} = 0.6957 \times 10^9 {\rm m}.
\label{eq:mass-of-sun}
\end{equation}
The near equality of these two radii could be just a coincidence. 

But, for a designer attempting to turn the sun into a super-accelerator, it is a lucky coincidence.
It is a choice that the actual designer of the solar system, whoever she or he might have been,
might have made in designing the solar system.  It means that,
\emph{with gravity alone}, protons (and other nuclear particles of any energy per nucleon from 1\,keV up,
at least, to $10^6\,GeV$ and possibly to the energy of the highest energy gamma ray ever unambiguously
detected, can be captured onto
circular orbits by the sun's gravity, within a sphere whose radius is only ten times the radius
of the sun.  This is by virtue of the replacement of mass by energy in Newton's gravitational formula;
or, more explicitly, by the factor of $\gamma_2$ appearing in Eq.~(\ref{eq:Newton-gravitation-law-a}).

This does not mean that every such high energy particle would be captured. It is only nuclear particles
traveling (transversely), almost along circular orbits that could stay captured indefinitely.  Spiraling in,
most particles heavier than protons, and many protons, would have already fallen into the sun.
Also there would be no force strong enough for long enough to turn around  protons with significantly
positive radial velocity component aimed away from the sun.

Though demonstrating the physical possibility of ultra-high energy cosmic rays is important as proof
of principle, for purposes of actually calculating cosmic ray fluxes the magnetic force is too strong to be
neglected, especially if its sign is not constructive.  For motion remote from the sun it is
magnetic forces that become dominant. This would be far too complicated to be addressed in the present paper,
but formalizing the superposition is outlined next.

\subsection{Magnetically assisted ${\rm G}\&{\rm M}$ energy acceleration }
Though the demonstration that gravitational force, alone, is sufficient to explain the existence
of highly energetic cosmic rays, the actual situation is further complicated by magnetic forces.
This is especially true for the treatment of charged particles more distant from the sun than
have been discussed so far.

For the superposition of magnetic and gravitational bending we need to derive the analogs
of Eqs.~(\ref{eq:BendFrac.2}) which apply to the superposition of magnetic and electric bending.
A gravitational analog to the Lorentz force law is, with $\hat{\bf r}=\hat{\bf x}$,
\begin{align}
{\bf F}_{GM} &= -G\frac{m_1\gamma_2m_2}{r^2}\hat{\bf x} + qc\hat{\pmb\beta}\hat{\bf z}\times B_0\hat{\bf y}\\
            &= \Big(G\frac{m_1\gamma_2m_2}{r^2} + q\beta cB\Big)\ (-\hat{\bf x}),
\label{eq:F-GM.1}
\end{align}
where ${\bf r}\equiv{\bf x}$, ${\bf v}$, and ${\bf B}$ are mutually perpendicular vectors.
(As stated previously, particle~1 is heavy, particle~2 light). The only change introduced by
the revision of Newton's gravitational formula is the presence of the factor $\gamma_2$ in
Eq.~(\ref{eq:F-GM.1}).

We wish to mimic the $\eta_E$
and  $\eta_M$ partitioning, defined in Eq.~(\ref{eq:BendFrac.2}).
for which commensurate factors, both electromagnetic, are isolated.

Unfortunately, in the ${\rm G}\&{\rm M}$ case we are not at liberty to choose the magnitudes
of either the gravitational or magnetic force, nor the sign of the magnetic field. Clearly we
want both forces to be centripetal in order to produce a circular orbit around
the sun or a planet. For guaranteed centripetal orbit, the sign of $B$ must be positive,
meaning the correct rotation direction needs to be taken. With $q$ necessarily positive in
Eq.~{\ref{eq:F-GM.1}) to produce constructive bending we require $B=|B|$. 

We therefore define $\eta^{**}_G$ and $\eta^{**}_B$, where the overhead $^{**}$ symbols provide
warning that the fractions do not sum to 1. We therefore define bending fractions
\begin{equation}
\eta^{**}_G = G\frac{m_1\gamma_2m_2}{r^2},\quad \eta^{**}_M = q\beta|cB(r)|,
\label{eq:BendFrac.GM}
\end{equation}
where the overhead $^{**}$ symbols warn that the fractions do not sum to 1.

The most important case concerns the bending of a nuclear particle around the sun or a planet,
such as Jupiter, on a circle of radius $r_0$, in which case, for small radial deviation $\delta r$,
the bending fractions can be expressed as
\begin{equation}
  \eta^*_G \approx  G\frac{m_1(1-2\delta r/r_0)}{r_0^2}\gamma_2(t)m_2,\quad
  \eta^*_M = q\beta|cB(r)r_0|. 
\label{eq:BendFrac.etaG}
\end{equation}
where a common factor of $r_0$ has been included and where the only significant time dependence
is carried by the factor $\gamma_2(t)$, caused by the longitudinal Parker electric field component
acting on a captured particle being accelerated.  Expressed in this way, these factors are useful
in cases where one or the other, gravitational or magnetic, bending is dominant.

\subsection{Improving the performance of solar particle acceleration}
Continuing our effort to design, or rather to understand the design, of cosmic accelerators,
the essential ingredient is the longitudinal electric field component $E_{\parallel}$ that is
causing the beam particle energy ${\mathcal{E}_2}(t)$ to increase with time;
\begin{equation}
\mathcal{E}_2(t) = \mathcal{E}_{2\,\rm inj} + q_2E_{\parallel}t.
\label{eq:Accel.1}
\end{equation}
At this point we make a few significant observations:
\begin{enumerate}
\item
  We have been assuming the Parker longitudinal electric field component $E_{\parallel}$ is constant,
  causing the beam energy to increase monotonically.  This would seem to violate the
  E\&M requirement that the line integral of static longitudinal electric field must vanish.
\item
  The task seems, therefore, to be impossible. But wait, the Sun's electric field is not
  constant; it oscillates with a period of 22 years.  Perhaps, or even probably, this adequately
  overcomes the item~1 constraint. In any case, the direction of stable rotation around the sun
  certainly must reverse every 11 years. During reversals, when the magnetic field vanishes,
  any stored particle will almost certainly be dumped.
\item
  We are striving to accelerate a beam, say a proton beam, from an energy, of say $10^{-2}$\,GeV,
  to an energy of, say, $10^6$\,GeV, which is just slightly higher than the highest
  energy cosmic ray proton energy plotted on the right hand graph in
  Figure~(\ref{fig:Simpson-Cosmic-Rays-annotated}).  This is eight orders of magnitude acceleration.
\item
  Previously, in Section~(\ref{sec:Rough-values}), based on the known magnetic field at the
  surface of the sun it was estimated that protons could be accelerated from 2\,MeV 
  to an energy of energy 200\,GeV in a single revolution, an acceleration
  by a factor of five orders of magnitude. Though impressive, this is not even close to
  eight orders of magnitude. We are still missing three orders of magnitude.
\item
  The whole point of a ground based particle accelerator is to increase the energy a little
  bit each turn, for many turns. All we need is for the cosmic ray beam to stay captured
  by the sun for a thousand turns. This would seem to be trivially easy, considering that beams
  traveling at (almost) the speed of light have been stored in ground based accelerators, with
  tiny apertures, for day-long runs.
\item
  Up until now there has been one remaining show-stopper.  Even though the  proton mass
  (expressed as an energy) is only 1\,GeV, the gravitational force, as calculated by Newton's
  original gravitational formula, is woefully too weak to keep captured protons of such high momentum.
\item
  The point of the entire present paper, has been that all we need to assume is that Newton's
  gravitational law has the proton mass replaced by the proton energy. This provides
  the missing gravitational bend strength factor needed to keep the protons, and other nuclear
  particles captured for at least one thousand turns.
\item
  In short, under the special conditions of massive bodies acting on atomic scale particles, gravity
  is not as weak as has been assumed until now.
\item
  In actuality, in real life, even assuming that Newton's law needs energies, not masses,
  the situation is more complicated than has so far been assumed.
\item
  However, we need not require 100 percent beam capture, nor perfect beam flux retention
  through the acceleration process.  According to Figure~(\ref{fig:Simpson-Cosmic-Rays-annotated})
  the extracted beam flux at maximum energy is less by 12 orders of magnitude at maximum compared
  to minimum energy.
\end{enumerate}

\subsection{Stochastic acceleration of solar system cosmic ray particles}
The primary importance of magnetic fields in the solar system is not inside the sun or planets;
it is in the free space external to these bodies.  There is a strong tendency
for massive atoms to fall back into the sun or one of the planets. In equilibrium, it is predominantly
protons, electrons, and alpha material (predominantly $\alpha$-particles and deuterons) that continue
to circulate freely and longer, in more or less isotropic directions. Though all alpha material particle
orbits of the same momentum are nearly the same, they are predominantly of very low energy, and,
consequently of low momentum.  Meanwhile electrons and nuclear particles will have merged into atoms
that are captured in the atmospheres of planets.

It is predominantly higher momentum protons, deuterons and $\alpha$-particles that continue to circulate
freely. In equilibrium these particles will have gravitated toward the most massive objects in
the solar system, namely either the sun or Jupiter.

Quoting from reference~ \cite{Agle-Juno}, ``NASA’s Juno mission to Jupiter made the first definitive detection beyond our world of an internal magnetic field that changes over time, a phenomenon called secular variation. Juno determined the gas giant’s secular variation is most likely driven by the planet’s deep atmospheric winds.''\cite{Agle-Juno}  Also `` The field rotates with the approximately 9 hour rotational period of the planet.''.

\emph{Since these changes in magnetic field are fractionally quite small we will, nevertheless, ignore them, and treat the magnetic field of Jupiter (unlike the sun) as constant in time, rotating along with Jupiter.}
\begin{figure}[hbt!]
  \centering
  \includegraphics[scale=0.5]{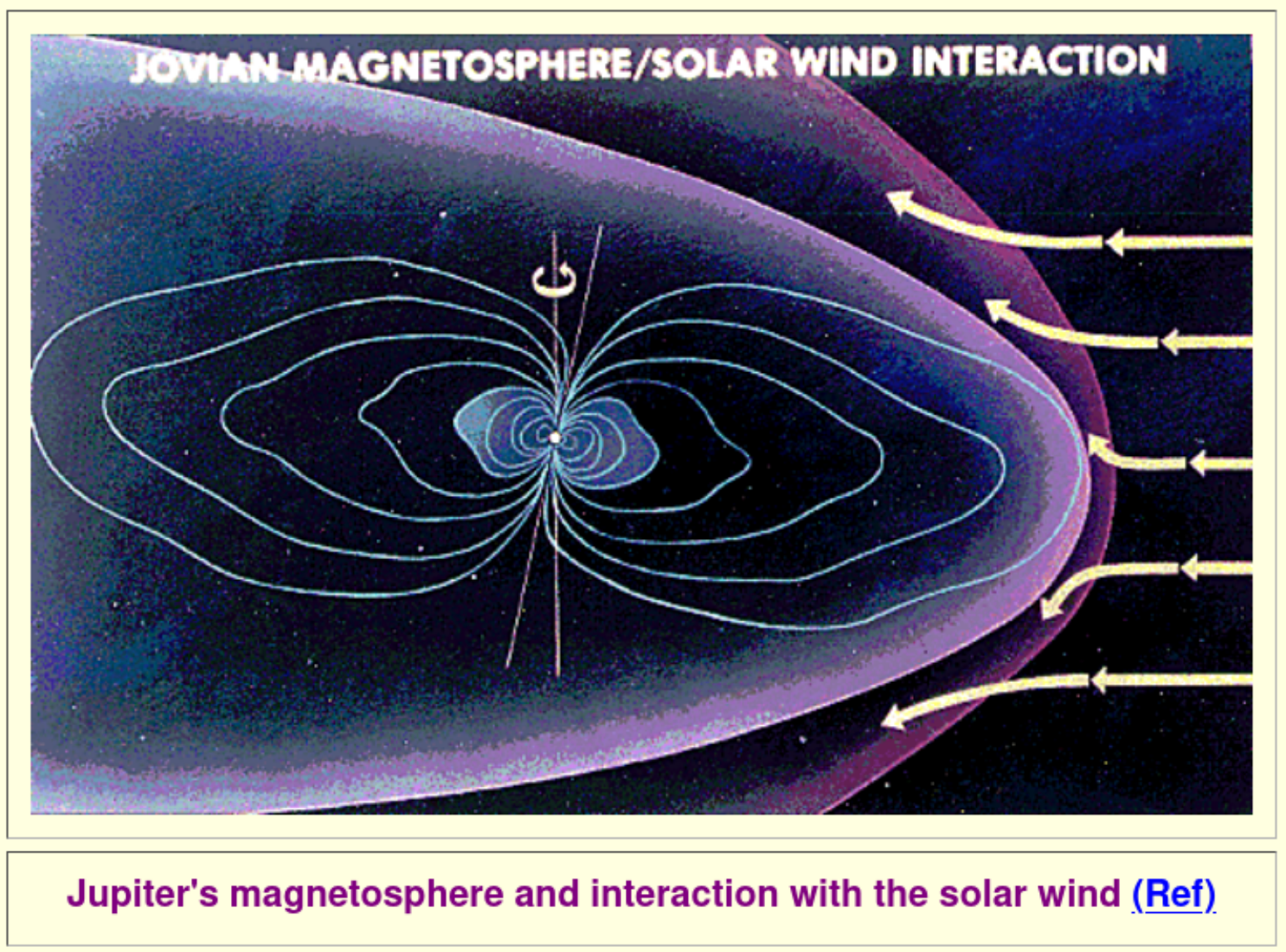}
  \caption{\label{fig:Jupiter_magnetic_field_contours} Jupiter magnetic field contours, copied from reference\ \cite{Agle-Juno}.
 }
\end{figure}

When produced in standard terrestrial alpha radioactive decay, alpha particles generally have a kinetic energy of about 5 MeV
and a velocity in the vicinity of $0.04\,c$. At this speed the time of flight of an alpha particle from Jupiter to the sun is
approximately 0.0007/0.04 = 0.02 years, or about 7 days.  At higher energy or lower mass this time would, of course, be less.

The ``open solar flux'' (OSF) represents the variable heliospheric magnetic flux. The ``galactic cosmic ray'' (GCR) measure
represents cosmic ray intensity near Earth.  Quoting Koldobsky\cite{Koldobskiy}, ``The flux of \emph{galactic} cosmic rays (GCRs)
outside the heliosphere is generally assumed to be constant at the time scales shorter than a hundred thousand years.'' as
well as ``The GCR flux variability is known to be delayed with re-respect to solar activity, leading to the
so-called “hysteresis” effect of the phase shifts in the development of the 11-year solar cycle in both indices.''
Here the \emph{galactic} qualifier has been emphasized, if only to stress that the present paper is concerned only with
solar cosmic rays.

The long-term variability of GCRs is routinely monitored by a global network of ground-based neutron monitors (NMs) that
are sensitive to the nucleic component of the cosmic-ray-induced atmospheric cascades and located, as a worldwide network,
around the globe since the 1950s.

Again quoting Koldobsky\cite{Koldobskiy}, ``All of these three indices appear highly coherent at a timescale longer than a few
years with persistent high coherence at the timescale of the 11-year solar cycle. The GCR variability is delayed with respect
to the inverted SSN (solar sunspot number) by about eight 27-day Bartels [a measure of solar activity rotation cycles] is
1/2 of a year on average, but the delay varies greatly with the 22-year cycle, being shorter or longer around positive A+
  or negative A-.''

  Time Lag measures of SSN, OSf, and NM from 1950 to 2020 are platted in Figure~(\ref{fig:Cosmic-ray-solar-cycle}),
  along with the original caption. Frequency compositions are commented upon in the new caption.
\begin{figure}[hbt!]
\centering
\includegraphics[scale=0.47]{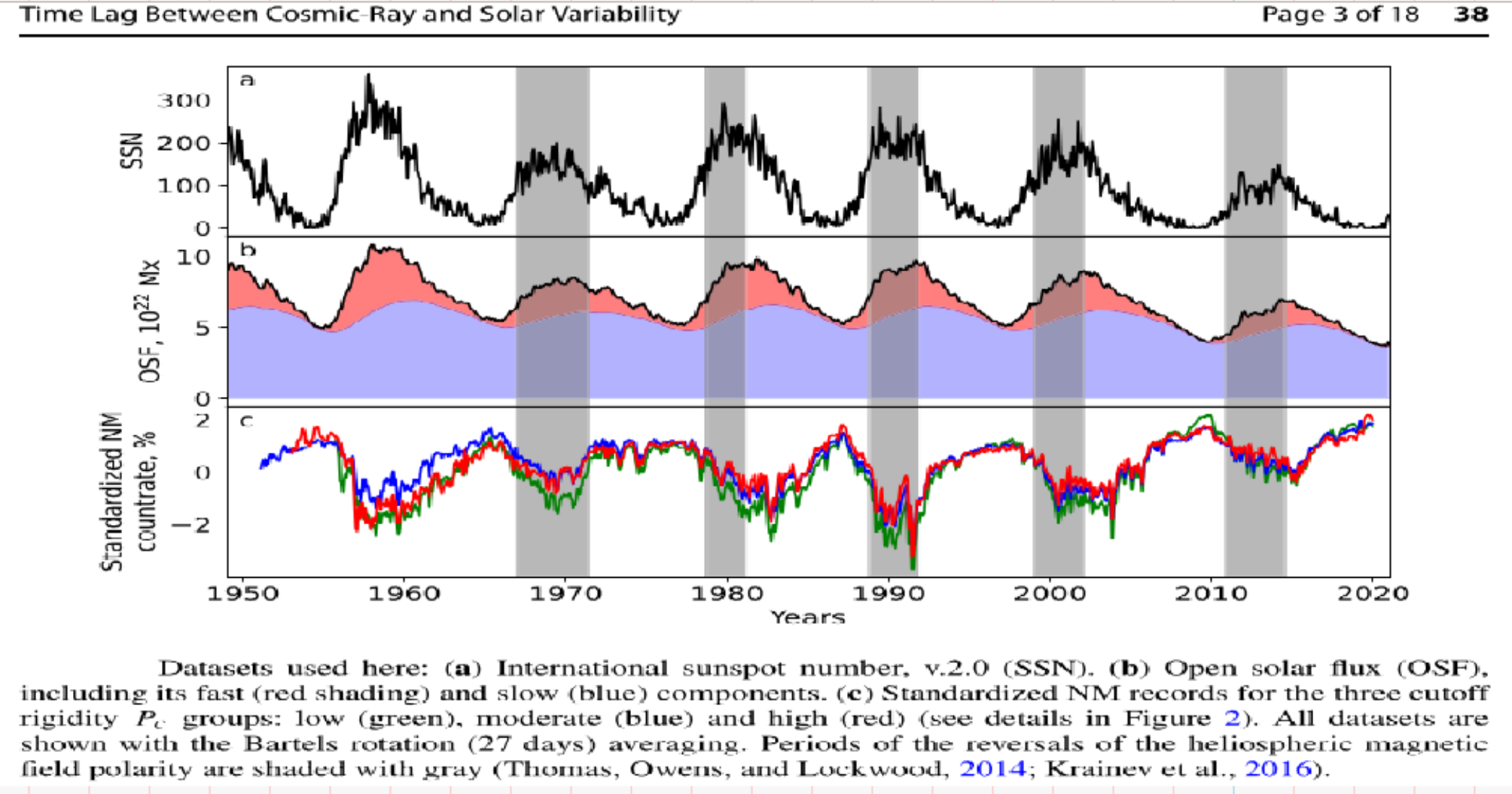}
\caption{\label{fig:Cosmic-ray-solar-cycle}Time Lag comparisons, 1950-2020, of sunspot number (SSN),
  open solar flux (OSF) and cosmic ray measure (NM). {\bf (a)} Notice that, unlike the upper two solar plots, the
  cosmic ray measure remains within a $\pm 2$-percent range. As regards signs, the upper two solar plots remain faithfully
  in phase. So also do the cosmic ray deviations. But maximum solar flux coincides with reduced cosmic ray rates and
  minimum solar flux coincides with increased cosmic ray rates. {\bf (b)} But notice also from the top two figures,
  that the OSF is a superposition of an AC signal of significantly variable amplitude, or different frequency component,
  and a DC signal. {\bf (c)} In a DC sense, it might be said that, on the average, the OSF and the cosmic ray flux
  are proportional.  {\bf (d)} Though opposite in phase, so also are the SSN's and the cosmic ray deviations.
}
\end{figure}

The NM data is especially relevant for the ``double slingshot'' mechanism to be described next.

\subsection{``Double slingshot'' mechanism for solar cosmic ray production}
This section proposes in semi-quantitative terms a ``double slingshot'' mechanism for generating cosmic rays
within the solar system.  To begin, the sun's extracted solar wind serves as multi-particle injector, onto Jupiter
as pre-accelerator, of nuclear isotopes onto orbits in the equatorial plane above the equator.  Only orbits on the side
of Jupiter for which the force is centripetal will be (temporarily) captured. 
Here the magnetic field is parallel to Jupiter's axis of rotation, with a magnitude of approximately 4.2 Gauss at the surface.
Other rough parameter values have been provided in the previous Section~\ref{sec:Rough-values}. Measured cosmic ray
energy dependence of nuclear isotopes has been plotted above in Figures~(\ref{fig:Simpson-Cosmic-Rays-annotated}) and
(\ref{fig:Simpson-Cosmic-Rays-reactions}).
\begin{figure}[hbt!]
  \centering
  \includegraphics[scale=0.44]{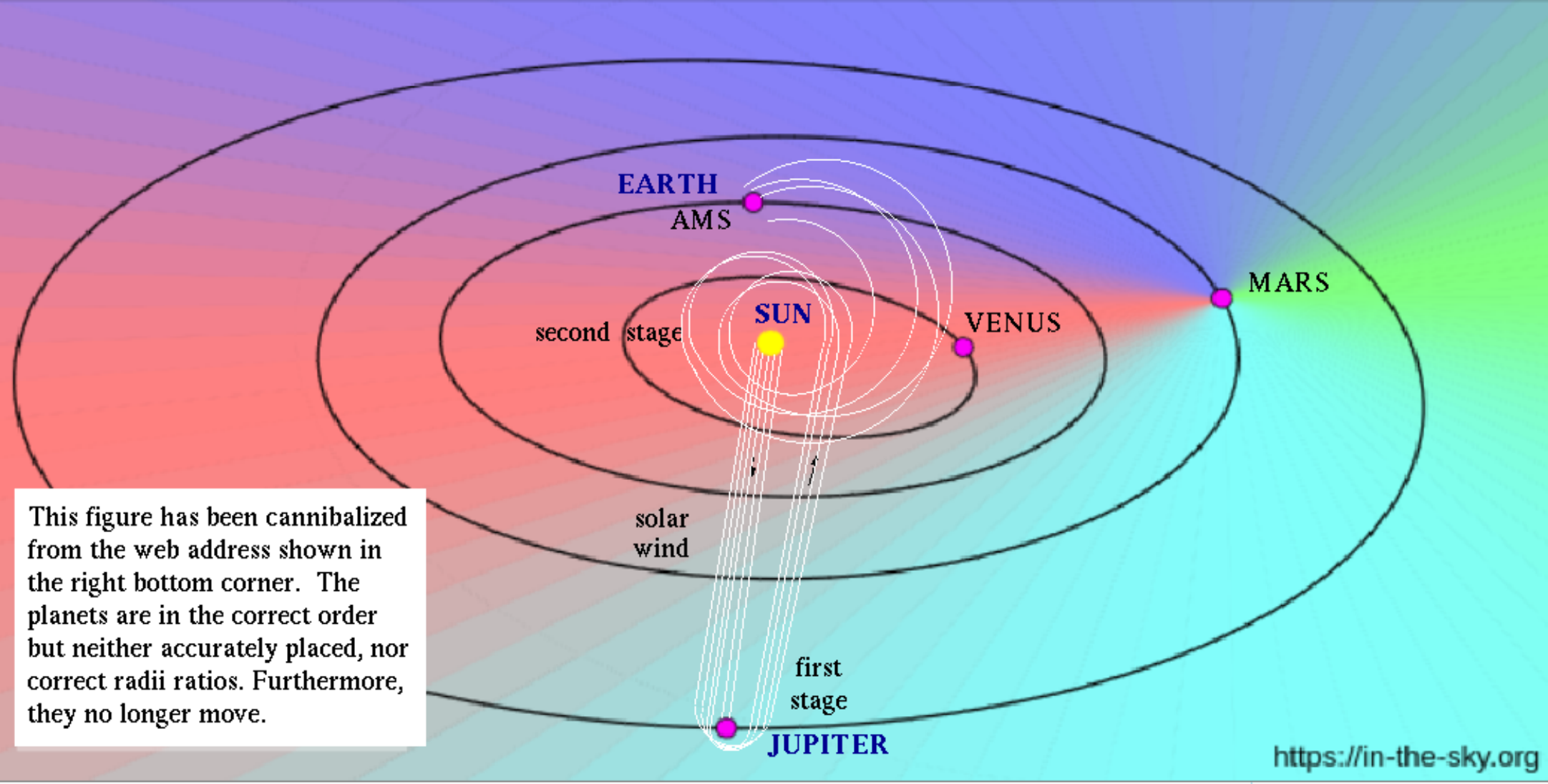}
  \caption{\label{fig:Double-slingshot}Artist's conception of two stage ``Double-Slingshot''
    solar system production of cosmic rays, impinging, for example, upon the
    Alpha Magnetic Spectrometer (AMS) detector\cite{Particle-Data-Group-Chap-29-Cosmic-Rays}
    aboard the International Space Station (ISS).  As indicated in the corner note, planetary distances
    from the sun are much distorted.  Mainly, the radius of Jupiter has been reduced, relative to the other
    planets, including earth, by roughly a factor of two.  Slingshots first around Jupiter, then around the
    sun, produce acceleration factors discussed in Section~(\ref{sec:Rough-values}). Of the six orbits shown
    leaving the sun towards Jupiter, two could potentially be detected near Earth by the AMS detector.
 }
\end{figure}

The shape of Jupiter's magnetic field is essentially the same as that of the magnetic field of the sun,
which is shown in Figure~(\ref{fig:sun-interior-magnetic-field}).  After one or more partially
complete helical turns around Jupiter, the much accelerated isotopes will have extracted themselves into a fan of
energetic nuclear isotopes emerging isotropically in the two dimensional equatorial plane external to Jupiter.
This is illustrated in Figure~(\ref{fig:Double-slingshot}).

Even the least energetic of these isotopes aimed exactly toward the sun will arrive there in less than a day,
where they are injected into the sun's exterior magnetic field, which is playing its role as a circular accelerator.
Half of these injected nuclei will be deflected away from the sun, back into space. The other half will be attracted
toward the sun.  These will mainly crash back into the sun.  But a significantly large number will have low enough
momentum to commence expanding helical motion around the sun.

In this situation, these orbiting particles (probably predominantly protons, but doubly energetic deuterons,
and quadratically energetic $\alpha$-particles, and some even more energetic nuclear isotopes will continue cycling
helically away from the sun.  In this situation there is no way of telling (short of a careful modeling of the
overall process) the maximum energy these particles can obtain.  As a guess, one hopes/expects the energy distributions
will match those of various the measured distributions shown in Figures~(\ref{fig:Simpson-Cosmic-Rays-annotated}) and
(\ref{fig:Simpson-Cosmic-Rays-reactions}).

It is not ruled out for the occasional nuclear isotope to make more than one passage through one or the other
(or both together) of the slingshots. At least in principle this could account for even the highest energies shown in
Figure~(\ref{fig:Simpson-Cosmic-Rays-annotated}).

\subsection{Start-up of Sun-SR, the sun's equatorial storage ring}
Already in Figure~(\ref{fig:sun-interior-magnetic-field}), there was displayed a toroidal region with
transverse area half as big as
the sun itself, shaped much like the magnetic field of a ground based particle accelerator, such as the LHC.
Furthermore the more or less constant vertical magnetic field is not unlike the field of a weak focusing
storage ring. As shown in Figure~(\ref{fig:Parker-solar-wind} there is a ``constant'' (modulo fluctuations
and 22 year period, sinusoidal variation) ``longitudinal'' electric field. Particles with velocities directed
more or less parallel to the electric field, will be accelerated.
Figure~(\ref{fig:Beam-falling-in-views}) shows views of particles falling gravitationally in toward the sun.

Though much like ground-based storage rings, there are, of course, many features distinguishing
Sun-SR from them.  \emph{The most important of these differences is that, unlike a ground-based
accelerator, the bending field does not ramp up to match the increasing momenta of the charged particles.}
Less important, the acceleration is ``DC'' rather than ``AC-RF''.

The duration of a ``long run'' in Sun-SR cannot exceed 10 years; the direction of any established
``beam'' not able to survive the periodic reversal of the sun's magnetic field.  As with land-based rings,
the details of start-up are also likely to be quite messy.  For example, start-up may be synchronized with
unpredictable sun-spot activity.  Let us, therefore, skip to a conjectural low energy start-up condition, with
specified proton beam of definite energy and current, circulating stably in a circular ring of radius minimally
greater than the radius $r_{\rm sun}$ of the sun. The centripetal bending to establish this condition is
produced by the ``constructive'' superposition of gravitational and magnetic bending.  To start, we will
assume the bending is primarily gravitational, and neglect the magnetic bending.

The absence of bend field ramping seems, definitively, to make the sun a very poor accelerator. What makes this
incorrect is the factor $\gamma_2$ in the (as-modified form) of Newton's gravitation
formula~(\ref{eq:Newton-gravitation-law-a}). Without this factor the Sun-SR ring would be useless.  But an
accelerator designer could only dream of a bending field that automatically adjusts the bending curvature of
all particles to match their instantaneous bending curvature in exact synchronism with every particle's
current ``stiffness''.

\emph{It is this feature which has motivated my (here repeated) revision of the Newtonian Gravitation formula.}
\begin{equation}
F = G\frac{\mathcal{E}_1\mathcal{E}_2}{r^2} \approx  G\frac{m_1\gamma_2m_2}{r^2},
\label{eq:Newton-law-mod-bis} 
\end{equation}
This fully relativistic feature kicks in fully, however, only once any magnetic bending has become
negligible, and most of the surviving particles have become almost fully relativistic.  Conditionally,
Eq.~(\ref{eq:Newton-law-mod-bis} has solved the problem of how to continue to accelerate relativistic
particles to sufficiently high  energies for momentum and energy to be equal ($\beta=1,\,c=$1).
\begin{figure}[hbt!]
  \centering
  \includegraphics[scale=0.53]{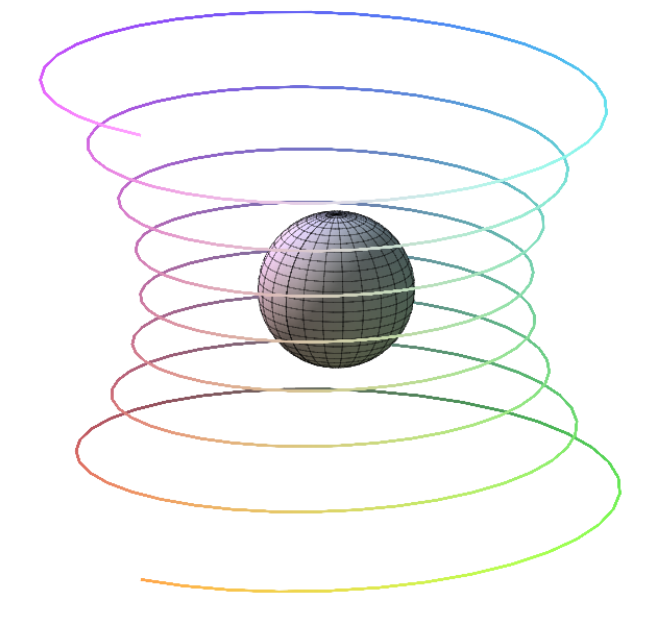}
  \includegraphics[scale=0.39]{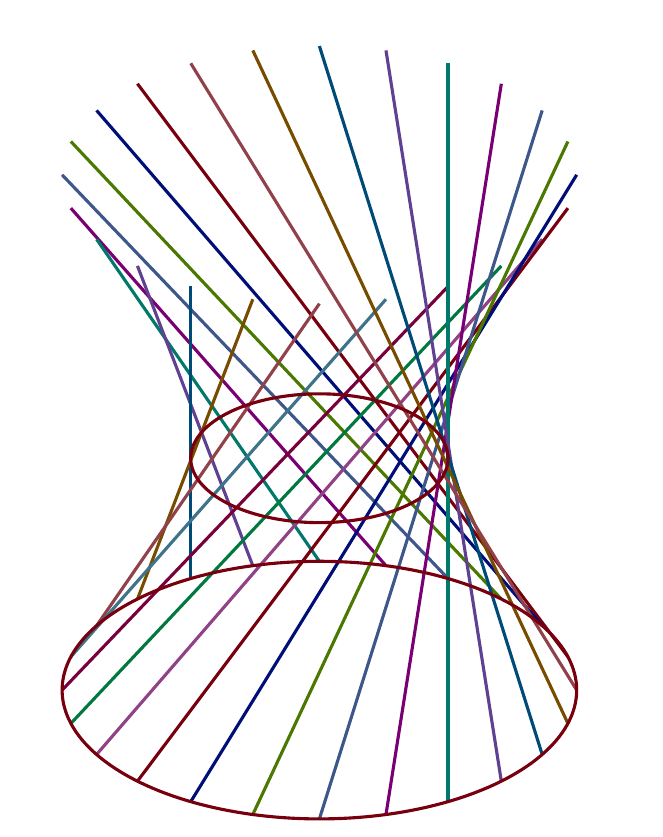}
  \includegraphics[scale=0.44]{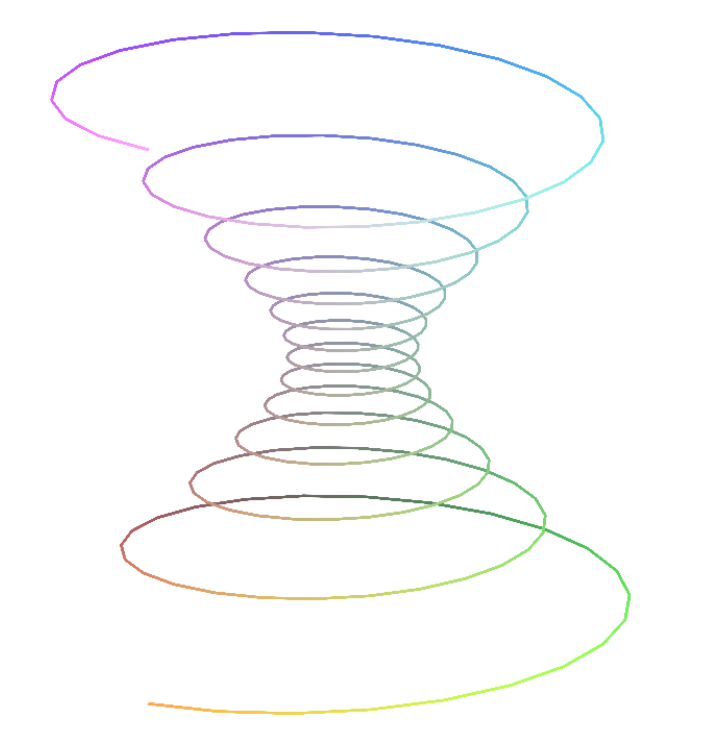}
  \caption{\label{fig:Beam-falling-in-views}Views of selections of particles falling in toward the sun.
     {\bf Left: }Particle being accelerated.  {\bf Center: }Particles grazing the sun.
    {\bf Right: }Particle captured by the sun.} 
\end{figure}

\clearpage

\section{Rear-end colliding beam storage rings}
\subsection{Diverse particle types circulating in a common magnetic ring}
As explained in previous sections,
it is possible, in an all-magnetic ring, for two different particle types having the same magnetic
rigidity, such as (Z=1, A=2) deuterons and (Z=2,A=4) alpha particles to co-rotate on
circular orbits of nearly identical radius $R$, in spite of the alpha particle charge
being greater by a factor of two. But, at least at non-relativistic velocity, this is compensated
by the doubled alpha particle mass. 
With particle momentum $p=m\gamma\beta c$,
this establishes the near equality of their ``B-rho'' magnetic
rigidities, defined by
\begin{equation}
  B\-\rho = (Am_U)\gamma\beta c/(Ze) = pc/e,
  \label{eq:B-rho.1}
\end{equation}
where $B$ and $\rho$ are the common magnetic field and orbit curvatures.
\footnote{For bridging low energy and high energy physics there are serious notational complications. especially concerning energy definitions and symbols. Not counting weak interactions and neutrinos, which are discussed only secondarily in this paper, nonrelativistic kinematics, such as $E=mv^2/2$, along with masses retained to the highest possible precision, would be adequate for the low energy nuclear physics. Yet fully relativistic formulas are needed, and used, for accelerator physics, such as for high precision orbitry and nuclear spin precession; that is, $\mathcal{E}\equiv E_{\rm rel}=\sqrt{p^2c^2 +m^2c^4}$, along with $E_{\rm rel}=mc^2 + KE \equiv \gamma mc^2$, where for semi-relativistic nuclear isotopes $\gamma$ is not much greater than 1, and $\beta=v/c$ is always significantly less than 1. Momentum $p$, as used in Eq.~\ref{eq:B-rho.1} is the true relativistic momentum. This means that Eq.~\ref{eq:B-rho.1}, which is perhaps the most important formula in the paper, is valid both classically and relativistically. The other concerns the very meaning of energy, irrespective of its dimensions and units, which also complicates the understanding of the Boltzmann relationship between energy and temperature. This subject is considered briefly in Sections~\ref{sec:TempEnergy} and Appendix~\ref{sec:Units}.}
Because of binding energy effects the particle velocities of different alpha materials
are slightly different. This means there are many possible pairs of different nuclear
isotope that can co-circulate simultaneously and engage in rear-end collisions..
In this paper the isotope combinations for which Eq.~(\ref{eq:B-rho.1}) is almost exactly
satisfied, are being emphasized. For mnemonic purposes every one of these isotopes can
be referred to as a ``multi-deuteron'', with N=P=A/2.  This class of nuclear isotope includes
``multi-$\alpha$'' particles, which can now be referred to as
``better-behaved multi-deuterons''.
    
The result of all this is to make rear-end collisions between different
alpha medium types far more common than would apply in an uncorrelated Boltzmann
plasma distribution. In the present context, what makes
these collisions special is that they occur in a frame of reference that is moving
with (weakly)-relativistic velocities, either in the laboratory
or in the cosmos.

As a consequence, the $Q$-values (i.e. the sum of their kinetic energies in the center of mass (CM)
system) can be small; say hundreds of keV, comparable, for example, with the Coulomb barrier which
is impeding their collisions. For a given particle type, say type 1, there may or may not be isotope
energy of type 2 matched to some required $Q_{12}$-resonance value.  In other words, it may or may
not be possible to optimize input conditions, for example to tune the initial conditions for $Q_{12}$ to satisfy
some intended resonance condition.

It is, by now, well established  that (in a laboratory storage ring) that this limitation can be overcome
by the superposition of electric bending.  For now we have only established the
efficacy of magnetic rigidity, as it equates particle momentum with field-radius product,
irrespective of their specific $Z$, $A$, $m$, $p$, $B$, and $\rho$ values.

\subsection{Simultaneous beam types in ${\rm E}\&{\rm M}$ storage ring}
I.A. Koop\cite{Koop}, in 2013, suggested the use of a storage ring with superimposed electric and
magnetic bending, for the purpose of testing time reversal violation in the form of a non-vanishing proton
electric dipole moment (EDM). Since then, many variants of this proposal, with various theoretical embellishments, as
well as many experiments demonstrating successful implementation of essential ingredients have been published,
including those in the following list.  
\cite{RT-JT-AGS-Analogue} 
\cite{RT-ICFA} 
\cite{RT-CLIP-PTR}
\cite{RT-PREDOM-ELECTRIC}
\cite{RT-pp-Wave-Particle} 
\cite{RT-Snowmass-Seattleinser} 
\cite{CYR} 
\cite{RT-CLIP-PTR-conformal}
\cite{Eversmann}
\cite{Hempelmann}
\cite{Rathmann-Kolya-Slim}
\cite{Slim-Rathmann}
\cite{RT-p-h-doubly-magic}
\cite{Talman-Nikolaev}
\cite{KolyaFrankPaulo}
Of these, reference~\cite{CYR}, though somewhat outdated, is the most comprehensive concerning predominantly
electric  storage ring proton EDM measurement practicalities;
references~\cite{Eversmann} and \cite{Hempelmann} describe high precision storage spin tune control in COSY;
reference~\cite{RT-ICFA} derives equations needed for dual-beam, MDM comparator, frozen beam spin tune control,
and introduces compromise quadrupoles for dual beam optical focusing. Reference~\cite{RT-p-h-doubly-magic}
describes doubly-magic circulating $p$ and $h$ beams for precision measurement of the proton-helion EDM
difference (with canceling systematic error).
Reference~{\cite{RT-PREDOM-ELECTRIC} contains by far the most comprehensive discussion of predominantly
electric storage rings with superimposed magnetic bending; compromise quadrupoles are explained,
along with transverse and longitudinal beam dynamics; also included are 
tables of precise nuclear anomalous magnetic moment $G$-values, usefully expressed as rational fractions,
with advice concerning their extraction from $g$-values.                     

As yet, no such E\&m storage ring project has been set in motion.

In a storage ring with predominantly electric bending, to represent a small part of the required bending
force at radius $r_0$ being provided by magnetic bending while preserving the orbit curvature,
we define ``electrical and magnetic bending fractions'' $\eta_E$ and $\eta_M$ satisfying
$$\eta_E + \eta_M=1, \hbox{\ where,\ say,\ } |\eta_M/\eta_E| < 1/3. $$
This perturbation ``splits'' a unique velocity solution into two separate velocity solutions. This enables
two different beam particle types to co-circulate at the same time, with different velocities.

As the (intentional) consequence, there are periodic ``rear-end'' collisions between two particles of different type
co-moving with different velocities in the storage ring (which is, of course, stationary in the laboratory).
It can be arranged for the CM KEs to be in
the several 100\,keV range, comparable with Coulomb barrier heights, even though all incident and scattered
particles have convenient laboratory KEs, two orders of magnitude higher, in the tens or hundreds of MeV range.

For these collisions to occur always at the same storage ring location the beams need to be bunched, and the ratio of
beam velocities $vratio=v_1/v_2$ needs to be rational, expressible as the ratio of two (not necessarily positive)
integers. This same condition enables both beams to be bunched by a single RF cavity, as illustrated in
Figure~\ref{fig:7-8-actual}, for two values of $vratio$.

\begin{figure}[hbt]
\centering
\includegraphics[scale=0.34]{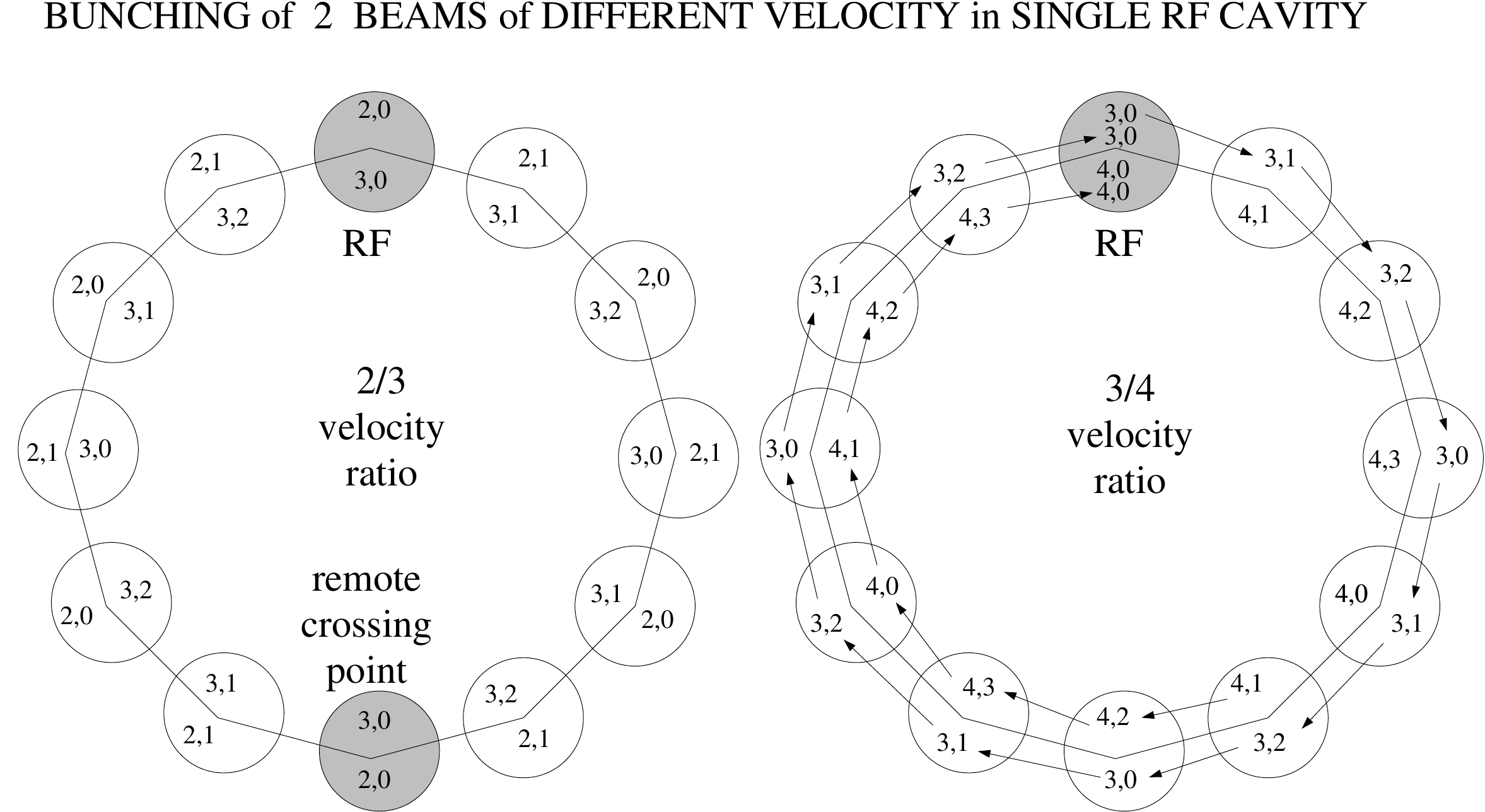}    
\caption{\label{fig:7-8-actual}Stable RF buckets for different  velocity ratio beams.} 
\end{figure}

\subsection{``${\rm E}\&{\rm M}$ rear-end'' storage ring collider}
Figure~\ref{fig:PTR-layout-Toroidal8_102p2-mod}, on the left, provides a layout diagram of a proposed 
storage ring, PTR,  with predominantly electric bending and superimposed magnetic bending.
``Compromise quadrupoles'' (for which electric and magnetic quadrupoles are only approximately
superimposed) are shown expanded in the inset diagram.  A perspective view of one PTR bend sector is
shown on the right.

\begin{figure}[hbt!]
\centering
\includegraphics[scale=0.13]{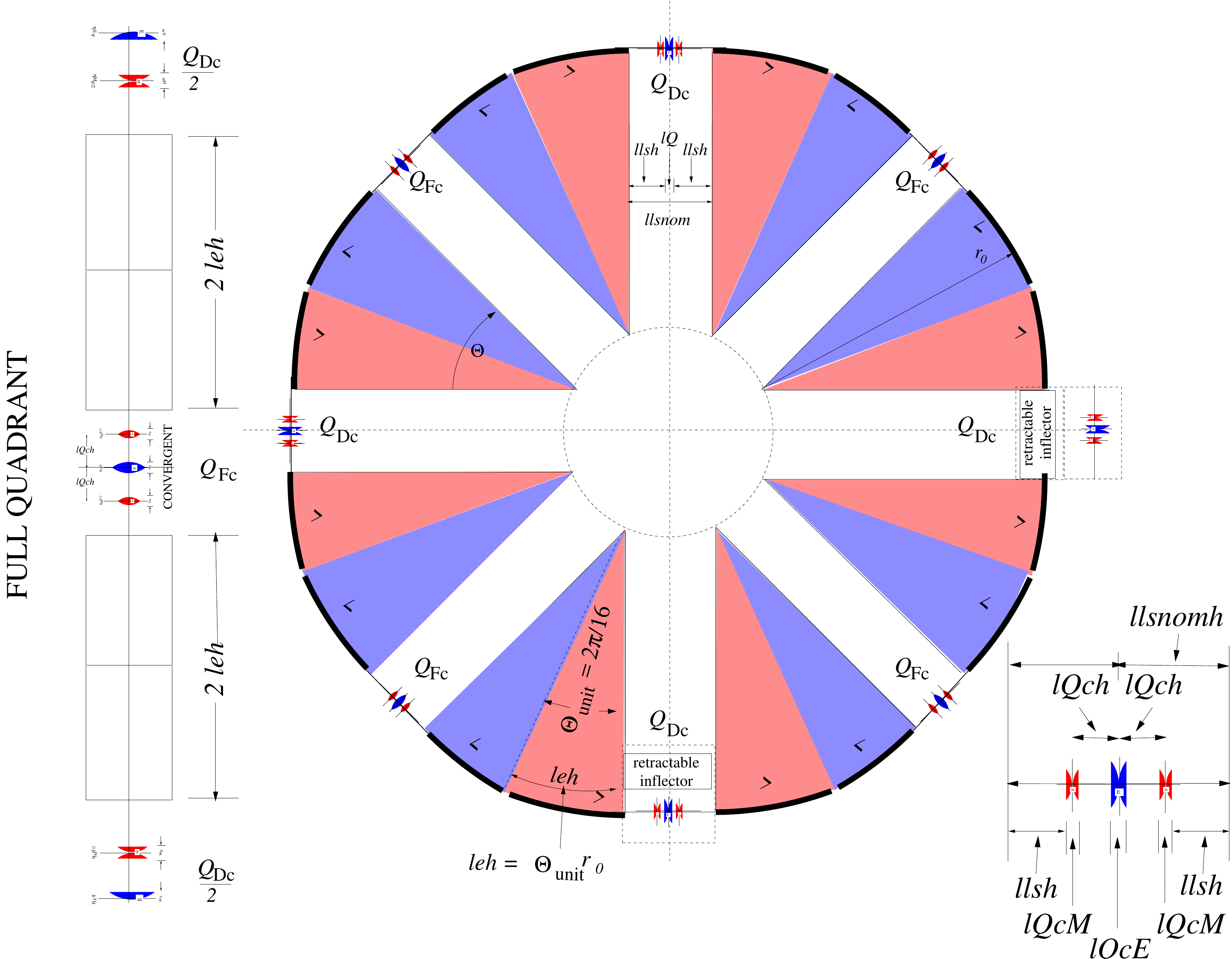}
\includegraphics[scale=0.33]{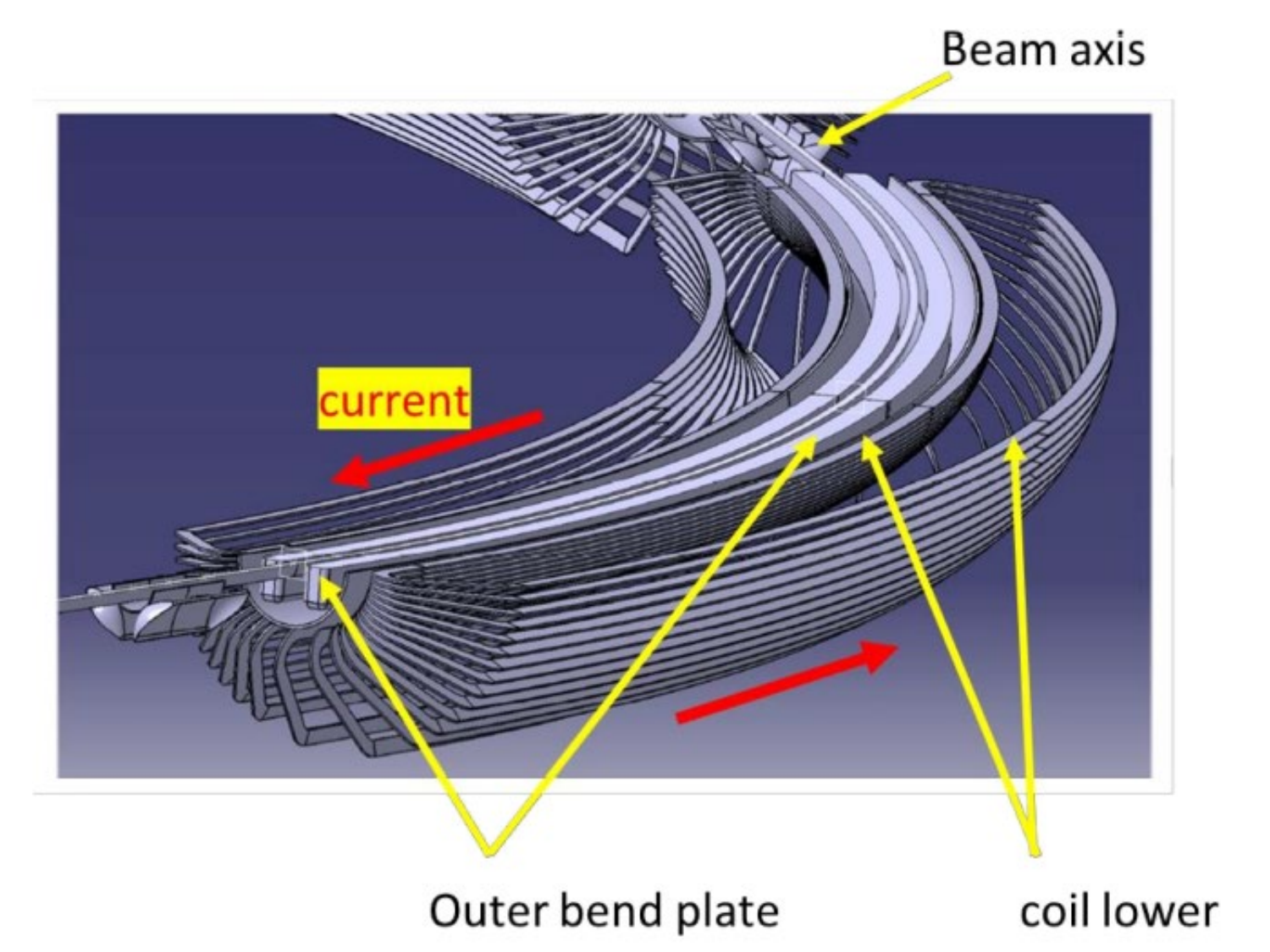}
\caption{\label{fig:PTR-layout-Toroidal8_102p2-mod}
{\bf Left:\ }
Lattice layout for PTR, the proposed nuclear transmutation storage ring.
``Compromise quadrupoles'' are shown inset.  The circumference has been taken to be 102.5\,m, but 
the entire lattice can be scaled, e.g. to reduce peak field requirements. 
{\bf Right:\ }
Perspective sector mock-up of one PTR sector.  Cos\,$\theta$-dipoles surround the beam tube, within 
which are the capacitor plate electrodes.  The superimposed coil design is due to Helmut S\"oltner.
}
\end{figure}
PTR optical $\beta$-functions, with super-periodicity 8, are shown in Figure~\ref{fig:etaE-dependence}.
With magnetic bending to be treated as perturbation, this figure represents almost vanishing
magnetic perturbation; $\eta_E=0.9999$, $\eta_M=0.0001$.

The toroidal shape parameter $m_{\rm nom}$,
which expresses the radial power law dependence of the electric field in the form $E(r) = -E_0/r^{1+m}$, 
is taken to be $m_{\rm nom}=0.32349\approx1/3$.  The $m$-value for perfectly cylindrical electrodes would
be $m=0$, for which there would be no vertical focusing. As indicated within the figure, the $m_{\rm nom}$
value has been tuned to make horizontal and vertical tunes differ by exactly 1---onto
the so-called ``difference resonance''.  This is a rare ``good'' resonance.

In the Bruno Touschek designed AdA, an $e_-,e^+$  storage ring prototype, AdA was first commissioned in
Frascati, then transferred to Orsay, in France. The first electron-positron collisions in a laboratory were
observed in 1963-1964 at the Laboratoire de l'Accélérateur Linéaire d'Orsay.  This ring was tuned to run
exactly on the difference resonance.
\begin{figure}[htb!]
\centering
\includegraphics[scale=0.90]{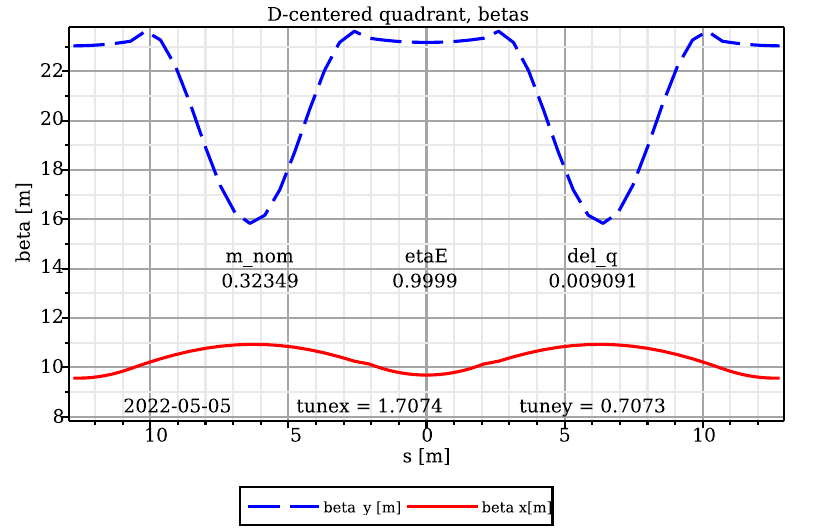}
\caption{\label{fig:etaE-dependence}Refined PTR tuning, with quad strengths and 
$m_{\rm nom.}$ (adjusted to 0.32349) for (distortion-free) equal-fractional-tune, $Q_x=Q_y+1$, 
operation on the difference resonance. Not counting geometric horizontal focusing,
thick lens pole shape horizontal and vertical focusing strengths are then identical.
Mnemonic: $m_{\rm nom.}$=1/3.}
\end{figure}

A ``second-stage'' racetrack-shaped, PTR rear-end Collider storage ring is shown
in Figure~\ref{fig:PTR collider optics}. For increased luminosity this ring
has two round beam, low beta intersection regions (IR), with low-$\beta$ optics,
for increased luminosity. ``Compromise triplets''s are used to produce the $\beta$-squeeze optics at the IRs.

\begin{figure}[htb]
\centering
\includegraphics[scale=0.13]{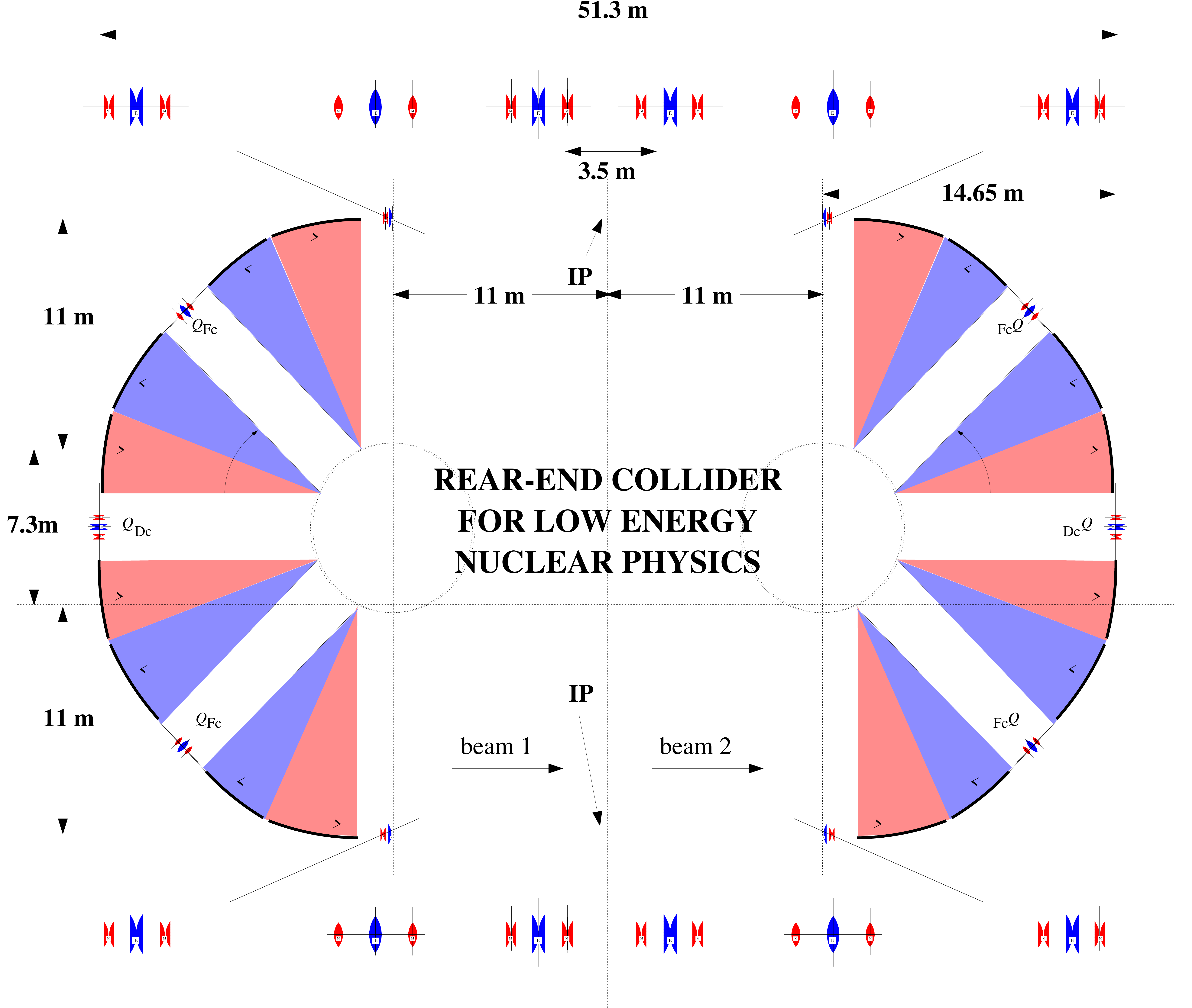}
\caption{\label{fig:PTR collider optics} Beta-squeezed round beam collider schematic.
The (seemingly quadrupole triplets) are actually ``compromise quadupole''
singlets---a central strong thin lens electrical quadrupole, say focusing, is
centrally superimposed on a weak ``thick'' quadrupole consisting of two identical
thin magnetic quadrupoles of the same focusing polarity.
This produces the four identical  DFD ``compromise triplets'' shown.
The compromise quadrupoles are powered separately with the same, or approximately the
same, $\eta_E$ and $\eta_M$ fractional strengths as the arcs.}
\end{figure}

\subsection{Predominantly magnetic ${\rm E}\&{\rm M}$ storage ring}
Rather than the predominantly electric ${\rm E}\&{\rm m}$ ``rear-end collider I have been promoting for
several years for laboratory-based astrophysics research, Jim Ritman has suggested, as an interim measure,
implementing rear-end nuclear transmutation collisions in a conventional, all magnetic storage ring.  This
could be implemented much more quickly and cheaply than an E\&m collider could be. Furthermore, the multi-$\alpha$
and multi-deuteron terminology introduced in this paper makes it easy to select isotope pairs that can co-circulate
in an all-magnetic ring, from those pairs which cannot.

Agreeing with this concept, I have come to believe that it is sensible to start all-magnetic, while preparing
for general ${\rm E}\&{\rm M}$ bending with arbitrary superposition of magnetic and electric bending.

Electron, nuclear-isotope pairs can also co-circulate simultaneously in the same storage ring,
but this is only possible with superimposed ${\rm E}\&{\rm M}$ bending. An important candidate channel for
laboratory-based nuclear astrophysics requiring ${\rm E}\&{\rm M}$ bending would be, for example,
$ e^- + helion \rightarrow triton + \nu $.  This has been described in greater detail in previous reports.

\subsection{Initial state control in ${\rm E}\&{\rm M}$ storage rings}
This paper is also concerned with replicating in the laboratory nuclear transmutation processes actually occurring
in the sun and other stars. \emph{Though insufficiently strong to alter gross solar kinetic particle distributions,
magnetic fields in the sun are amply strong to influence the kinematics of individual nuclear transmutation events.}
Such effects can be studied using co-traveling beams of different type in ${\rm E}\&{\rm M}$ storage rings.
This allows storage ring studies to more accurately represent initial state conditions, including spin control
along with exothermic adjustment to match actual transmutations in stars.  This is not possible using fixed
(or essentially fixed) targets in the laboratory. 

An essential distinction between astrophysical and laboratory-based nuclear physics investigations concerns the
degree to which ambient magnetic fields influence solar processes.  Astrophysically, the magnetic fields
cannot be controlled but, based on theory, may be inferred over long histories.  In the laboratory, precise
electric and magnetic fields can be applied conveniently.

All of the considerations mentioned so far, for kinetics in the presence of magnetic fields,
imply orbits containing short helical arc segments  with cyclotron radii and frequencies
that depend on isotope mass A and charge Z, as well as on kinetic energy KE. There will be a special concentration
of the so-called $\alpha$-material particle types, for which A=2Z, starting with deuterium, for which the orbits
are all the same, independent of A. Protons, as always, will provide the predominant concentration.

A point made repeatedly in this paper is that rear-end collisions have low CM energies that range continuously
from below to above the classically imposed Coulomb energy barrier that can be overcome only by quantum
mechanical tunneling. To investigate the issues just discussed in laboratory experiments,
a superimposed ${\rm E}\&{\rm M}$ storage ring is required.

To make this final point more explicit, lengthy tables of nuclear process collision rates
taken from two of the most prominent historical sources, Burbidge, Burbidge, Fowler, and
Hoyle (B${}^2$FH),\cite{B2FH} and Zobrov and Zeldovich\cite{Zobrov-Zeldovich} are copied
in Appendix~(\ref{fig:Dolgov-Zeldovitch-1981}).

In principle all of the processes exhibited in these tables can be measured in the ${\rm E}\&{\rm M}$
rings described in this paper.  Several processes with two available particle types as incident beams
have been analyzed in this paper.  Other than these processes, the kinematic practicality of the processes
in the appendix have not been individually checked.  Establishing and solving the quartic kinematic equation
is automatic, but, especially for very unequal incident masses, all of the kinematically interesting nuclear
processes may not be achievable in a single ring.

\subsection{Precision beam energy control and IP collision centering\label{sec:Precision}}
Almost every nuclear transmutation process discussed in this paper can be made capable of
measuring the process spin dependence. Since full discussion of this capability would increase
the length and complexity of this paper, only a few important features are mentioned in this section.

In all cases discussed the incident state contains two different particle types. In most cases
the beams consist of different nuclear isotopes. Preferably, neither beam would be
spinless, but this would not apply to alpha particles, as the most important example.
This case is not discussed.  Neither will cases with electron or positron as incident beams.
Certainly electron beams can be polarized, but the issues are complicated.

For now, we consider only the case of both beams being nuclear isotopes with neither spinless.
Most of the kinematic tables have columns labeled ``Qs1'' and ``Qs2'' which are the spin tunes
of the two incident beams.  Though these spin tune parameters may not be much implicated
in the physics, they are religiously included in already-too-complicated tables, as a reminder of
their critical experimental importance in controlling the kinematics of the experiments.

With superimposed E\&M bending it is always possible for one of
the beam spins to be ``globally frozen'' meaning Qs=0.  As a nominal starting condition
beam-1 will always be frozen; Qs1=0. Since the second beam can be ``locally-frozen'',
meaning that Qs2 is a non-zero integer, this too, will be a starting condition for beam-2.

Scanning the tables in this paper, it can be seen that both of these conditions, typically,
need to be sacrificed. In most cases, these formal starting constraints are incompatible with
the process being studied. Usually it is the requirement that rear-end collisions occur at a an IP
intersection point where a collision detector is located.  This condition is usually
incompatible with either beam spin being frozen.

For precision spin control it is useful for one or both beam spins to be frozen, since
this guarantees high precision beam control and reset-ability. But the absence of frozen
spins is not a ``show stopper'', provided that the spins can be measured and phase locked.
This capability, pioneered at the German Cooled Synchrotron (COSY), discussed in papers
listed in thebibliography;\cite{Eversmann}\cite{Hempelmann}.

Using simultaneous helion and triton beams as example,
Table~\ref{tbl:BendParms-PTR.1} exhibits kinematic parameters necessary for
IP collision point centering, by showing the dependence of
$bratio = \beta_2/\beta_1$ on KE1, the helion KE. For multiplication factor $t=3$, the second to
last column shows the product $t\times bratio$.  The middle row shows, for KE1=31.7\,MeV,
that $t\times bratio =  4.00015$, which is close enough to 4 to be phase-locked to exactly 4.

Not explained in the caption to
Table~\ref{tbl:BendParms-PTR.1}, is the fact that, in order to roughly center the table, a coarse
adjustment of the electric field away from the value at which the helion spin is frozen (i.e. $Q_{s1}=0$)
has been required. In this case the fractional electric bending factor $\eta_E$ has been increased by a
factor \emph{fudge}=1.1, and $Q_{s1}$ has changed accordingly. The \emph{modified value} of $\eta_{M1}$
is given in column 5 and, as always, the modified value of $\eta_{E1}$ is given by $1-\eta_{M1}$.
The bending fractions for beam~2 are adjusted accordingly.

Other important parameters not listed in Table~\ref{tbl:BendParms-PTR.1}, are magnetic field
$B_0$=-0.03069\,Tesla, and spin tunes $Q_{s1}$=0.42306, and $Q_{s2}$=-3.73310.  As explained
in an appendix discussing unit conversions, the use of ``accelerator units'' (in the present paper)
is straightforward for electric fields (which are measured in GeV/m rather than eV/m) but
confusing for magnetic fields. For this reason the magnetic field strength is usually conveyed
in this paper by the magnetic
fraction etaM1 of the magnetic bend for beam 1, which in this case is helions. The identity
$\eta_M + \eta_E=1$ is always true; either sign can be negative, but the sum is always 1.
Though beams 1 and 2 share the same magnetic field,
their magnetic bend fractions (necessarily) differ: $\eta_{M1}\ne\eta_{M2}$.

Explained elsewhere in the paper are the reasons why spin tunes $Q_s$ are shown to such high precision
in some tables in spite of the fact that this makes the tables long and indigestible,
\begin{table}[htb]\scriptsize
\caption{\label{tbl:BendParms-PTR.1}Fine-grain scan to center the collision point for co-traveling  KE1=31.7\,MeV helion 
kinetic energy and 17.702\,MeV triton KE.  The bend radius is $r_0=11$\,m.   CM quantities are indicated
by asterisks (*). Q12 is the sum of CM kinetic energy values (which is comparable with the Coulomb barrier potential).}
\centering
\begin{tabular}{c|cc|cc|cc|ccc|c|c}
\toprule
   bm & beta1 &  KE1  &     E0   &   etaM1   &  beta2  &    KE2 &  beta*  &     M*  &  Q12   & t,t*bratio & bm \\ 
   1 &        &  MeV  &   MV/m   &           &         &    MeV &         &    GeV  &  keV   &  3         &  2 \\ \midrule
   h & 0.1443 & 29.7  &  3.96487 &  -0.47620 &  0.1082 & 16.582 & 0.12628 & 5.61826 & 945.8  & 4.00148    &  t \\ 
   h & 0.1467 & 30.7  &  4.10054 &  -0.47724 &  0.1100 & 17.142 & 0.12836 & 5.61829 & 977.2  & 4.00081    &  t \\ 
   h & 0.1490 & 31.7  &  4.23635 &  -0.47828 &  0.1117 & 17.702 & 0.13041 & 5.61832 & 1008.6 & 4.00015    &  t \\ 
   h & 0.1513 & 32.7  &  4.37230 &  -0.47932 &  0.1135 & 18.262 & 0.13243 & 5.61835 & 1040.0 & 3.99948    &  t \\ 
   h & 0.1535 & 33.7  &  4.50839 &  -0.48036 &  0.1152 & 18.822 & 0.13441 & 5.61838 & 1071.3 & 3.99882    &  t \\ 
\bottomrule
\end{tabular}
\end{table}

With velocity ratio 4/3, multiplying by t=3 produces the central entry in the second last column,
which is close enough to 4 for exact phase locking.  While the triton bunch makes three complete revolutions the
helion bunch makes four.  The  electric field value is 4.23635\,MV/m (common, obviously, to both beams).
The fractional bending factor of the helion beam is $\eta_{M1}=-0.47828$.  Being negative, the helion bending 
superposition is ``destructive'', with $\eta_E=1.4782$.
  
Incident $t$ and $h$ beams can both be nearly 100\% polarized. Since both magnetic dipole moments (MDM)
are known to 9 decimal points, and their Larmor precessions can be stabilized to the same accuracy,
the initial state 4-momenta will be known and reproducible to the same accuracy.  In the overall CM, the hadron-jet
and lepton-jet 3-momenta are equal (but opposite). Explanation of the ``jet'' terminology follows.

For elastic scattering processes (1) and (2) (in the following section) the kinematics is highly over-determined.
And, in any case,
the output angles can also be measured quite accurately.  For $\beta$-decay processes, for example to determine
the neutrino mass, it is useful to interpret processes (3) and (4) using perfectly collimated incident
state hadron jet,
(quite tightly collimated) final state hadron jet, and {\bf almost isotropic} final state lepton jet.
The precision with which the neutrino mass can be measured depends on the accuracy with which the jet kinematics 
can be reconstructed.

\subsection{``Elastic/pseudo-elastic'' nuclear collisions}
Consider two body elastic and inelastic collisions
$$ t + h \rightarrow n_3 + n_4 + \dots $$
where, for simplicity, the incident nuclei have been taken to be 
   $n_1 = t (triton) \equiv {}^3H_1^+\equiv 3H$ and
   $n_2= h (helion) \equiv {}^3He_2^{++}\equiv 3He$.
and $n_3$ and $n_4$ are the same, or other long-lived isotope nuclei, 
such as $p = {}^1H_1^+$, $d={}^2H_1^+$, or $\alpha={}^4He_2^{++}\equiv 4He$, etc. 

Specializing further, consider elastic two body strong nuclear interaction processes,
$$ t + h \rightarrow t + h \qquad (1)$$
$$ t + h \rightarrow d + \alpha \qquad (2)$$
or inelastic, weak nuclear reactions (i.e. $\beta$-decay processes),
$$ t + h \rightarrow t + (t + e^+ + \nu) \qquad (3)$$
$$ t + h \rightarrow (h + e^- + \nu) + h \qquad (4)$$
formally treated as two-body processes. The output state can be treated as two-body
(as delineated by parentheses), with the parenthesized combination being the second body.

For now we consider neutrino mass determination in process (3).
All such nuclear scattering events can be studied in the PTR ring,
with $t$ and $h$ beams of different velocity, co-circulating in the same 
direction, at the same time.

Faster beam bunches (say helions) will ``lap'' the slower triton bunches
regularly.
Conditions can be controlled such that the resulting rear-end collisions all
occur within detection apparatus recording scattering events 
(1) through (4) at an intersection point (IP), such as those shown in Figure~\ref{fig:PTR collider optics}.

Uniquely special and important about these events is that they occur in a reference
frame that is moving with semi-relativistic velocity in the laboratory.

\section{Direct storage ring nuclear transmutation}
\subsection{Tantalizing compound nucleus ${}^8_4{\rm Be}^*$  \label{sec:8Be.1} }
Consider the nuclear transmutation channel into a compound nuclear state,
$$ {}^6_3{\rm Li} + {}^2_1{\rm H} \rightarrow {}^8_4{\rm Be^*}  \rightarrow ? .$$
with the release of 22.3\,MeV energy. It has already been noted that the ${}^8_4$Be isotope of beryllium
is unstable, with a lifetime of $81.9\times10^{-18}$\,s. One might have  anticipated  ${}^8_4$Be as being
simply formed from two $\alpha$-particles.  This calls into question the virtue of our favored
``$\alpha$-nuclear-material''. As a building material, beryllium $\alpha$-material hardly seems promising.
Like bricks, two alphas needs mortar for permanence.

Quoting from Wikipedia under ``The short 8Be lifetime'' 8Be
has important ramifications in stellar nucleosynthesis as it creates a bottleneck in the creation of heavier
chemical elements.  The properties of 8Be have also led to speculation on the fine tuning of
the universe, and theoretical investigations on cosmological evolution had 8Be been stable.''
\footnote{In the context of the present paper, interpreting the exceedingly short lifetime of beryllium-8
as ``fine-tuning of the universe'' seems unnecessarily dramatic, as carbon can be produced without the
intervention of beryllium.}

Figure~\ref{fig:p-11B-resonance} shows the narrow $p+11B\rightarrow\alpha_1+\alpha_2+\alpha_3$
3-$\alpha$, measured resonant cross section of a 3-$\alpha$ state at the quite low energy of 149\,keV.
This shows a weakly, but  unambiguously exothermic resonance producing three stable alphas.
But, since the total final state rest energy exceeds the rest energy of any carbon isotope, this channel cannot
contribute to he nucleosythesis of carbon.
 \begin{figure}[hbt!]
   \includegraphics[scale=0.53]{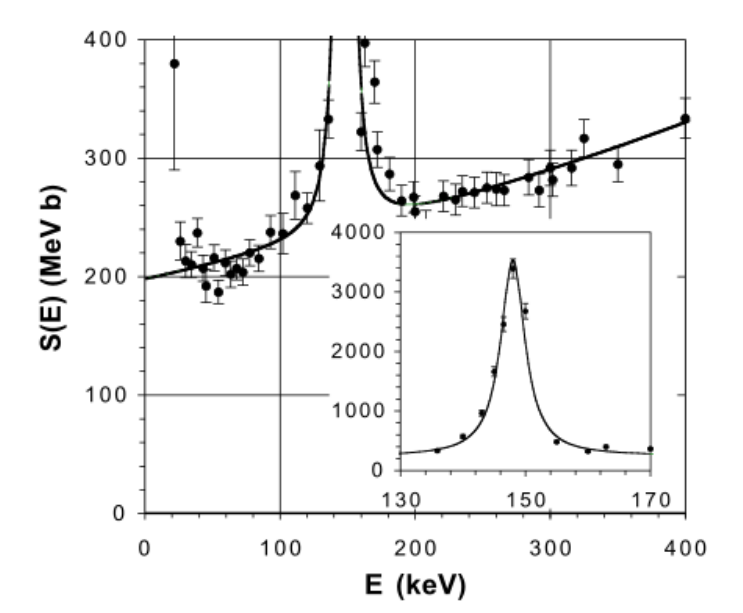}
   \includegraphics[scale=0.30]{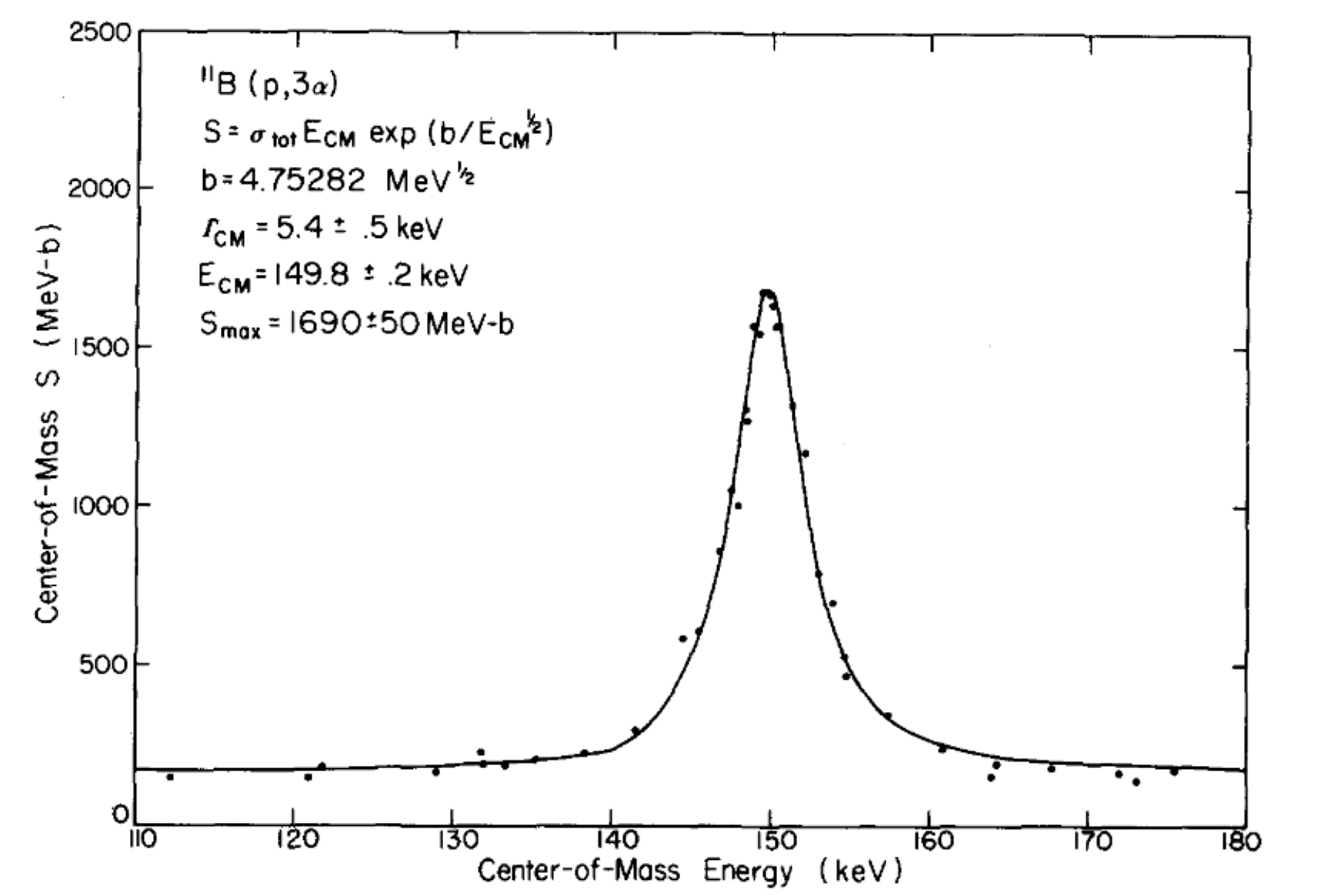}
   \caption{\label{fig:p-11B-resonance} {Experimental observation of the resonant production of
       an N=3 example of $\alpha$-matter; $ p + 11B \rightarrow \alpha_1 + \alpha_2 + \alpha_3. $}.
     The figure on the right is a blown up version of the inset on the left. which shows the
     energy dependence of the resonant cross section, as extracted from the raw data on the left.
     This is an example of direct experimental observation of an ``exothermic'' above threshold
     nuclear reaction.
   }
 \end{figure}

 Quoting again, not quite verbatim, from Wikopedia, ``the nuclear fusion reaction of
 two helium-4 nuclei produces beryllium-8, which is highly unstable, and decays back into
 smaller nuclei with a half-life of $8.19x10^{-17}$\,s, unless within that time a third alpha
 particle fuses with the beryllium-8 nucleus to produce an excited resonance state of
 carbon-12, called the Hoyle state, which nearly always decays back into three alpha
 particles, but once in about 2421.3 times releases energy and changes into the stable base
 form of carbon-12.''

 This issue has been investigated in the laboratory using the channel illustrated in
 Figure~\ref{fig:Abbrev-8Be-energy-diag}, copied from Sweeney and Marion\,\cite{Sweeney-Marion}, 
 \begin{figure}[hbt!]
   \includegraphics[scale=0.4]{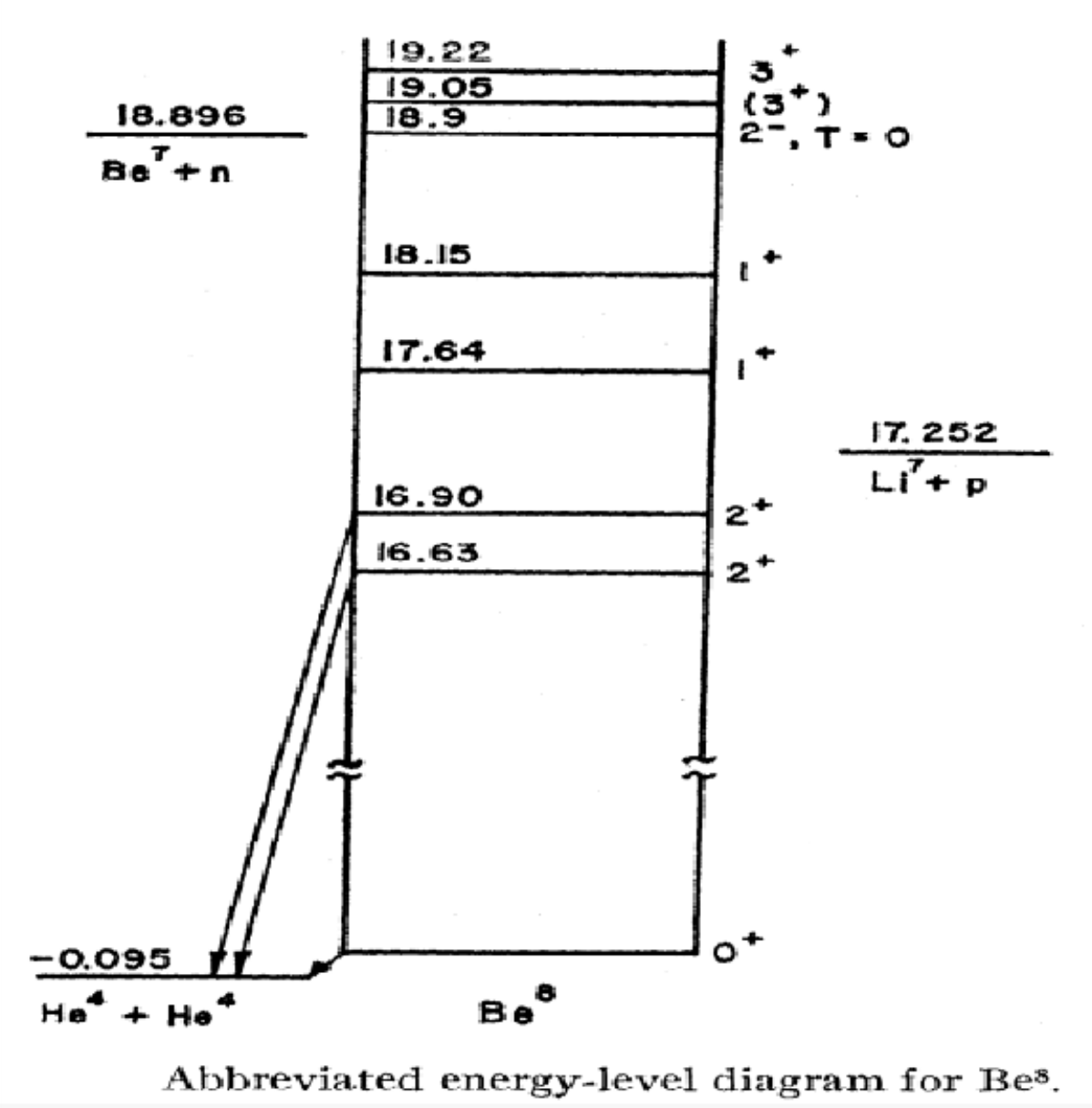}
   \includegraphics[scale=1.2]{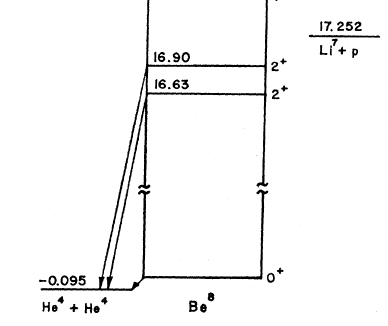}   
   \caption{\label{fig:Abbrev-8Be-energy-diag}Marion abbreviated 8Be energy level diagrams.
     This is an example' of an endothermic, below threshold or ``off-the-mass-shell'' process,
     that can only be accessed by extrapolation, as explained by Gamow.   }
 \end{figure}
 Much of what is known about solar nucleosynthesis has been inferred by extrapolation such as
 is exhibited in Figure~\ref{fig:Abbrev-8Be-energy-diag}. A virtue of rear-end collisions is that
 a process that is endothermic in one direction is exothermic in the other.  All example processes
 in the present paper are exothermic.  The term ``direct'', as applied to nuclear processes,
 has the same meaning.  It is the superimposed E\&M bending that makes this always possible.
 
 Numerous channels, similar to the process shown in Figure~\ref{fig:p-11B-resonance},
 supporting the commonality of alpha media, follow:
\begin{align}
   p + {}^{7}N   \rightarrow & 2\alpha,     & {}^{8}O  \\
   p + {}^{9}Be  \rightarrow & 2\alpha + d, & {}^{10}Ne \\
   p + {}^{11}B  \rightarrow & 3\alpha,     & {}^{12}Mg \\
   p + {}^{13}Ne \rightarrow & 3\alpha + d, & {}^{14}Si \\
   p + {}^{15}P  \rightarrow & 4\alpha,     & {}^{16}S  \\
   p + {}^{17}Cl \rightarrow & 4\alpha + d, & {}^{18}Ar \\
   p + {}^{19}K  \rightarrow & 4\alpha,     & {}^{20}Ca \\
   p + {}^{21}Sc \rightarrow & 4\alpha + d, & {}^{22}Ti \\
   p + {}^{23}V  \rightarrow & 4\alpha,     & {}^{24}Cr \\
   \notag
\end{align}

Further discussion of nucleosynthesis of carbon is deferred to Section~\ref{sec:13C},
following discussion of  the laboratory availability of Lithium isotope beams.
\clearpage

\subsection{Laboratory availability of intense light nuclear isotope beams }
\begin{figure}[hbt!]
  \centering
\includegraphics[scale=0.4]{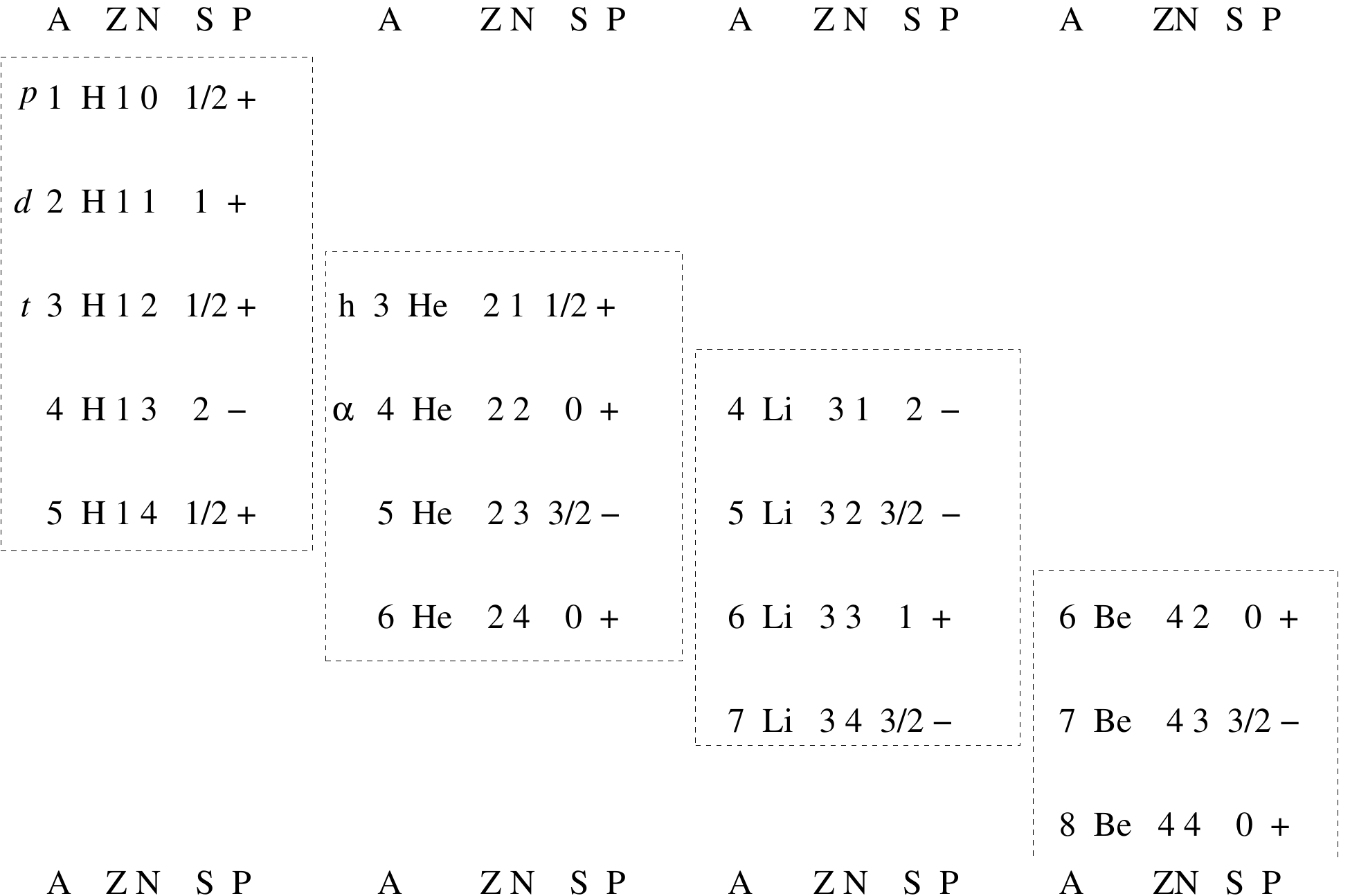}
\caption{\label{fig:Low-mass-candidates}
Figure showing a table provided by D. Raparia,\cite{Raparia} of side-by-side low mass ($0<Z<5,\ A<9$)
candidates for free electron or positron capture $\beta$-decay'' transmutations.  Every entry
(except 8Be) has at least a multi-year lifetime and can be linac-accelerated to 60\,MeV/nucleon energy
to produce mA-level average current beams ($0.6\times10^{16}\,p$\,/s) with 1/2 percent energy spread.}
\end{figure}

\subsection{Storage ring study of
   ${}^6_3{\rm Li} + {}^2_1{\rm H} \rightarrow {}^8_4{\rm Be^*}\rightarrow {}^4_2{\rm He} +  {}^4_2{\rm He}$
process}{\label{sec:8Be.2}}
The curiously short lifetime of ${}^8_4{\rm Be^*}$ was discussed in section~\ref{sec:8Be.1}.
This channel plays an important role in what has become the ``standard model' of the nucleogenesis
of carbon in the sun. \footnote{As it happens, one of the processes in this solar nucleogenesis chain
is the source of
the neutrinos whose production rate was measured (in the famous Ray Davis experiment) to be too low
(by roughly a factor of three). This has, more recently, been accepted as evidence for neutrino
regeneration which (in this case) causes neutrinos to ``vanish'', or rather, convert, into a
neutrino to which the Davis detection process is ``blind''.}
A pictorial representation of this model (copied from Wikipedia) is shown on
the left in Figure~\ref{fig:Hoyle-carbon-wki-model}
   \begin{figure}[htb!]
   \centering
   \includegraphics[scale=0.5]{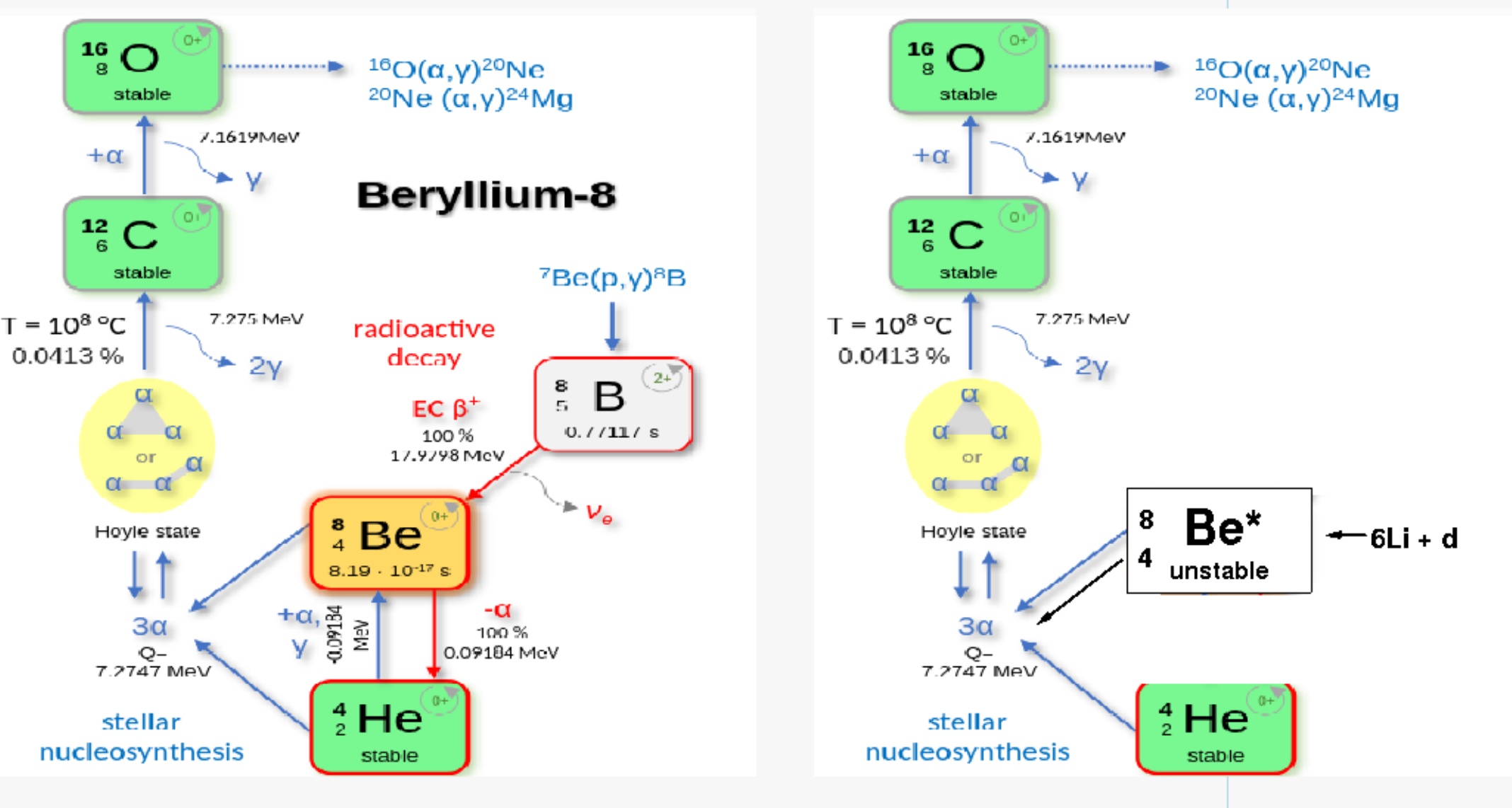}
   \caption{\label{fig:Hoyle-carbon-wki-model}Figure copied from Wikipedia, showing, on the left,
     the theoretically calculated chain of nuclear processes by which lighter nuclear isotopes
     are converted to carbon and then oxygen. On the right is shown a candidate replacement (proposed in
     the present paper) for replacement of what may be the weakest link in this chain. More speculative
     than serious, this measurement proposal is intended to support the laboratory capability of measuring
     nuclear processes in moving frames of reference.
   }
   \end{figure}
   On the right of the same figure is a representation of essentially the same model with the single
   alteration that the ${}^8_4{\rm Be^*}$ is produced by ${}^6_3{\rm Li} + {}^2_1{\rm H}$
   rather than involving boron. This replacement channel is presumably inferior since, otherwise it
   would have become part of the standard model. But this channel is itself interesting because of the
   importance of lithium in the evolution of light isotopes in the sun, and for which
   the interpretation of traditional experimental results remains murky\cite{Taskaev-Bikchurina}.
   Important channels include
   \begin{align}
   6Li + d &= 2\ alphas + 22.38 MeV;\\
   6Li + d &= n + 7Be + 3.385 MeV;\\
   6Li + d &= p + 7Li + 5.028 MeV;\\
  \end{align}
  Parameters for an E\&M storage ring investigation  are given in
  Table~\ref{tbl:BendParms-6Li-d.1}.  The compound nucleus velocity in
  this table is labeled $\beta^*$, with value 0.06, for the central,
  15 MeV 6Li kinetic energy.
  
  Also given are the spin tunes Qs1 and Qs2, of the incident state 6Li and d beams.
  As discussed previously, these spin tunes play play two significant roles.
  Since their anomalous magnetic moments are known to 8 or 9 decimal
  places, and the spin tunes can be measured to the same accuracy, the
  process kinematics can be established with high precision and restored
  to the same precision in multiple runs over days or months.
  Precise initial state process spin control is the second role.
  
  Furthermore this process serves as an introduction to the lithium to carbon
  alchemy promised in the title to this paper. By accelerating both incident
  lithium beams to sufficiently high energy the direct transmutation into 13C carbon
  becomes exothermic.  Parameters for this process are given in
  Table~\ref{tbl:BendParms-7Li-6Li.2} on the sideways page.   

 A screenshot from the MAPLE program that evaluates the electric and magnetic
 fields of co-rotating beams by solving a quartic equation is shown in
 Figure~\ref{fig:6Li+d-alphas-eq-graph}.
 In general, a quartic equation has four roots, of which zero, two, or four are real.
 In this case it can be seen either by the factorization or by the plot of the
 left hand side of the equation, that all four roots are real. The quantity ``x''
 that is plotted actually stands for the momenta (expressed as a signed kinetic energy
 in MeV).

Being negative the first (and negative) root corresponds to beam 2, in this case deuterons,
counter-rotating.  The remaining three roots correspond to the deuterons co-rotating.
In Table~\ref{tbl:BendParms-6Li-d.1} parameters for the lowest energy co-rotating
beam are shown.  Triple rows are shown in each case, to cover a range in which beam
crossings at a fixed ring location can be phased-locked to the value of ``turns'' $t$,
that is shown in the second to last column. Rows in the other table on the same rotated
are tripled for the same reason, but the process is different and the beam energies are
significantly higher.

\begin{figure}[htb!]
\centering
\includegraphics[scale=0.45]{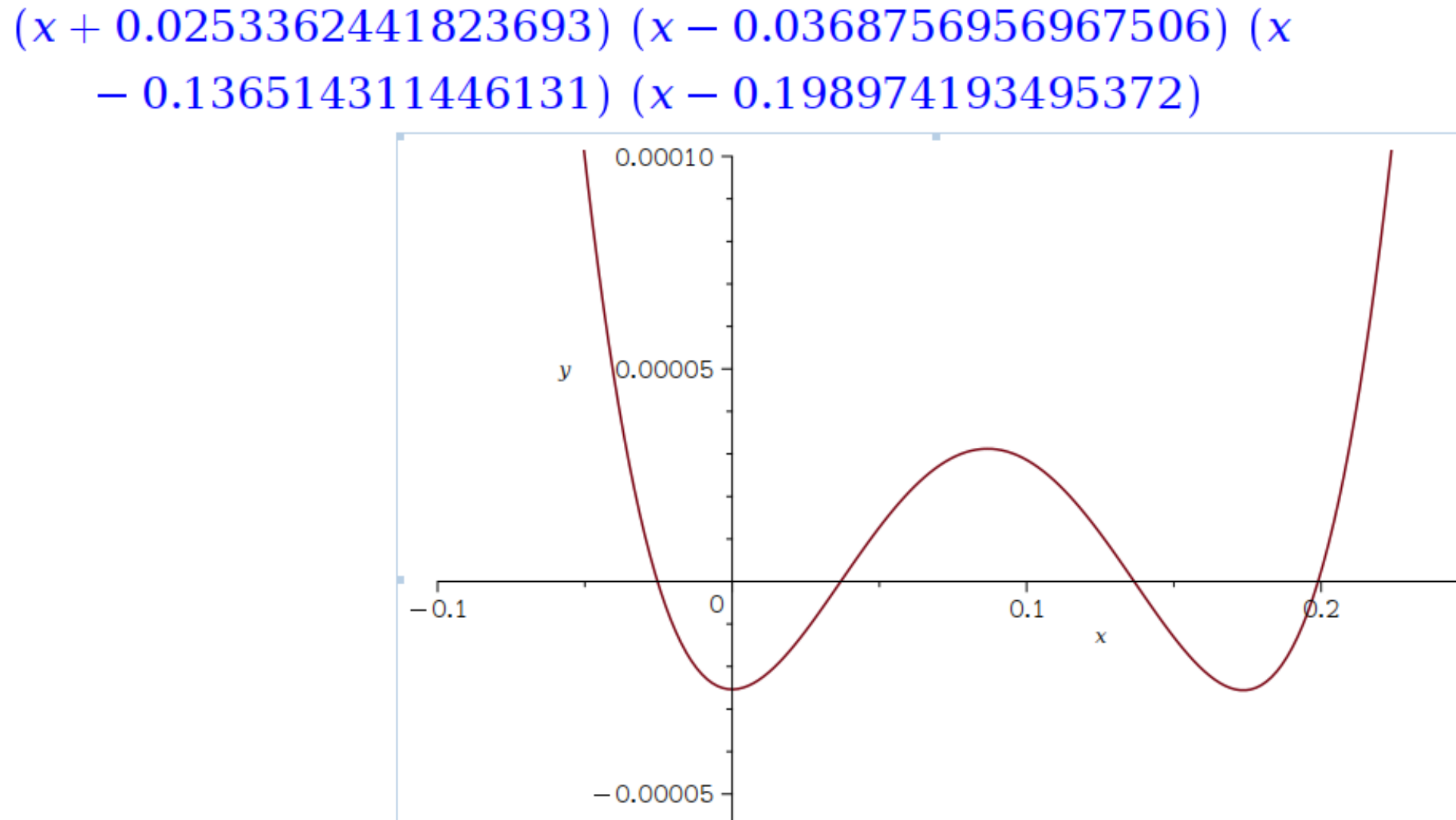}
\caption{\label{fig:6Li+d-alphas-eq-graph} Screenshot of MAPLE worksheet
evaluating the roots of the quartic equation whose solutions establish
mutually consistent parameters for the stable storage of two beam types.
In this case $x$ is the momentum (expressed as energy $pc$ in accelerator
units) of a deuteron co-circulating along with a 15 MeV 6Li nuclear isotope.
The only negative root is counter-rotating  and the other three roots are
co-rotating. This data corresponds to the central row in
Table~\ref{tbl:BendParms-6Li-d.1} and to the final state angle and kinetic
energy graphs in Figure~\ref{fig:7Li-6Li-to-13C-gam-KE-angl-lab.1}.}
\end{figure}

   \clearpage

 \subsection{Lithium to carbon alchemy; ${\rm 6LI} + {\rm 7Li} \rightarrow {\rm 13C} + \gamma$  \label{sec:13C}}
 Though lithium was present very early in the history of the sun, it has never figured prominently
 in the building up in the sun of the elements in the periodic table.  The terrestrial abundance ratio
 of 6Li/7Li is, 7.5/92.5, and the Q-value for the process
         $${\rm 6LI} + {\rm 7Li} \rightarrow {\rm 13C} + \gamma$$
 is 25.9\ MeV.  Though the cross section for the process is significantly large, the process is
 heavily suppressed by the Coulomb barrier repulsion between triply charged nuclei, especially for
 nuclear interactions between isotopes distributed isotropically in a Maxwell-Boltzmann gas.

 The present paper is primarily concerned with nuclear processes produced in an E\&M storage ring, which
 may better represent the building up of the periodic table in a solar environment in which magnetic
 bending is superimposed on gravitational bending alters the process probabilities significantly. 
 
Calculated kinematic variables for a range of energies for 6Li and 7Li circulating in the same direction
in an E\&M storage ring are shown in Figures~\ref{fig:7Li-6Li-to-13C-gam-KE-angl-lab.1} and
\ref{fig:7Li-6Li-to-13C-gam-KE-angl-lab.2}.  It can be seen from
Figure~\ref{fig:7Li-6Li-to-13C-gam-KE-angl-lab.2} that the process is kinematically accessible for beam energies
much lower than 100\ MeV.  Table~\ref{tbl:BendParms-7Li-6Li.2} displays initial state parameters for 
the processes shown in these figures.

The incident state in this reaction contains 6 protons and 7 neutrons,
with total rest energy of approximately 13 GeV.  Absent the infusion of a prodigious amount of energy,
and violation of other conservation laws as well, any possible final state produced by the strong nuclear force must
have the same number of protons and neutrons.  This is the exact composition of carbon-13. 

This is the nucleogenesis channel process under
consideration.  What is proposed is to accelerate both beams to kinetic energies, up to, say 100\ MeV.  Though this
is an energy too great to be ignored in nuclear physics, it is an amount of energy insufficient to produce
another nucleon.

\begin{figure}[hbt!]
  \centering
  \includegraphics[scale=0.55]{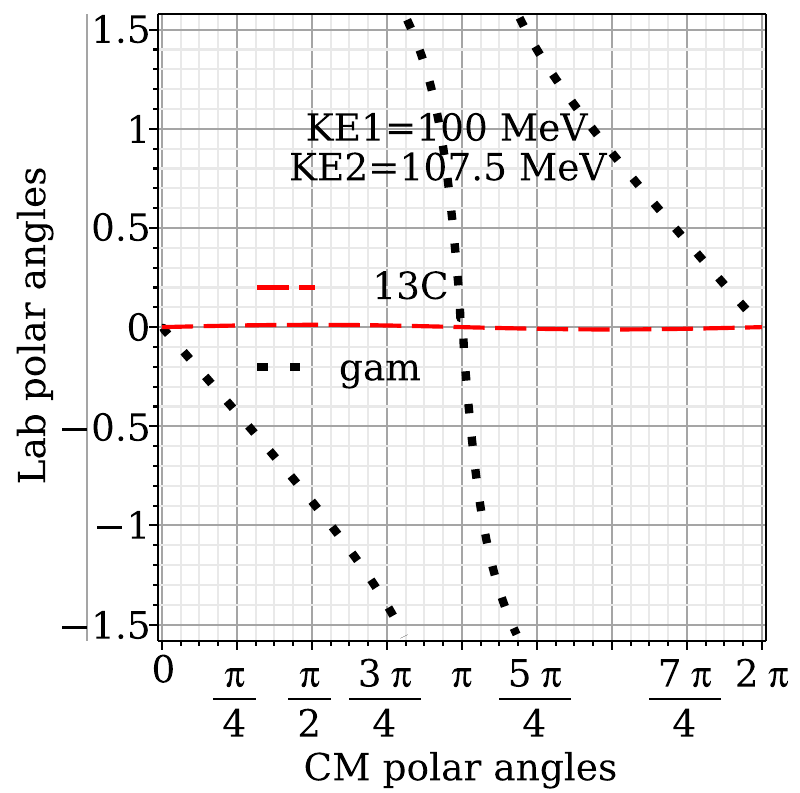}
  \includegraphics[scale=0.55]{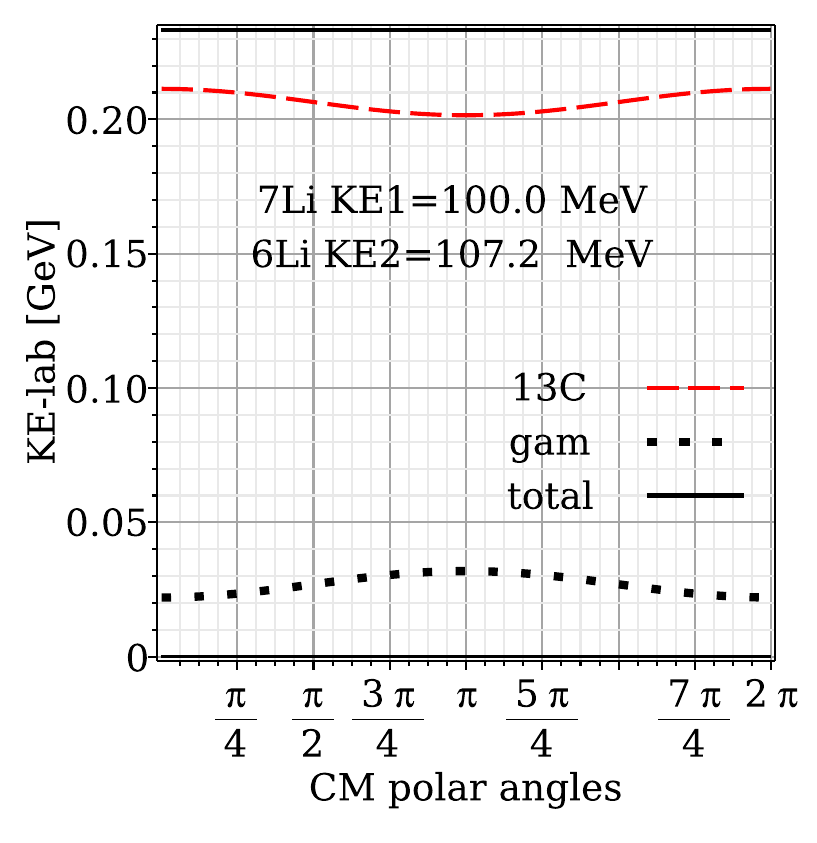}
\caption{\label{fig:7Li-6Li-to-13C-gam-KE-angl-lab.1}
  Graphs of final state angular and kinetic energy as functions of center of mass angle for the
  process ${\rm 6LI} + {\rm 7Li} \rightarrow {\rm 13C} + \gamma$. The incident kinetic energies are
  high enough for the process to be exothermic; i.e. both final state KE's are positive.
  Note, in the figure on the left, that, at the (admittedly-coarse) vertical scale of the graph,
  the produced carbon-13 isotope never leaves the beam axis. 
}
\end{figure}

\begin{figure}[hbt!]
  \centering
 \includegraphics[scale=0.55]{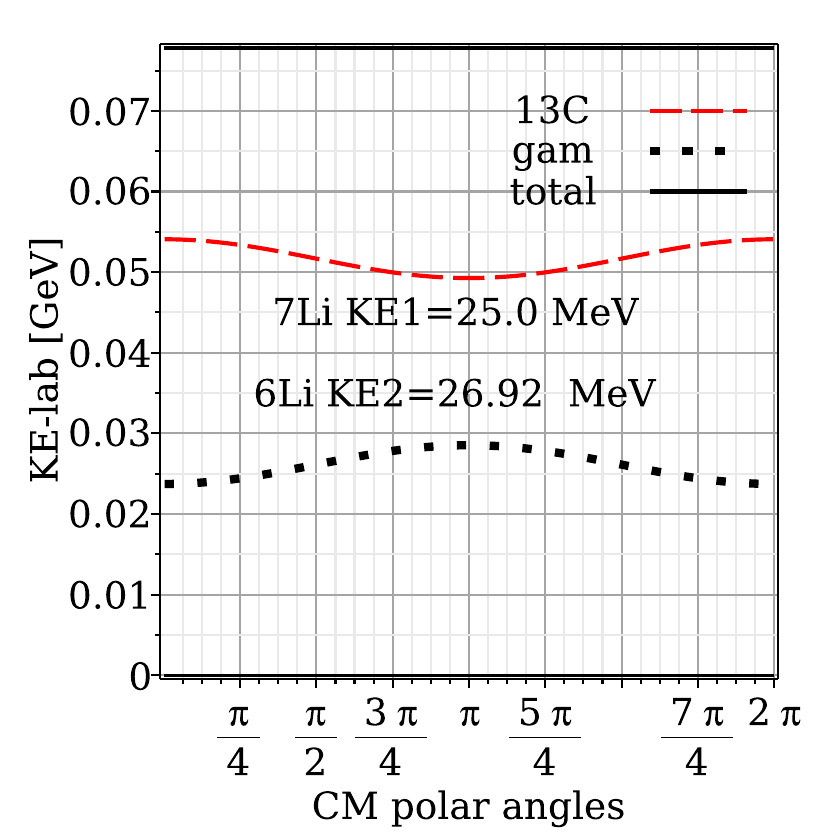}
 \includegraphics[scale=0.55]{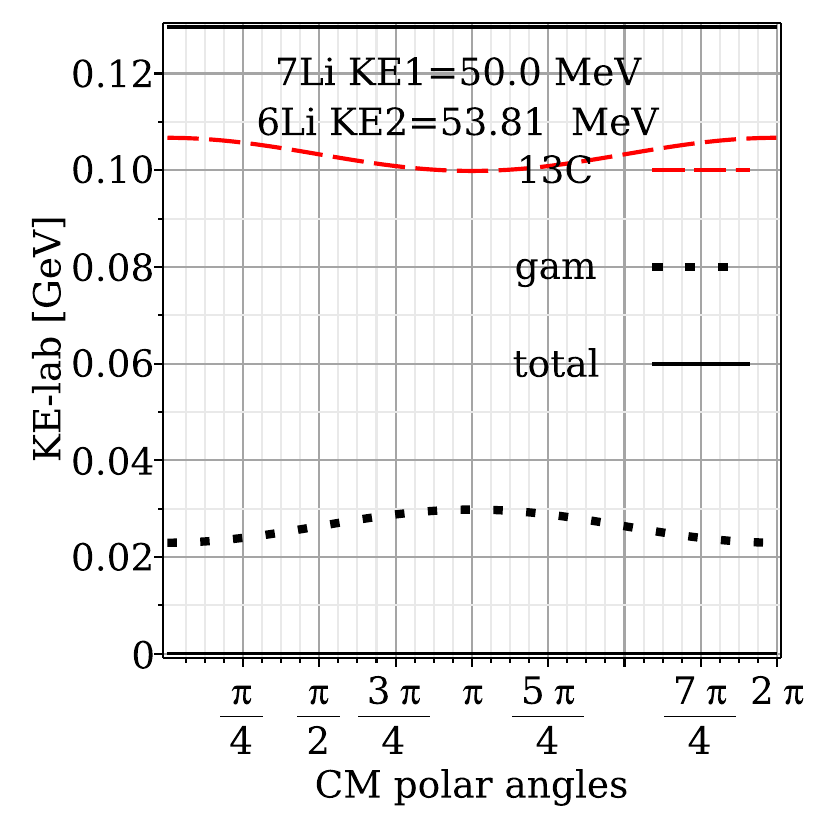}
 \includegraphics[scale=0.55]{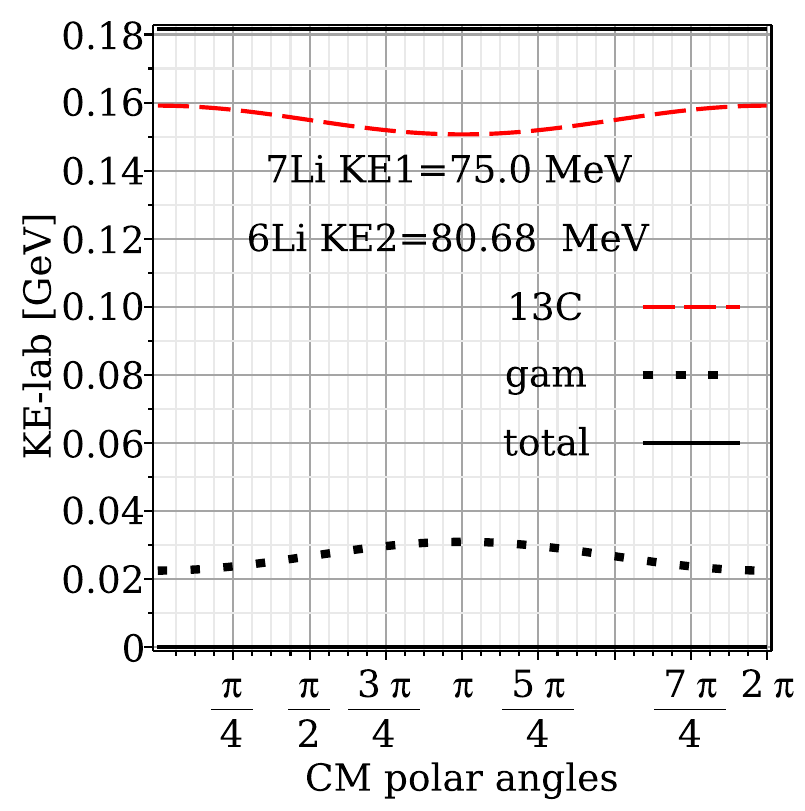}
 \includegraphics[scale=0.55]{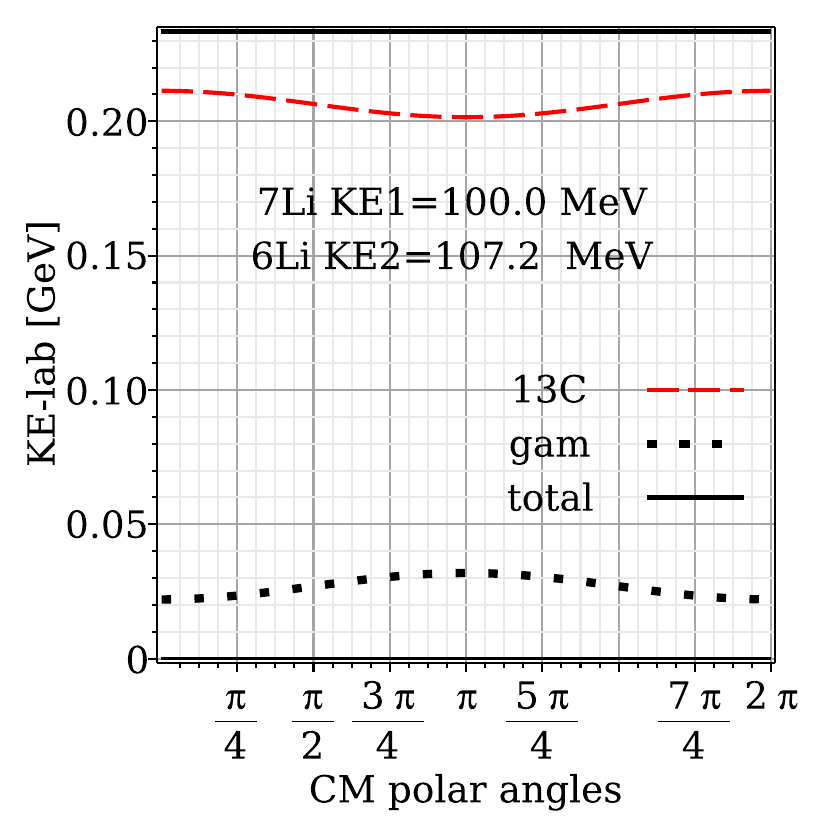}
\caption{\label{fig:7Li-6Li-to-13C-gam-KE-angl-lab.2}
  Graphs of kinetic energy as functions of center of mass angle for the
  process ${\rm 6LI} + {\rm 7Li} \rightarrow {\rm 13C} + \gamma$. In all cases
  The incident kinetic energies are high enough for the process to be exothermic;
  i.e. both final state KE's are positive.}
\end{figure}

\clearpage

\begin{sidewaystable}
  \caption{\label{tbl:BendParms-6Li-d.1}Fine-grain scans for 10, 15, and 20 MeV kinetic energies for the  6Li-d process.
    The integer factors ``t'', different for each triplet of rows, have been adjusted so that each triplet range includes the
    case in which rear end collisions recur at same ring location.}
\begin{center}
\begin{tabular}{c|ccc|cc|ccc|cccc|cc} \toprule  \small
   bm &  beta1 &    Qs1 &    KE1 &     E0 &  etaM1 & beta2  &    Qs2 &   KE2  & beta*  & gamma* &    M*  & Q12*   & t,t*bratio &   bm  \\
    1 &        &        &    MeV &   MV/m &        &        &        &    MeV &        &        &    GeV &    keV & intfac=?   &  2 \\ \midrule
  6Li & 0.0600 &  0.037 & 10.071 & -0.164 &  1.268 & 0.0161 &  3.073 &  0.242 &  0.049 &  1.001 &  7.466 &   1357 & 14.924 &     d \\ 
  6Li & 0.0602 &  0.037 & 10.152 & -0.165 &  1.268 & 0.0161 &  3.135 &  0.242 &  0.049 &  1.001 &  7.466 &   1372 & 14.984 &     d \\ 
  6Li & 0.0604 &  0.037 & 10.233 & -0.166 &  1.268 & 0.0161 &  3.199 &  0.242 &  0.049 &  1.001 &  7.466 &   1386 & 15.043 &     d \\ \midrule
  6Li & 0.0734 &  0.037 & 15.107 & -0.246 &  1.269 & 0.0197 &  3.071 &  0.364 &  0.060 &  1.002 &  7.467 &   2033 & 14.902 &     d \\ 
  6Li & 0.0737 &  0.037 & 15.228 & -0.248 &  1.269 & 0.0197 &  3.134 &  0.364 &  0.060 &  1.002 &  7.467 &   2056 & 14.962 &     d \\ 
  6Li & 0.0740 &  0.037 & 15.349 & -0.250 &  1.269 & 0.0197 &  3.198 &  0.364 &  0.060 &  1.002 &  7.467 &   2078 & 15.021 &     d \\\midrule
  6Li & 0.0847 &  0.037 & 20.142 & -0.328 &  1.269 & 0.0228 &  3.070 &  0.486 &  0.069 &  1.002 &  7.467 &   2709 & 14.880 &     d \\ 
  6Li & 0.0850 &  0.037 & 20.304 & -0.331 &  1.269 & 0.0228 &  3.133 &  0.486 &  0.069 &  1.002 &  7.467 &   2739 & 14.939 &     d \\ 
  6Li & 0.0853 &  0.037 & 20.466 & -0.334 &  1.269 & 0.0228 &  3.197 &  0.486 &  0.070 &  1.002 &  7.467 &   2769 & 14.999 &     d \\ 
\bottomrule
\end{tabular}

\caption{\label{tbl:BendParms-7Li-6Li.2}Fine-grain scan, for 25, 50, 75, and 100 MeV kinetic energies for the 7Li-6Li process. }
\begin{tabular}{c|ccc|cc|ccc|cccc|cc} \toprule MeV 
   bm &  beta1 &    Qs1 &    KE1 &     E0 &  etaM1 &  beta2 &    Qs2 &   KE2  & beta*  & Q34* &    M*  & Q12*   & t,t*bratio &   bm  \\ 
    1 &        &        &    MeV &   MV/m &        &        &        &    MeV &        &    MeV &    GeV &     keV &     t=19   &     2 \\ \midrule
  7Li & 0.0875 &  0.610 & 25.176 &  0.555 &  0.635 & 0.0977 & -0.460 & 26.917 &  0.092 &  1.004 & 12.136 &   158.00 &  8.961 &   6Li \\ 
  7Li & 0.0879 &  0.610 & 25.379 &  0.560 &  0.635 & 0.0977 & -0.460 & 26.917 &  0.092 &  1.004 & 12.135 &   147.33 &  8.997 &   6Li \\ 
  7Li & 0.0882 &  0.610 & 25.582 &  0.564 &  0.635 & 0.0977 & -0.461 & 26.917 &  0.093 &  1.004 & 12.135 &   137.03 &  9.033 &   6Li \\  \midrule
  7Li & 0.1234 &  0.612 & 50.355 &  1.117 &  0.633 & 0.1376 & -0.461 & 53.809 &  0.130 &  1.009 & 12.136 &   313.83 &  8.970 &   6Li \\ 
  7Li & 0.1239 &  0.612 & 50.762 &  1.126 &  0.633 & 0.1376 & -0.462 & 53.809 &  0.130 &  1.009 & 12.136 &   292.47 &  9.005 &   6Li \\ 
  7Li & 0.1244 &  0.612 & 51.170 &  1.135 &  0.633 & 0.1376 & -0.463 & 53.809 &  0.131 &  1.009 & 12.136 &   271.85 &  9.041 &   6Li \\  \midrule
  7Li & 0.1508 &  0.615 & 75.535 &  1.685 &  0.630 & 0.1679 & -0.463 & 80.677 &  0.159 &  1.013 & 12.136 &   467.54 &  8.978 &   6Li \\ 
  7Li & 0.1514 &  0.615 & 76.149 &  1.699 &  0.630 & 0.1679 & -0.464 & 80.677 &  0.159 &  1.013 & 12.136 &   435.45 &  9.014 &   6Li \\ 
  7Li & 0.1520 &  0.615 & 76.766 &  1.713 &  0.630 & 0.1679 & -0.465 & 80.677 &  0.159 &  1.013 & 12.136 &   404.50 &  9.050 &   6Li \\  \midrule
  7Li & 0.1736 &  0.617 & 100.718 &  2.260 &  0.627 & 0.1932 & -0.465 & 107.521 &  0.183 &  1.017 & 12.136 &   619.16 &  8.986 &   6Li \\ 
  7Li & 0.1743 &  0.617 & 101.541 &  2.279 &  0.627 & 0.1932 & -0.466 & 107.521 &  0.183 &  1.017 & 12.136 &   576.32 &  9.022 &   6Li \\ 
  7Li & 0.1750 &  0.617 & 102.368 &  2.298 &  0.627 & 0.1932 & -0.467 & 107.521 &  0.183 &  1.017 & 12.136 &   535.01 &  9.058 &   6Li \\ 
\bottomrule
\end{tabular}
\end{center}
\end{sidewaystable}

\clearpage
   
For elastic scattering processes (1) and (2) the kinematics is highly over-determined. And, in any case,
the output angles can also be measured quite accurately.  For $\beta$-decay processes, for example to determine
the neutrino mass, it is useful to interpret processes (3) and (4) using perfectly collimated incident state hadron jet,
(quite tightly collimated) final state hadron jet, and {\bf almost isotropic} final state lepton jet.
The precision with which the neutrino mass can be measured depends on the accuracy with which the jet kinematics 
can be reconstructed.


\section{Other laboratory-based nuclear investigations}
\subsection{``Two-body'' jet kinematics}
Experimentally, since the unscattered triton, call it $t_1$, and the scattered triton $t_2$ are nearly parallel,
with accurately known momenta, the 4-momentum of the final state hadron jet will be known to high precision. 
Our measured $e^+$ energy and direction provide sufficient further information to establish the entire kinematics
with good precision.

Assuming optimal control of initial spin orientations, and optimal final state kinematic
reconstruction, the precision of neutrino mass determination is expected to be approximately $\pm 10$\,eV.
Though very good, this is inferior to the KATRIN neutrino mass determination precision by more than an order of magnitude
On the other hand, the storage ring event-by-event characterization is vastly superior.
Though not every event detection is fully inclusive, event characterization is too detailed for any experimental
background events to survive.  In particular, there is no molecular beam ``spectator'' electron to confuse the event
interpretation.

The storage ring weak interaction cross section, which is proportional to the lepton energy,
can be much greater than for terrestrial beta decay.  A sample of two body kinematics for the process
    $$ (t + h) \rightarrow (t_1 + t_2) + (e^+ + \nu) $$
is exhibited in Figure~\ref{fig:IN-jet-H-jet-L-jet-head-on-reflected}.
\begin{figure}[hbt!]
\centering
\caption{\label{fig:IN-jet-H-jet-L-jet-head-on-reflected}. 
Two-body jet scattering representation of the $ (t + h) \rightarrow (t_1 + t_2) + (e^+ + \nu) $
nuclear transmutation process.}
\end{figure}

At the low energies of nuclear physics, a convenient rule of thumb for two body kinematics is that, while the
center of mass momentum is dominated by the hadrons, (in this case incident state $(t + h)$ and 
final state $(t_1 + t_2)$), the kinetic energy is carried primarily by the light particles (in this case $(e+ \nu)$).
The 3-momenta of all three of the jets lie in the same plane.  The incident jet is truly ``jet-like''---the 3-momenta
share the same beam axis.  The final state hadron jet is also quite tightly collimated---laboratory separation angle
several degrees.  But the lepton jet could scarcely be less ``jet-like''---in their own CM the electron and neutrino 
momenta are isotropic and back-to-back. In the laboratory frame the same is only more or less true.

\subsection{Electron-induced, inverse triton beta decay}
Continuing to consider only helions and tritons, one can identify their possible initial and final state weak interaction
channels.  Summing subscripts and superscripts to check the equations, the basic tritium $\beta$ decay channel is
     $$ t \rightarrow h + e^- + \nu,\qquad {}_3{\rm H}^1 \rightarrow {}_3{\rm He}^2 + {}_0e^{-1} + \nu.  \qquad(5)$$
or, replacing $\nu$ on one side by $\nu$ on the other, with input/output interchange,
    $$ h + e^- \rightarrow t + \nu,\qquad {}_3{\rm He}^2 + {}_0e^{-1} \rightarrow  {}_3{\rm H}^1 + \nu,\qquad(6)$$
$$ \hbox{or\quad} h + e^- \rightarrow p + d + e^- + \nu + \nu,\quad {}_3{\rm He}^2 + {}_0e^{-1} \rightarrow
            {}_1{\rm H}^1 +  {}_2{\rm H}^1 + {}_0e^{-1} + \nu + \nu.\quad(7) $$
What makes these equations noteworthy for the negative electron/helion initial state, is that the possible
final states containing stable nuclei can be produced only by weak interactions. Of course there can also be final
states including $\gamma$'s produced by electromagnetic forces.

\subsection{Co-traveling stored beams of opposite sign}
Table~\ref{tbl:BendParms-PTR.1} described simultaneously-stored triton and helion beams.
For helion kinetic energies roughly twice as great as tritium energies these isotopes can co-
or contra-circulate compatibly---with velocities related by integer ratios as desired.

Here we contemplate the possibility of negative electrons and positive helions co-circulating in the same direction,
in spite of their opposite signs.  As before, the purpose is to enable
rear-end collisions in a moving frame of reference. Now, however, since there is no Coulomb barrier to overcome, the
motivation is different.  In this case the Coulomb force is attractive.  Presumably, this is helpful for the collision
cross section.  This may also relax the ring space charge acceptance limitation.  The primary benefit, though, is that
the electron energy can be increased indefinitely, with the weak interaction rate increasing proportionally.

A strong motivation for enabling rear-end collisions is to influence the detectable weak interaction rate by suppressing
competing inelastic channels other than electron capture (EC).  This consideration does not apply in the present case.
As well as being very small, neutrino cross sections
are proportional to the laboratory neutrino energy. Zuber's ``Neutrino Physics'' book,\cite{Zuber} provides total 
neutrino cross sections;
\begin{align}
\sigma(\nu N)     &= (0.677 \pm 0.014) \times 10^{-38}\,{\rm cm}^2 \times E_{\nu}/({\rm GeV})\notag\\
\sigma(\bar\nu N) &= (0.334 \pm 0.008) \times 10^{-38}\,{\rm cm}^2 \times E_{\nu}/({\rm GeV})
\label{eq:neutrino-total-cross-sections}
\end{align}
where $N$ stands for $p$ or $n$ or their average.  Surely one wants the neutrino energies to be as large as possible,
consistent with the maximum achievable electric field (which is inversely proportional to the ring bending radius). 

Here we consider electron-induced, inverse triton beta decay, $ h + e \rightarrow t + \nu $, 
in which, rather than waiting for tritium atoms to $\beta$-decay, we use storage ring two body scattering interaction
to ``reincarnate'' tritons from helions. (Or speed up the EC-decay of tritons into helions.)
Unlike terrestrial $\beta$-decay, for which the energy required
to fuel the transformation is derived from an inner-shell stored electron of, perhaps, 100\,eV, a storage ring can
provide arbitrarily large energy, with correspondingly large production rate.

Table~\ref{tbl:GOLDEN-EC-2} exhibits kinetic parameters for the process
                       $$ e^- + h \rightarrow t + \nu\qquad(8)$$
with incident electron KE value of 92.1\,MeV, which is some 6 orders of magnitude greater than the incident
energy fueling terrestrial $\beta$-decay.  Recall, however, that the tritium $\beta$-decay half-life is 12.3 years.
So some such large multiplicative factor is needed to produce a usefully large storage ring data rate.

Returning temporarily to accelerator physics, as explained previously, a quartic Newtonian orbit equation is
satisfied by the
design orbit of every circular storage ring with arbitrarily superimposed E\&M bending. One might suppose that a
predominantly magnetic ring, like an $e^+/e^-$ collider, would be needed to store positive and negative beams at
the same time.  This is not correct in our case, however.  For one thing, to obtain rear-end collisions both beams
must travel in the same direction.  What makes this possible, with predominantly electric bending in the present case,
is that electrons are four orders of magnitude lighter than helions.   

For the storage to succeed, a negative ($e^-$) beam and a positive ($h$) beam have to circulate in the same direction.  
This means the magnetic field bending has to be destructive in its bending effect on the $h$ beam.  So, the electric field 
has to be stronger than it would be with no magnetic field.  Because of its charge being +2, the helion does have a two-fold 
advantage as regards the electric bending force, and it is also slow, not very responsive to the wrong sign magnetic 
field.

The magnetic field must be constructive in its effect on the electron beam.  Fortunately, the electron velocities are 
four times greater than the helion velocities.  As a result, the (constructive) magnetic force on the electrons is 
far more effective than is the (destructive) magnetic force on the helions.
 
The net superposition, in spite of being  destructive in both cases, produces centripetal bending in both cases,
as regards the superposition of electric and magnetic contributions. The helion bending remains, therefore, predominantly
electric, but stronger than would be needed to store only the helions, and the electrons respond mainly to the
magnetic field.

In the context of the present paper, in contrast with terrestrial $\beta$-decay, the primary distinction is that 
the initial electron is under direct experimental control, both in kinetic energy and spin orientation.
This causes initial state visualization to be more particle-like than wave-like. 

There are other considerations that makes the $e^- + h \rightarrow t +\nu$ channel 
more useful than the $e^+ + t \rightarrow h +\nu$ channel.  One is that intense polarized helion beams are 
already available---for example at BNL\cite{Zelenski-He3}. More important is that electron beams are easy,
positron beams are difficult.

We conclude, therefore, that negative electrons and positive helions can engage indefinitely in regular rear-end
collisions in a storage ring.  What remains is to demonstrate the kinematics in more detail, as in
Figure~\ref{fig:3He-e_Lab_Lab_angs_vs_CM_angle}.

\begin{figure}[hbt!]
\centering
\caption{\label{fig:3He-e_Lab_Lab_angs_vs_CM_angle}{\bf Left:\ }Plot of lab polar angles vs CM polar angles, for the final state of
the reaction $e^- + h \rightarrow t + \nu$.  Both incident beams are at (0,0) in this plot. 
Notice that, viewed in the laboratory, the produced tritium nuclei are reasonably well collimated, while the (invisible) 
scattered neutrinos are more or less isotropic.
{\bf Right:\ }Plot of final state kinetic energies vs CM polar angles, for $e^- + {\rm 3He} \rightarrow {\rm triton} + \nu$.}
\end{figure}

\begin{table*}[htb]\scriptsize
\caption{\label{tbl:GOLDEN-EC-2}Field strengths, kinematic data, and spin tunes for the reaction $e^- + h \rightarrow t + \nu$.}
\setlength{\tabcolsep}{3pt}
\centering
\begin{tabular}{c|ccc|cc|ccc|cccc|c|c} \hline \scriptsize
  bm &  beta1 &    Qs1 &   KE1  &   E0   &   etaM   & beta2 &  Qs2  &  KE2   & beta* & gamma* &   M*  &   Q12*  & 4*bratio & bm  \\ 
   1 &        &        &   MeV  &   MV/m &          &       &       &  MeV   &       &        &  GeV  &    MeV  &          & 2 \\ \hline
   h & 0.2500 & -0.864 &  92.09 &  9.238 & -0.12138 & 1.00  & 0.324 & 145.12 & 0.285 & 1.043  & 2.919 & 110.120 & 0.99994  & e \\ 
   h & 0.2500 & -0.864 &  92.10 &  9.240 & -0.12139 & 1.00  & 0.324 & 145.14 & 0.285 & 1.043  & 2.919 & 110.131 & 0.99999  & e \\ 
   h & 0.2500 & -0.864 &  92.11 &  9.241 & -0.12140 & 1.00  & 0.324 & 145.15 & 0.285 & 1.043  & 2.919 & 110.142 & 1.00004  & e \\ 
\hline
\end{tabular}
\end{table*}

An unsolved mystery concerns the five or six orders of magnitude deficiency of ``light elements'',
Li, Be, and B, solar system \cite{Rolfs-Rodney}\cite{Cameron-Fowler-1971}\cite{Simpson}
as compared to their quite copious cosmic abundance. Copying the Cameron, Fowler,
1971 abstract\cite{Cameron-Fowler-1971}:

Some consequences are discussed of the possibility that helium-burning shell flashes in advanced
stages of stellar evolution occasionally induce complete convection of the outer
envelope down to the
helium-burning shell. If the hydrogen mixing is relatively small for the first $10^7$ seconds, the result may be
the production of large amounts of heavy elements by the s-process. When complete mixing commences,
the 3He in the envelope will be converted to 7Be, and the subsequent delayed electron capture to form 7Li
may allow enough lithium to remain near the surface to account for the very large lithium abundances in
some S and carbon red-giant stars. On this basis the 7Li/6Li ratio in these stars should be quite large
($>$100).''


\subsection{Storage ring ``induced $\beta$-decay'' candidates}
There are many low energy nuclear $\beta$-decays channels, other than $t+h$, that are subject to the same
treatment as discussed so far.   
Figure~\ref{fig:Low-mass-candidates} lists all or most of the storage ring induced ``$\beta$-decay'' candidates
for nuclear mass numbers less than A=8. All side-by-side pairs preserve mass number A and can be traversed in either
direction. The $t+h$ case discussed previously in this paper and listed in the third row of Figure~\ref{fig:Low-mass-candidates}
is only the lowest mass possibility.  

This paper has so far emphasized weak interaction $\beta$-decay processes.
A major concern for these experiments is that, currently poorly known, the event rates will be quite low.
Concentration on these processes has been quite recent, with most of the preceding electric storage ring development
aimed at measurement of electric (EDM) and magnetic (MDM) dipole moments.

Investigation of other low energy nuclear physics issues will also be made possible, for example as needed for astrophysical
purposes, such as star formation, neutron stars, and other dense objects.  The data rates for all such investigations will
be comfortably high even for highly specialized cases. This can be expected to enable the production of precise spin
dependence measurements.

One such area for investigation concerns the composite structure of nuclei.  It has already been explained why a storage
ring with predominantly electric bending enables the experimental investigation of spin-dependence that has previously
been  impossible. 

In rear-end collision of helions and tritons, there are two possibilities, helions rear-end tritons or vice versa.
In standard quantum physics, as interpreted in the CM system, these processes are identical.  But, in classical physics,
to the extent the particles are composite, internal symmetries can cause these processes to differ.

\subsection{${}^{12}C$ spallation production of Li, Be, and B}
\begin{align}
\hbox{hot\ origin: \quad} \alpha + h         & \rightarrow {}^{7}{\rm Be} + \gamma     \qquad   &                \notag\\
\hbox{cool\ follow\ up: \quad} e^- + {}^{7}{\rm Be} & \rightarrow  {}^{7}{\rm Li} + \nu \qquad  &                 \notag\\
\hbox{eventual:  \quad}
p + {}^{12}C & \rightarrow {}^{11}{\rm B} + 2p                                 \qquad  & (Q = -16.0\,{\rm MeV}), \notag\\ 
            & \rightarrow {}^{10}{\rm B} + 2p + n                             \qquad  & (Q = -27.4\,{\rm MeV}), \notag\\
            & \rightarrow {}^{10}{\rm B} + {}^{3}{\rm He}                     \qquad & (Q = -19.7\,{\rm MeV}), \notag\\
            & \rightarrow {}^{9}{\rm Be} + 3p + n                             \qquad & (Q = -34.0\,{\rm MeV}), \notag\\
            & \rightarrow {}^{9}{\rm Be} + {}^{3}{\rm He} + p                 \qquad & (Q = -26.3\,{\rm MeV}), \notag\\
            & \rightarrow {}^{7}{\rm Li} + 4p + 2n                            \qquad & (Q = -52.9\,{\rm MeV}), \notag\\
            & \rightarrow {}^{7}{\rm Li} + {}^{4}{\rm He} + 2n                \qquad & (Q = -24.6\,{\rm MeV}), \notag\\
            & \rightarrow {}^{6}{\rm Li} + 4p + 3n                            \qquad & (Q = -31.9\,{\rm MeV}), \notag\\
            & \rightarrow {}^{6}{\rm Li} + {}^{4}{\rm He} + 2p + n            \qquad & (Q = -24.6\,{\rm MeV}), \notag\\
            & \rightarrow {}^{6}{\rm Li} + {}^{4}{\rm He} +  {}^{3}{\rm He}    \qquad & (Q = -24.2\,{\rm MeV}), \notag\\
\notag
\end{align}


\section{Recapitulation}
This paper has the multiple purposes of designing and promoting storage rings that enable rear-end collisions
between two different nuclear particle types as well as uncovering ways in which such storage rings can perform the
sort of controlled condition experimentation needed to confirm the validity of conjectured explanations for
astrphysical observations.  Though stellar temperatures cannot be replicated in the laboratory, a storage
ring can replicate (and arbitrarily increase) the ``effective temperature'' in the form of higher particle
energies.

Initially regarded as peripheral, the production of cosmic rays in nature became central to this paper, and
analogies between laboratory-based accelerators and astrophysical accelerators are stressed.  Especially stressed
has been the demonstration that revising Newton's gravitation formula by the replacement of masses by energies
makes it easier to understand how cosmic rays can be produced predominantly within the solar system.  The most
persuasive theoretical indication that this replacement of mass by energy may be physically correct comes
in the comparison of Eq.~(\ref{eq:Newton-kinetics-revised-b}) and Eq.~(\ref{eq:mass-of-sun}), both reeated here.
\begin{equation}
  r_{\rm A-lim} = \frac{G\,M_{\rm sun}}{2} = 6.6 \times 10^{9}\,{\rm m}.
\label{eq:Newton-kinetics-revised-b-bis}
\end{equation}
\begin{equation}
  r_{sun} = 0.6957 \times 10^9 {\rm m}.
\label{eq:mass-of-sun-bis}
\end{equation}
The replacement of mass by energy in Newton's formula leads to the prediction that sufficiently high
energy muclear particles of arbitrary mass number A, attracted gravitationally toeard the sun,
can follow stable circular orbits around the sun with radii in the (narrow) range between these two values.

Returning to laboratory-based astrophysics experiments, from a storage ring with superimposed electric
and magnetic bending, both incident particles can be accurately controlled, and all particles (with obvious
exceptions) in the final state can be accurately measured. This enables unprecedented laboratory-based
investigation of nuclear processes of astrophysical importance.

Design of a practical, superimposed ${\rm E}\&{\rm M}$ bending storage ring design is described.
It enables experiments going well beyond the fixed target experiments performed in the era initiated by
Rutherford up to the present. 

Questions concerning the influence of magnetic fields in the sun and in the laboratory are emphasized,
and mismatches between laboratory measurement and astrotheoretical modeling have been pointed out.
Issues associated with the acceleration of cosmic rays are introduced, but only at a preliminary level,
limited mainly to the issue of injection efficiency.

In the first paper addressing the source of cosmic rays, Fermi acknowledged the importance of the injection
issue, and confessed to the absence of any realistic suggestion concerning injection efficiency in his paper.
Though Fermi's name continues to come up in this context, he had the bad luck of dying a few years before Parker
(in Fermi's own laboratory) explained the solar wind, which, according to this present paper,
is significantly implicated in cosmic ray acceleration.

Fermi's analysis did address the complication that, for every acceleration mechanism, there is a matching
deceleration mechanism.  As well as being individually very weak, this cancellation is nearly perfect for
Fermi's Boltzmann gas statistics. Here it is pointed out that magnetic field present in the core of the
sun disrupt this cancellation, amplifying the importance of ``rear-end'' nuclear collisions. 

For the cosmic ray acceleration mechanism introduced in the present paper there is a potentially
matching cancellation.  A particle about to be captured into a circular orbit in the equatorial plane
of a star or planet can be centripetally deflected to be dumped to the nearby surface, or captured,
leading to temporary acceleration followed by extraction back into space, or centrifugally deflected, with
immediate return to space.  There would seem to be no significant natural cancellation of these outcomes.

From this perspective, Fermi's contribution can be better described as refining the problem of cosmic ray
production rather than as suggesting a route toward its solution.  The cosmic ray problem itself still exists,
but some progress toward its solution has been made since Fermi's time.

Investigation of weak interaction collisions involving neutrinos are also susceptible to storage ring based
experimentation.  Though solar temperatures cannot be matched in the lab, solar nuclear energies can be
exceeded by many orders of magnitude in an ${\rm E}\&{\rm M}$ storage ring,  Weak-interaction cross
sections increase proportionally with energy, potentially increasing data rate by a factor of the
order of a million compared to natural radioactivity.  In this way weak interaction rates can become (marginally)
respectable.

Nevertheless, because of their small cross sections, weak interaction studies need to wait for experience gained
with strong interaction studies. Meanwhile, the more accurate measurement of low energy nuclear processes can be
expected to provide more accurate input data than is presently available for astrophysical calculations concerning
solar evolution, black holes, white dwarfs, neutron stars, and other compact objects. \cite{Teukolsky-Shapiro}

In practice, any such facility would probably spend some years studying conventional nuclear physics
processes (for example their spin dependencies) before proceeding to neutrino physics.
Instead of the single electron signature of terrestrial tritium beta decay, the proposed weak interaction
processes, with pure initial states of known momenta and spin orientations, would produce clean, background-free
detection of all final state particles (with accurately-implied neutrino mechanical parameters) with
fully-determined kinematics and measurable polarizations. Just a few of many low energy nuclear measurement
possibilities have been described.

Another low energy nuclear process of significant importance, and not difficult, would be
the search for and detection of elastic scattering in which one spin flips while the other does not,
which would be a violation of time reversal invariance.


%

\appendix

\section{Cary-Brizard guiding-center adiabatic theory  \label{sec:Cary-Brizard}}
John Cary and Alain Brizard\cite{Cary-Brizard} (C-B) have produced a clear and comprehensive
modern Lagrangian and Hamiltonian classical mechanics phase space treatment of the guided field motion
of single particles.  This is exactly what is needed for the present paper. The full C-B treatment
covers quite general classes of magneto-hydrodynamics processes and is unnecessarily general for
our purposes. Fortunately, their paper also provides formulas that have been simplified to more
nearly match our needs.

For any serious application of the C-B formalism, a sensible procedure would be to first
study their paper in detail.  But, for a more casual understanding of the solar wind and
Fermi acceleration, a stripped-down, more elementary version might seem to be useful.
At the risk of introducing errors into their treatment, this is what is attempted in
this appendix.

I have tried to retain C-B notation, and, to the extent possible, copy their language
explanations.  At the same time, I have not hesitated to interject comments that
are especially appropriate in cases where I have specialized the original C-B formula
inappropriately, from their more general perspective.

This appendix therefore provides a ``Coles-Notes'' annotated version of part of the original paper.
Except for pointing out errors, any response generated by this appendix should refer
explicitly to the original C-B paper.

Of the 45 pages of the C-B original paper I have drawn on only a small fraction.  A sensible procedure might
be, first, to concentrate on the early parts of their paper and, only then read read my description
of formulas needed for the present paper.

A charged particle in a constant magnetic field
    ${\bf B} = B{\bf\hat b}$
moves along a helix, while conserving its kinetic
energy and, therefore, its speed
      $v = |{\bf v}|.$
In addition, the motion parallel to the magnetic field is uniform,
i.e., the velocity
      $v_{\parallel} = {\bf v}\cdot{\bf\hat b}$,
parallel to the magnetic field is constant, and so the perpendicular speed
     $v_{\perp} = |{\bf v}_{\perp}| = (v^2-v_{\parallel}^2)^{1/2}$
is also constant. The perpendicular motion for gyro-motion is confined
to a circle, whose gyration center remains on the same magnetic-field
line. The gyration frequency or gyro-frequency is given by
      $\Omega =\frac{eB}{mc}$
and the gyration radius or gyro-radius vector
      ${\pmb\rho}(v_{\perp},\zeta)={\bf\hat b}\times{\bf v_{\perp}}/\Omega$
depends explicitly on the gyration angle.

Alfvén showed that the magnetic moment
\begin{equation}
\mu \equiv
\frac{e}{mc} \oint \frac{d\zeta}{2\pi} \Big[m{\bf v} + \frac{e}{c}{\bf A}({\bf X} + \pmb\rho) \Big]
\cdot\frac{\partial \pmb\rho}{\partial\zeta} = \frac{mv_{\perp}^2}{2B}
 \qquad\qquad\qquad\qquad [CB 1.1]
\label{eq:C-B.1}
\end{equation}
is the adiabatic invariant associated with the fast gyro-motion of a charged particle (with mass $m$
and charge $e$ in a slowly varying magnetic field ${\bf B} = \nabla \times {\bf A}$ and the gyro-action
$J \equiv (mc/e)/\mu$ is canonically conjugate to the ignorable gyro-phase angle~$\zeta$.\footnote{From
the Carey-Brizard perspective, which emphasizes the guiding center dynamics, the gyro-phase angle~$\zeta$,
which provides the phase relationship of the actual particle transverse angle relative to the guiding
center is, for most purposes a nuisance to be suppressed.  For our purposes $\zeta$, is the essential
dynamical variable describing the nuclear ion's motion relative to a non-inertial reference frame. In other
words, what C-B average away is essential to us, and vice versa.}

From this adiabatic invariance and energy conservation, it follows that there must
be a parallel force due to the gradient of $\mu B$ (the perpendicular kinetic energy, taken with $\mu$
held constant. Alfvén also calculated the cross-field drifts due to the gradient of $B$ and the
magnetic-field-line curvature.  His results showed that the cross field drifts are smaller than $v_{\perp}$
by the ratio $\rho/L \equiv \epsilon$.
\footnote{The C-B quantity $\epsilon$ is an artificial parameter serving to retain flexibility in
the choice of ``small parameter'', depending on the application.  For our purposes $\epsilon$ is
somewhat anti-intuitive, since we are concentrating on the actual particle dynamics rather than the
guiding center. For our purposes $\epsilon$ should just be dropped; i.e. set equal to 1. For most
C-B purposes this would be utterly inappropriate.}

The formulation of mechanics follows from noting that the canonical equations of motion derive from
requiring stationarity of the action integral
\begin{equation}
   {\mathcal{A}} =  \int L({\bf x},{\bf\dot{x}},t) dt,
 \qquad\qquad\qquad\qquad\qquad\qquad\qquad\qquad  [CB 1.2]
\label{eq:C-B.2}
\end{equation}
where
\begin{equation}
  L \equiv {\bf p}({\bf x}, {\bf\dot{x}},t) - H({\bf x}, {\bf\dot{x}},t),
  \label{eq:C-B.3}
\end{equation}
with respect to virtual displacements, $\delta {\bf x}$ in configuration space.

The Lagrangian for a set of coordinates, ${\bf q}= (q^1,q^2, \dots q^N)$ is a
function $L({\bf q}, {\bf \dot q}, t)$ of the coordinates and their time derivatives.
This leads to the Lagrange equations,
\begin{equation}
  \frac{d}{dt}\Big(\frac{\partial L}{\partial {\dot q^i}}\Big) = \frac{\partial L}{\partial q^i} ,
    \qquad\qquad\qquad\qquad\qquad\qquad\qquad\qquad [CB 2.1]
  \label{eq:C-B.4}
\end{equation}
for the trajectory.  For charged particle motion in an electromagnetic field the Lagrangian in
Cartesian coordinates is
\begin{equation}
  L({\bf x}, {\bf \dot x}, t) = \frac{m}{2}|{\bf\dot x}|^2
                            + \frac{e}{c}{\bf x}\cdot{\bf A}({\bf x}.t) - e\Phi({\bf x},t),
  \qquad\qquad  [CB 2.2]
  \label{eq:C-B.5}
\end{equation}
in terms of the scalar potential $\Phi$ and the vector potential ${\bf A}$ which give the
the electric and magnetic fields
\begin{equation}
{\bf E} = -\pmb\nabla\Phi - c^{-1}\frac{\partial{\bf A}}{\partial t}
          \hbox{\ and\ }
{\bf B} = \pmb\nabla\times{\bf A}.
  \label{eq:C-B.6}
\end{equation}
\emph{Canonical momentum} ${\bf p}$ are defined with components
\begin{equation}
  p_i = \frac{\partial L}{\partial q^i} ({\bf q},{\bf\dot{\bf q}}, t) .
    \qquad\qquad\qquad\qquad\qquad\qquad\qquad\qquad [CB 2.3]
\label{eq:C-B.7}
\end{equation}
In this case a point in \emph{phase space} is determined by ${\bf q}$ and ${\bf p}$
rather than ${\bf q}$ and ${\bf\dot q}$.  The equations of motion are Hamilton's equations,
\begin{equation}
{\dot q^i} =  \frac{\partial H}{\partial p_i},  \hbox{\quad and \quad}
{\dot p_i} = -\frac{\partial H}{\partial q^i}, 
  \qquad\qquad\qquad\qquad\qquad\qquad [CB 2.4] \label{eq:C-B.8}
\end{equation}
where
\begin{equation}
H({\bf q}, {\bf p}, t) = {\bf p}\cdot{\bf \dot q}({\bf q}, {\bf p}, t) - L[{\bf q},({\bf\dot q},t),t))]
  \qquad\qquad\qquad\qquad [CB 2.5] \label{eq:C-B.9}
\end{equation}
Notice that this is a Legendre transformation, which means that, in this case, both
the function and the coordinates are being changed.

For charged particle motion in an electromagnetic field the Hamiltonian in
Cartesian coordinates is
\begin{equation}
H({\bf x}, {\bf p}, t) = \frac{1}{2 m}\Big|{\bf p} - \frac{e}{c}{\bf A}({\bf x},t)\Big|^2 + e\Phi({\bf x},t).
  \qquad\qquad\qquad\qquad [CB 2.6] \label{eq:C-B.10}
\end{equation}
and the canonical momenta are
\begin{equation}
p_i = \frac{\partial L}{\partial\dot x^i} = m\dot x_i + \frac{e}{c}A_i.
  \qquad\qquad\qquad\qquad \label{eq:C-B.11}
\end{equation}

\clearpage

\section{Relativistic Mechanics    \label{sec:RelativisticMechanics}}
\subsection{The relativistic principle of least action}
It is straightforward to generalize the principle of least action 
in such a way
as to satisfy the requirements of relativity while at the same time leaving
non-relativistic relationships (i.e., Newton's Law) valid when speeds are
small compared to $c$.
Owing to the homogeneity of both space and time,
the relativistically generalized action $S$ cannot depend on the
particle's coordinate 4-vector $x^i$. Furthermore it must be a
relativistic scalar since otherwise it would have directional 
properties, forbidden by the isotropy of space.

The action of a free particle 
(i.e., one subject to no force) is
\begin{equation}
S = (-mc)\int_{t_0}^t ds 
  = (-mc^2)\int_{t_0}^t \sqrt{1 - \frac{v^2}{c^2}}\,dt,
\label{eq:Relint.nine2} 
\end{equation} 
where the invariant interval $ds$ is the proper time multiplied by $c$.
As always, the dimensions of $S$ 
are momentum$\times$distance or,
equivalently, as energy$\times$time.
Though the first expression for
$S$ is manifestly invariant, the second depends on values of
$v\equiv|\dot{\bf x}|$ 
and $t$ in the particular frame of reference in which Hamilton's 
principle is to be applied. \emph{A priori} the
multiplicative factor could be any constant, but it will be seen 
shortly why the factor has to be $(-mc^2)$. 
The negative sign is significant. It
corresponds to the seemingly paradoxical result
that the free particle
path from position $P_1$ to position $P_2$ {\it maximizes} 
the proper time taken. Comparing with the standard definition of the
action in terms of the Lagrangian, it can be seen that the free particle 
Lagrangian is
\begin{equation}
{L}({\bf x},\dot{\bf x}) = -mc^2 \sqrt{1 - \frac{|\dot{\bf x}|^2}{c^2}}.
\label{eq:Relint.ten2} 
\end{equation} 
As always, the Lagrangian has the dimensions of an energy.

\subsection{Energy and Momentum}
In Lagrangian mechanics, once the Lagrangian is specified, the equations
of motion follow just by ``turning the crank''. Slavishly following
the Lagrangian prescriptions, the momentum ${\bf p}$ is \emph{defined} by
\begin{equation}
{\bf p} = \frac{\partial {L}}{\partial \dot{\bf x}}
      = \frac{m{\bf v}}{\sqrt{1 - v^2/c^2}}.
\label{eq:Relint.eleven2} 
\end{equation} 

\noindent
For $v$ small compared to $c$, this gives the non-relativistic result
${\bf p}\simeq m{\bf v}$. 
This is the relation that fixed the constant factor in
the initial definition of the Lagrangian.
Using Eqs.~(\ref{eq:Relint.ten2}) and 
(\ref{eq:Relint.eleven2}), one obtains the Hamiltonian $H$ and hence 
the energy $\mathcal{E}$ by
\begin{equation}
H = {\bf p}\cdot{\bf v} - {L}
      = \frac{m c^2}{\sqrt{1 - v^2/c^2}}.
\label{eq:Relint.twelve2} 
\end{equation} 
\index{${\cal E}_0=mc^2$, relativistic rest energy}
\index{${\cal E}$, relativistic energy}
\index{Relativistic!rest energy ${\cal E}=mc^2$}

\noindent
For $v$ small compared to $c$, and the numerical value of $H$
symbolized by $\mathcal{E}$, this gives 
\begin{equation}
\mathcal{E} \simeq \mathcal{E}_0 +\frac{1}{2} mv^2,
\label{eq:Relint.thirteen2} 
\end{equation} 
which is the classical result for the kinetic energy,
except for the additive constant  $\mathcal{E}_0=mc^2$,
known as the rest energy. An additive constant like this has no effect
in the Lagrangian description.
From Eqs.~(\ref{eq:Relint.eleven2}) and 
(\ref{eq:Relint.twelve2}) come the important identities
\begin{equation}
\mathcal{E}^2 = {\bf p}^2c^2 + m^2 c^4,\quad
{\bf p} = \frac{\mathcal{E} {\bf v}}{c^2}.
\label{eq:Relint.fourteen2} 
\end{equation}
For massless particles like photons these reduce to $v=c$ and
\begin{equation}
p = \frac{\mathcal{E}}{c}.
\label{eq:Relint.fifteen2} 
\end{equation} 
This formula also becomes progressively more valid for a massive 
particle as its total energy becomes progressively large
compared to its rest energy.
As stated previously, $m$ is the ``rest mass'', a constant
quantity, and there is no
question of ``mass increasing with velocity'' as occurs in some
descriptions of relativity, such as the famous ``$\mathcal{E}=mc^2$'', which
is incorrect in modern formulations.

Remembering to express it in terms of ${\bf p}$, the 
relativistic Hamiltonian is given by
\begin{equation}
{H} ({\bf p}) = \sqrt{p^2c^2 + m^2c^4}.
\label{eq:Relint.sixteen2} 
\end{equation} 

\subsection{Forced Motion}
We define a momentum 4-vector $p^i$ by
\begin{equation}
p^i = mu^i
 = 
\frac{m}{\sqrt {1 - v^2/c^2}}\,
\begin{pmatrix} c \\ {\bf v} \end{pmatrix}
 =
\begin{pmatrix}\mathcal{E}/c \\ {\bf p}\end{pmatrix}.
\label{eq:Relint.seventeen2} 
\end{equation} 
We expect that $p^ip_i$, the scalar product of $p^i$ with itself should,
like all scalar products, be invariant. 
The first of Eqs.~(\ref{eq:Relint.fourteen2}) shows this 
to be true;
\begin{equation} 
p^ip_i = \mathcal{E}^2/c^2 - p^2 = m^2c^2.
\label{eq:Relint.eighteen2} 
\end{equation} 
Belonging to the same 4-vector, the components of 
${\bf p}$ and $\mathcal{E}/c$ in different
coordinate frames are related according to the Lorentz transformation.

If the 4-velocity is to change, it 
has to be because force is applied to the particle.
It is natural to define the 4-force $G^i$ by the relation
\begin{equation}
G^i = \frac{dp^i}{ds/c}
    = \gamma\,\bigg(\frac{d\mathcal{E}/c}{dt},\frac{d{\bf p}}{dt}\bigg)^T
    = \bigg( \frac{{\bf F}\cdot{\bf v}/c}{\sqrt {1 - v^2/c^2}},
              \frac{{\bf F}}{\sqrt {1 - v^2/c^2}} \bigg)^T 
\label{eq:Relint.twentythree2} 
\end{equation} 
where 
\begin{equation}
{\bf F} = \frac{d{\bf p}}{dt}
\label{eq:Relint.twentyfour2} 
\end{equation} 
is the classically defined force. Since this formula is valid both
relativistically and non-relativistically it is the least accident-prone
3D-form of Newton's law. 
The energy/time component $G^0$ is 
related to the rate of work done on the particle by the external 
force.  Note that 
this component vanishes
in the case that $ {\bf F}\cdot{\bf v} = 0 $, as is true,
for example, for a 
charged particle in
a purely magnetic field.

\subsection{Hamilton-Jacobi Formulation}
All this is quite elementary, 
and it is all that one really needs to remember in order
to proceed with relativistic dynamics.
When the minimized function $S(x_0,t_0,x,t)$ is expressed as a function of
$x$ and $t$ at the upper spatial end point,
holding the lower end point fixed, $S$ satisfies the so-called ``Hamilton
Jacobi Equation.'' This equation is rarely used for single particle dynamics
but the following definitions (associated with the equation) are sometimes important:
\begin{equation}
{\bf p} = \frac{\partial S}{\partial {\bf x}},\quad
H = -\frac{\partial S}{\partial t}.
\label{eq:Relint.eight2} 
\end{equation} 

\noindent
Corresponding to the Hamiltonian of Eq.~(\ref{eq:Relint.sixteen2}), 
the Hamilton-Jacobi equation is 
\begin{equation}
\bigg(\frac{\partial S}{\partial t}\bigg)^2 =
c^2\bigg(\frac{\partial S}{\partial x}\bigg)^2 +
c^2\bigg(\frac{\partial S}{\partial y}\bigg)^2 +
c^2\bigg(\frac{\partial S}{\partial z}\bigg)^2 + m^2c^4 .
\label{eq:Relmech.18p}
\end{equation}

Since the relations~(\ref{eq:Relint.eight2}) can be 
derived for arbitrary $S$ purely from the
calculus of variations, without reference to the physical
interpretation of the quantities, they
must remain valid in relativistic mechanics. Nevertheless we
will re-derive these relations for practice in 
using abbreviated
manipulations for the calculus of variations.
The action
integral~(\ref{eq:Relint.nine2}), expressed
in terms of $ds=\sqrt{dx_idx^i}$, is
\begin{equation}
S = -mc\int_{P_0}^P\,\sqrt{dx_idx^i}.
\label{eq:Relint.eight2p} 
\end{equation} 
The variation $\delta S$ in the action that accompanies 
a variation
$\delta x^i(t)$ away from the true world trajectory, is what establish the
equations of motion. Here $\delta x^i(t)$ is an arbitrary function.
Variation of the integrand yields
\begin{align}
\delta \sqrt {dx_i dx^i}                   
 =\ & \frac{(\delta dx_i)dx^i + dx_i(\delta dx^i)}{2\sqrt{dx_i dx^i}} \notag \\
 =\ & \frac{dx_i}{ds} d\delta x^i
      = \frac{u_i}{c}\, d\delta x^i   \notag \\
 =\ & d\Big(\frac{u_i}{c}\,\delta x^i\Big) - \frac{du_i/c}{ds} \delta x^i\,ds.
\label{eq:Relint.nineteen2}
\end{align} 
\begin{figure}
\hspace{1cm}
\includegraphics[scale=0.5]{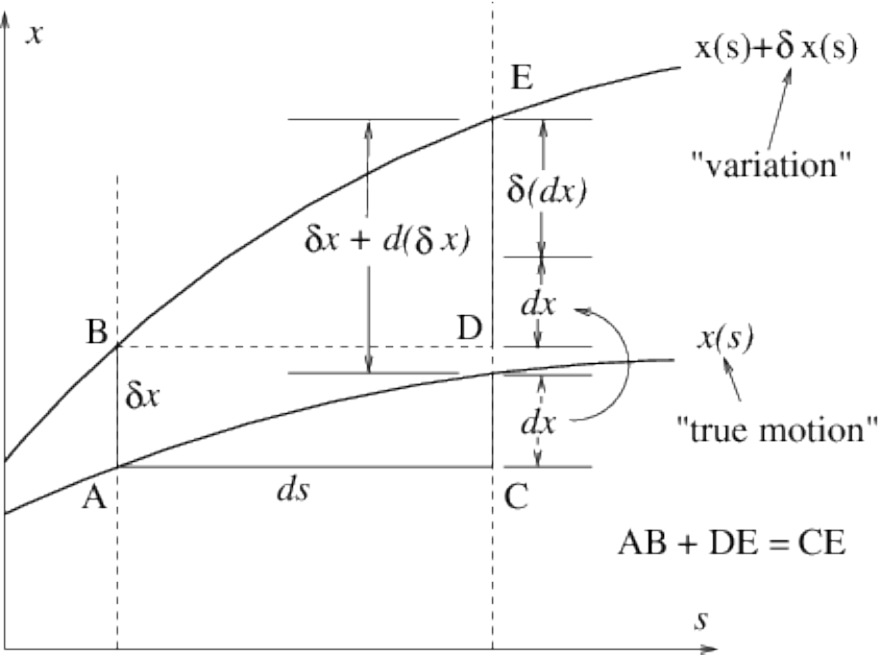}
\caption{\label{fig:Deltas}
The equation shown, AB+DE=CE, when expressed in terms of $dx$, $\delta x$,
$d\delta x$, and $\delta dx$, shows that $d\delta x=\delta dx$.}
\end{figure}

The relation $\delta dx= d\delta x$, whose validity is exhibited
in Fig.~\ref{fig:Deltas}, has been used.
The last line is preparatory to integration by parts.
The action variation is 
\begin{equation}
\delta S = 
-mc\int_{P_0}^P
\bigg(
d\Big(\frac{u_i}{c}\delta x^i\Big) - \frac{du_i/c}{ds} (\delta x^i) ds
\bigg).
\label{eq:Relmech.twenty}
\end{equation}
In this form
the upper integration limit can be held fixed or varied as we wish. 
If both end points are held fixed,
the first term in the integral vanishes 
vanishes, in which case, since the principle
of least action requires the vanishing of $\delta S$, and
since $\delta x^i$ is arbitrary, the vanishing of
the 4-acceleration $w_i = du^i/(ds/c)$ follows. This  
is appropriate for force-free motion.

When the upper end point of the integral in 
Eq.~(\ref{eq:Relmech.twenty}) is varied, but with
the requirement that the trajectory be a true one, then the 
second term in the integral vanishes, leaving
\begin{equation}
\delta S = -m u_i \delta x^i.
\label{eq:Relint.twentyone2} 
\end{equation} 
Substitution into  Eq.~(\ref{eq:Relint.eight2}) yields
\begin{equation}
-p_i = \frac{\partial S}{\partial x^i}
 =
-mu_i = \Big(-\frac{\mathcal{E}}{c}, {\bf p}\Big),
\label{eq:Relint.twentytwo2} 
\end{equation} 
which confirms the validity of those equations, 
remembering that the spatial covariant and contravariant 
4-vector components have opposite signs.

\clearpage

\section{Atomic and nuclear binding energy treatment \label{sec:Trbojevic}}
\begin{figure}[hbt!]
\centering
\includegraphics[scale=1.0]{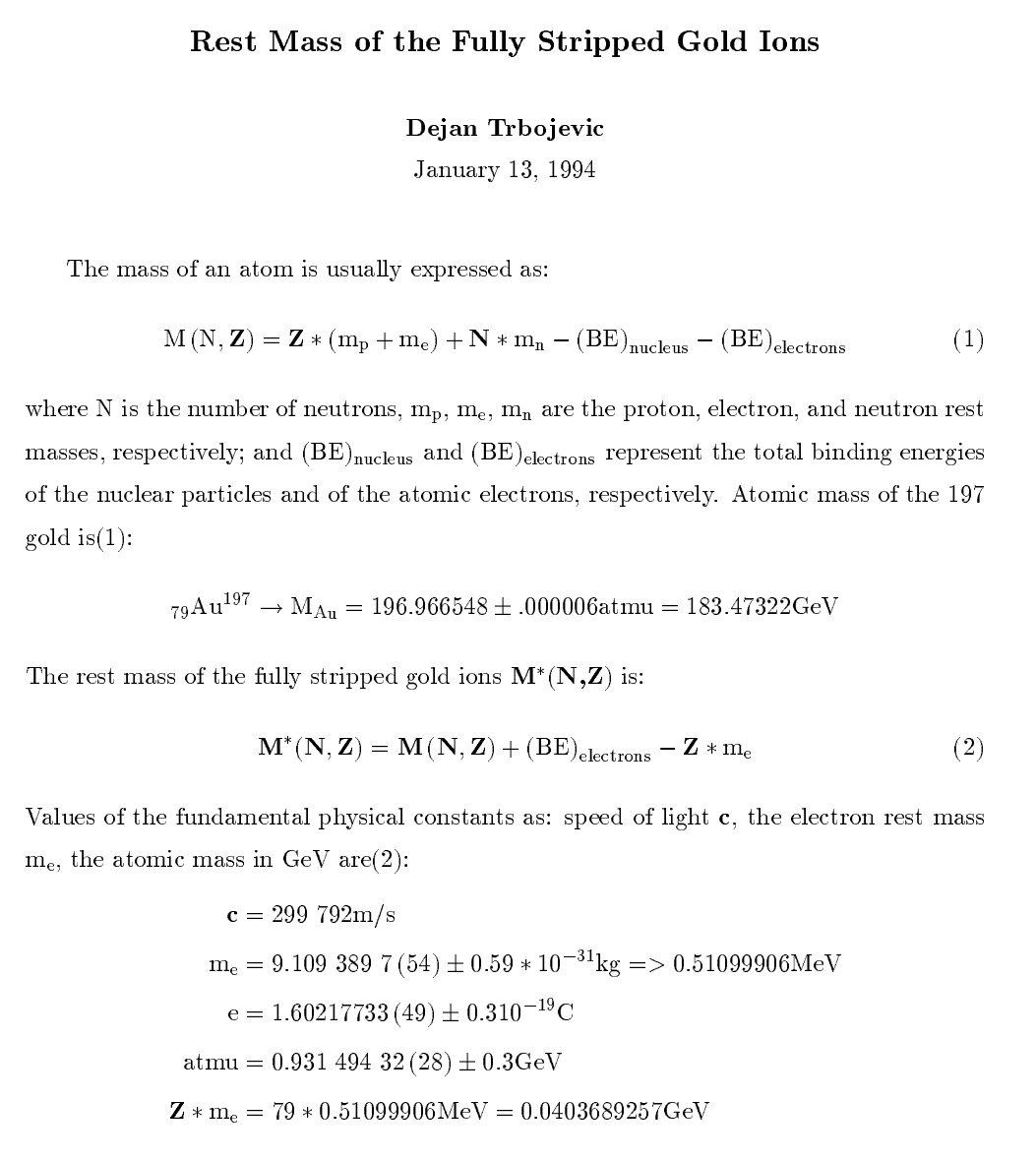}
\caption{\label{fig:RHIC_AP_20-Trbojevic}RHIC report by Dehan Trbojevic explaining
mass corrections associated with nuclear and atomic binding energies. Page~1.}
\end{figure}
\begin{figure}[hbt!]
\centering
\includegraphics[scale=1.0]{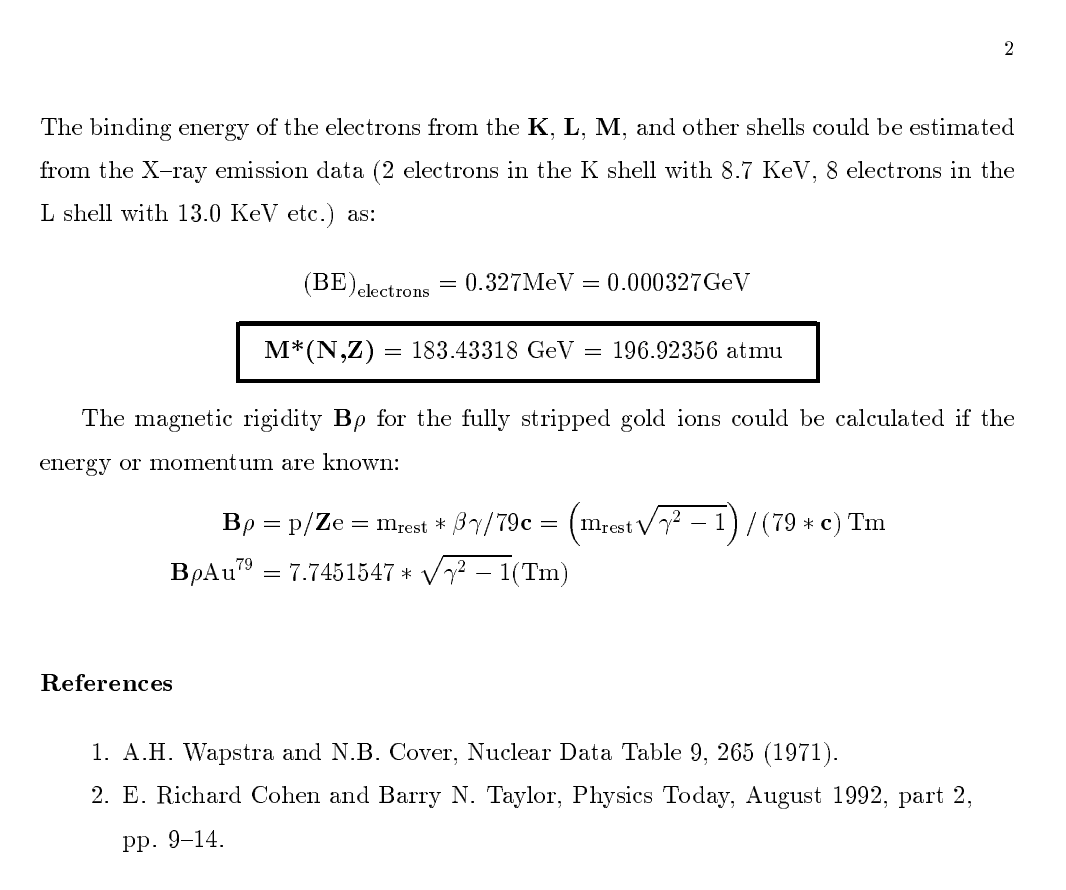}
\caption{\label{fig:RHIC_AP_20-Trbojevic}RHIC report by Dehan Trbojevic explaining
mass corrections associated with nuclear and atomic binding energies. Page~2.}
\end{figure}

\clearpage

\section{Wagoner, Fowler and Hoyle rate formulas\label{sec:Wagoner}}
\begin{figure}[hbt!]
\centering
\includegraphics[scale=0.45]{ 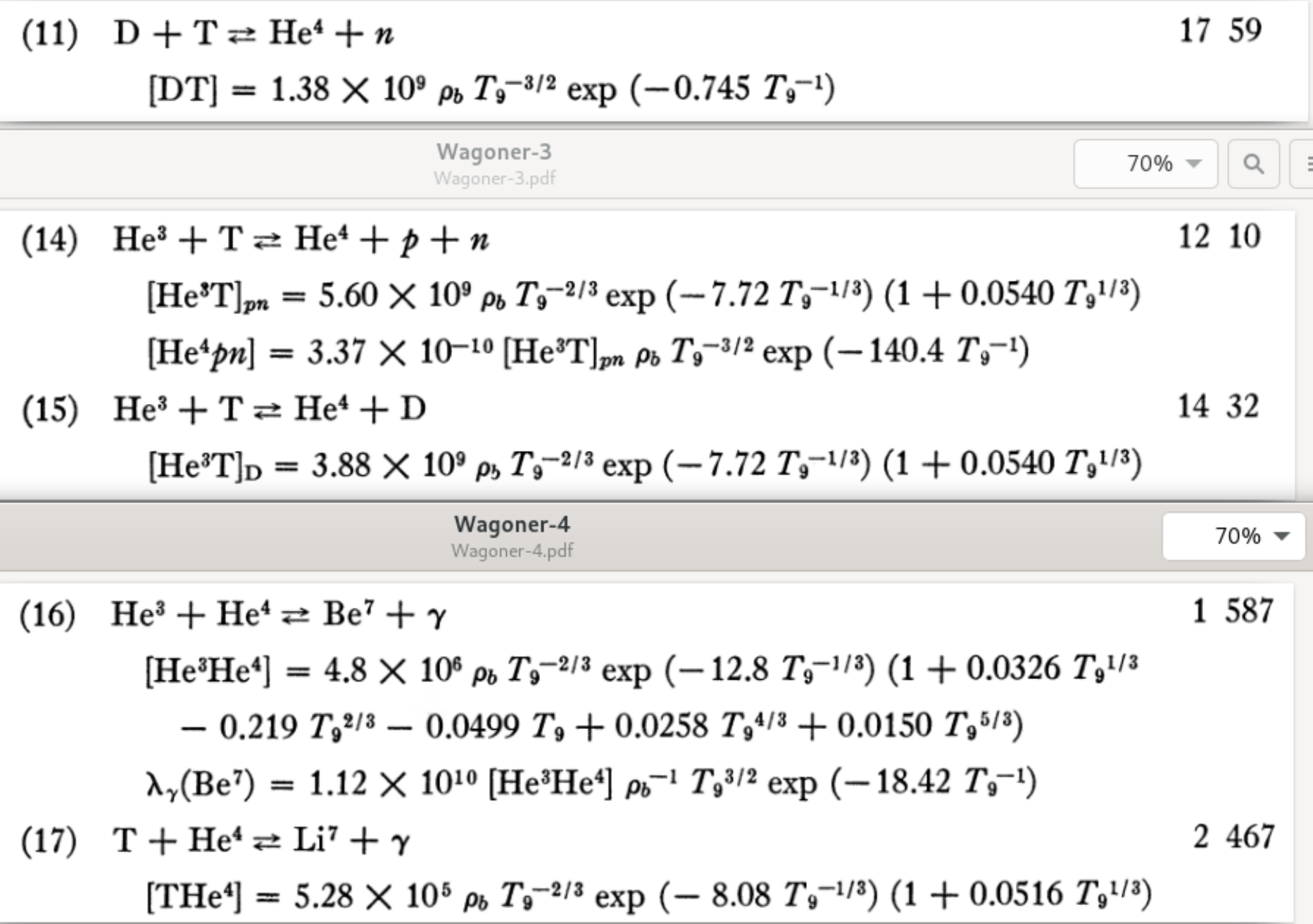}
\includegraphics[scale=0.45]{ 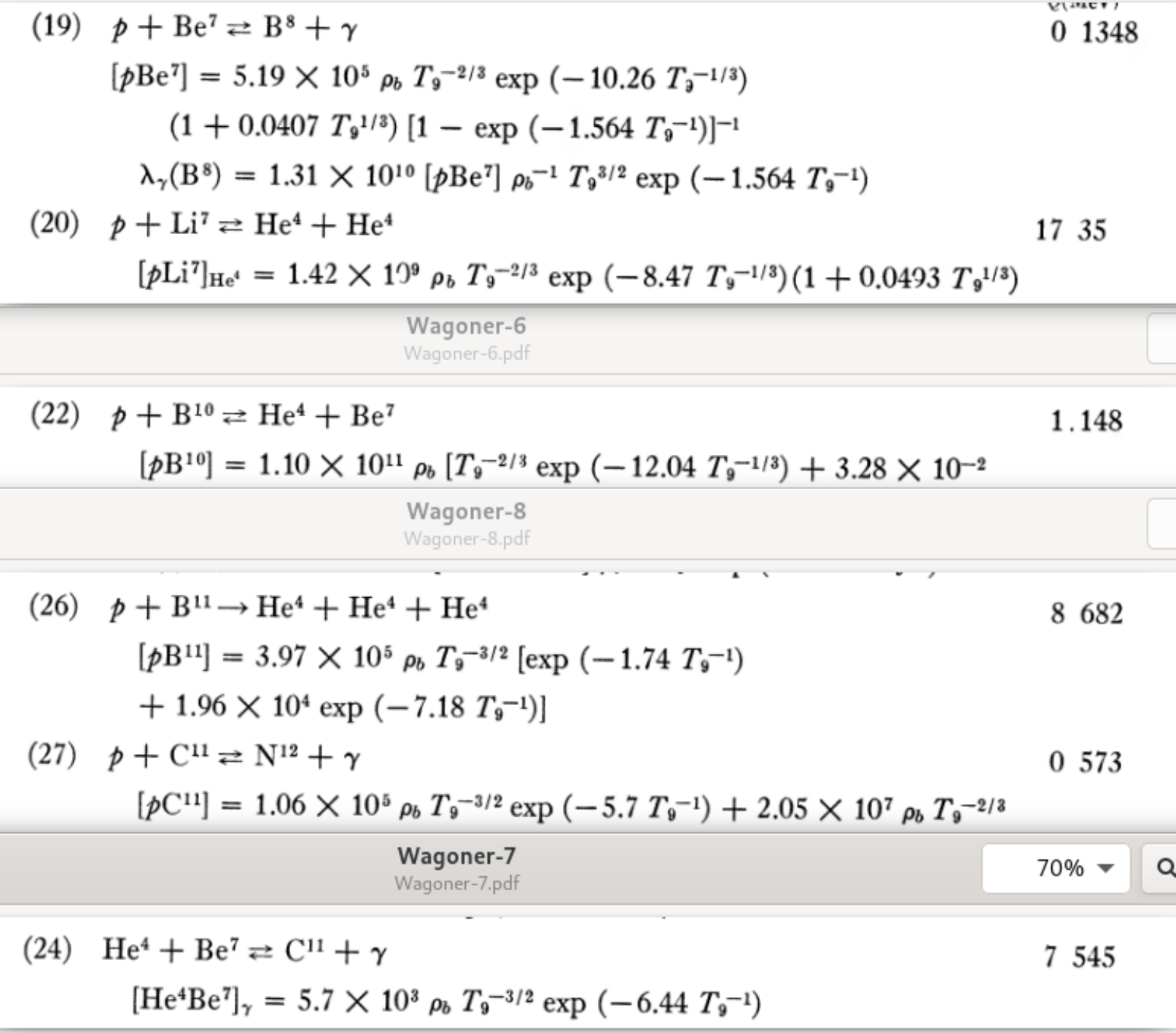 }
\caption{\label{fig:Wagoner-I-II} Wagoner et al. 1974 formulas. $Q$-values in MeV are shown on the right.
Temperature $T_g$ is defined in their Appendix A, Eq.~(A15); valid at very high temperature (Tg $>$ 10). }
\end{figure}
\begin{figure}[hbt!]
\centering
\includegraphics[scale=0.55]{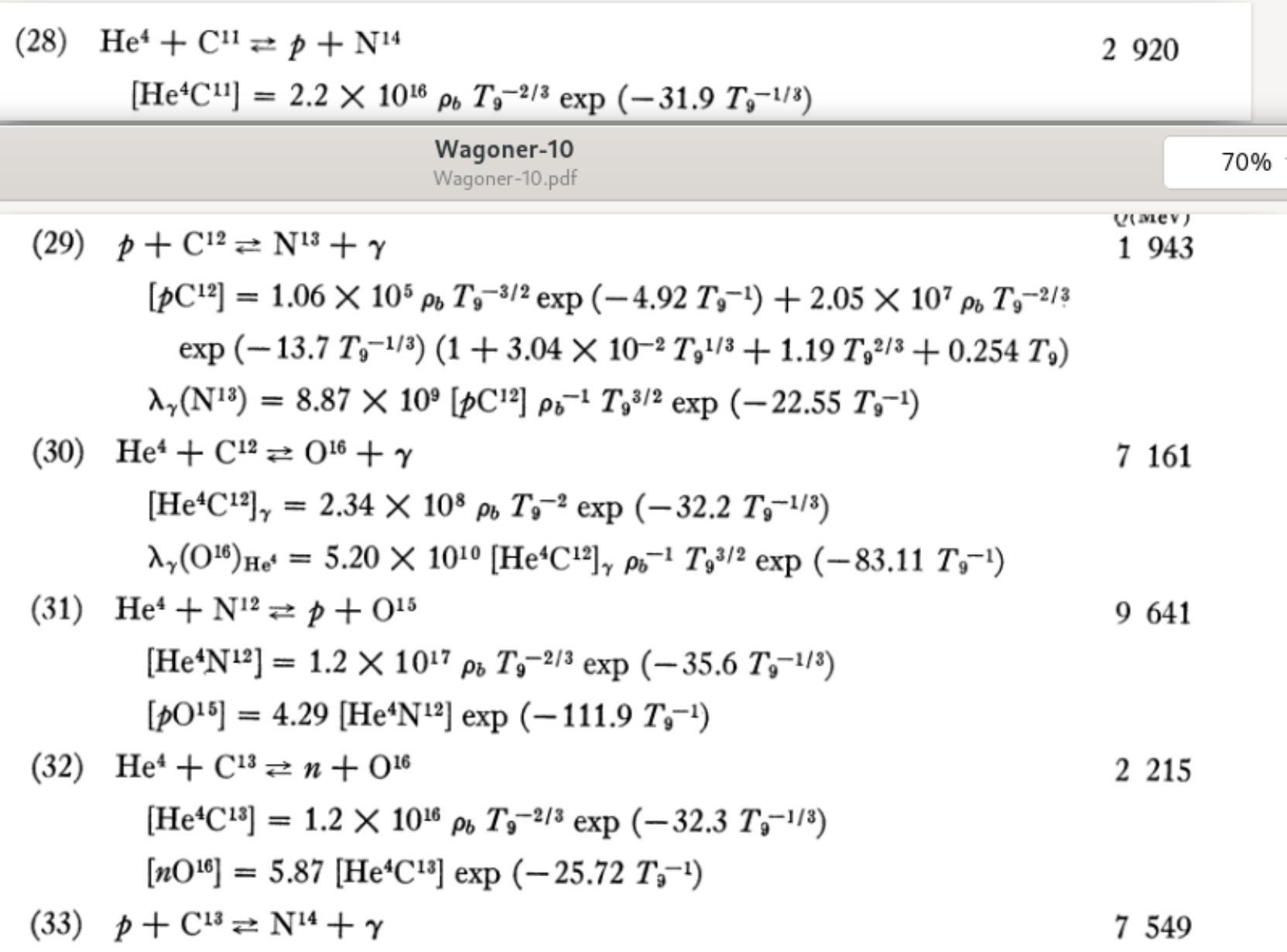}
\includegraphics[scale=0.55]{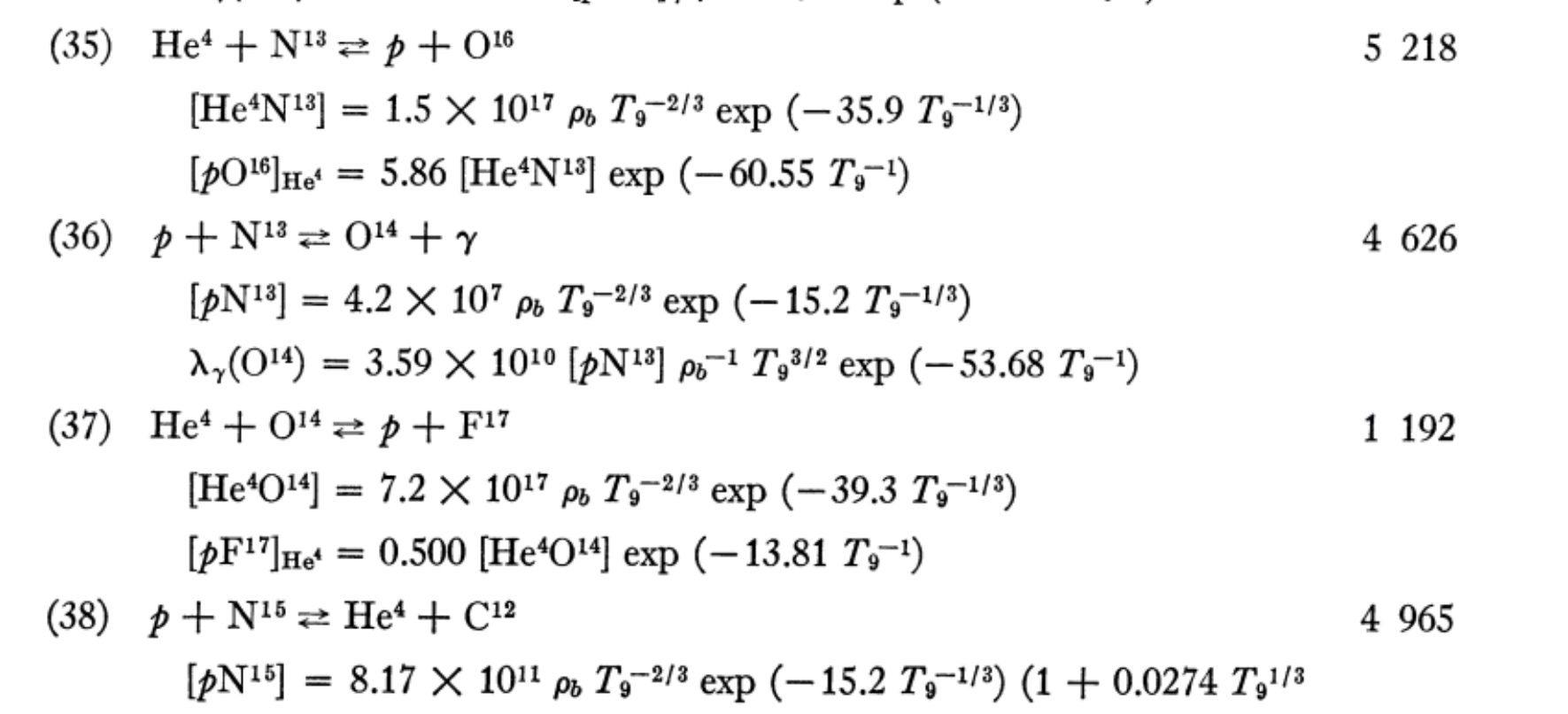}
\caption{\label{fig:Wagoner-I-II} Wagoner et al. 1974 formulas, continued.}
\end{figure}
\begin{figure}[hbt!]
\centering
\includegraphics[scale=0.53]{Dolgov-Zeldovich-1981.pdf}
\caption{\label{fig:Dolgov-Zeldovitch-1981}Dolgov and Zeldovich 1981 synthesis of Wagoner et al. 1974
  production rate predictions}.
\end{figure}

\clearpage

\section{Conversions to accelerator physics units \label{sec:Units}}
With just a single exception, ``accelerator physics units'' and MKSA units are interchangeable,
much in the same way that one meter is 100 centimeters.
The exception is that energies are measured in GeV, rather than eV. What makes this change
popular for high energy accelerator and particle physicists is that, within +/-15\%, the proton
mass (expressed as
rest energy), or by nuclear A-value, or by atomic mass units, is roughly equal to 1.

Another easiy-remembered practical result is the Bolzmann conversion from Kelvin temperature to
energy; an energy of 1\,MeV corresponds to a temperature of $1.16\times10^{10}$\,K.

Also electric fields are measured in GeV/m, rather than V/m, which is not inconvenient.
However, the units for most other physical quantities become obscure; especially the magnetic
field, because the magnetic force is proportional also to particle velocity. Along
with the following conversion tables, and checking that electric and magnetic fields
are close to what is expected, it should be possible to interpret all physical
parameters correctly.

Incidentally, the accelerator units are much like ``natural units'' in that
the speed of light is set equal to 1.  But, since the Planck (modified) constant $\hbar$,
is not set equal to 1 in accelerator units complicates this route. Nevertheless, for
analyzing weak interaction processes, natural units are greatly to be favored. 

For all tables in the present paper, electric and magnetic fields are superimposed, with
fractional electric bending fractions $\eta_M$ and $\eta_E$, which \emph{always} sum to 1;
             $$\eta_M + \eta_E=1.$$
To avoid the evaluation of magnetic fields in Tesla units, electric fields are given
in GV/m, while magnetic bending fields are conveyed by 
$\eta_{M1}$, the magnetic bending fraction for beam 1. It is important to realize that $\eta_{M2}$
is not the same as $\eta_{M1}$ and must not be used to obtain the (shared) magnetic bending field.
 
\begin{figure}[hbt!]
  \centering
  \includegraphics[scale=0.7]{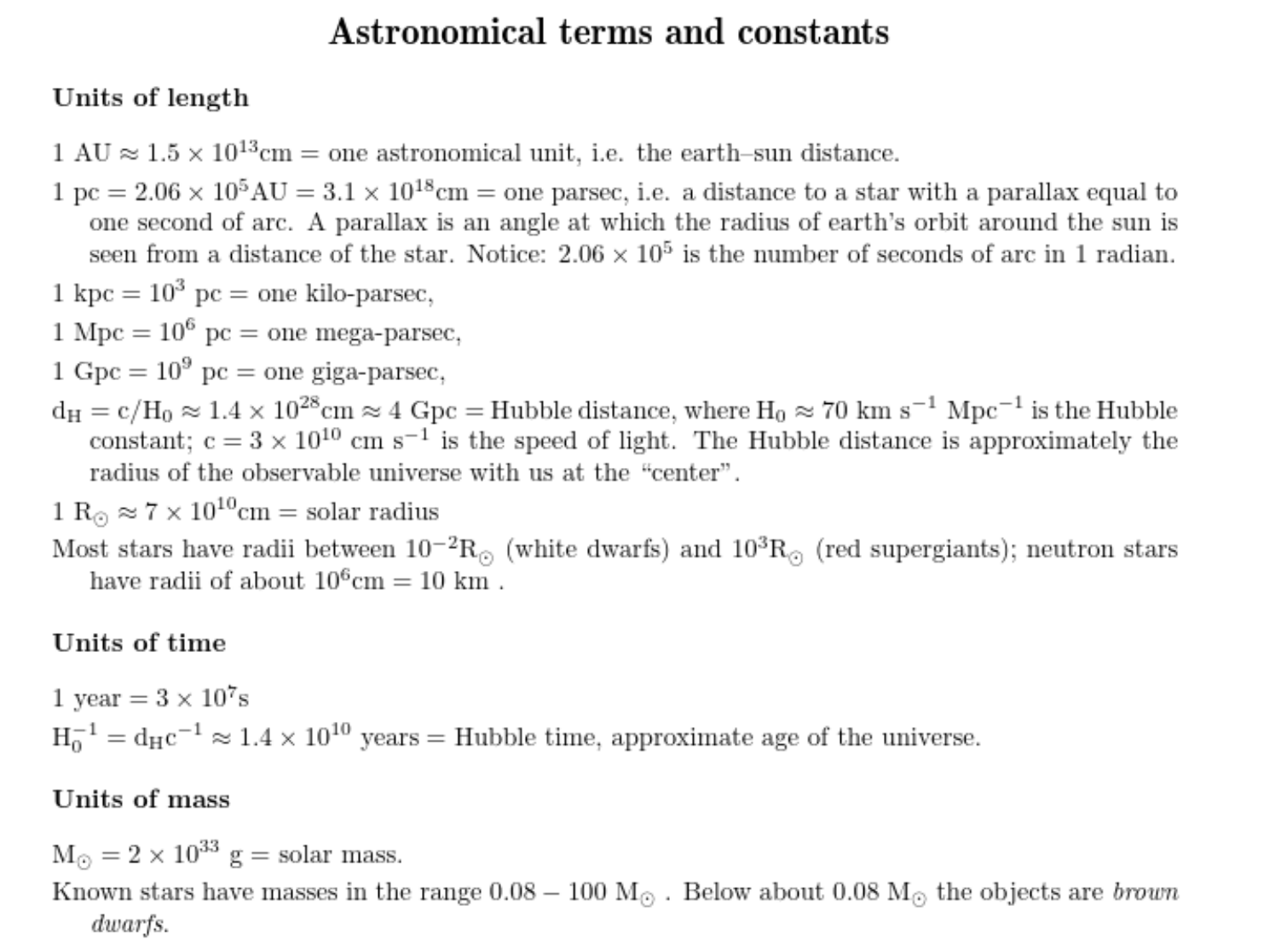}
\caption{\label{fig:Natural-Units}
 Astronomical terms and constants copied from www.astro.princeton.edu.}
\end{figure}

\begin{figure}[hbt!]
  \centering
  \includegraphics[scale=0.6]{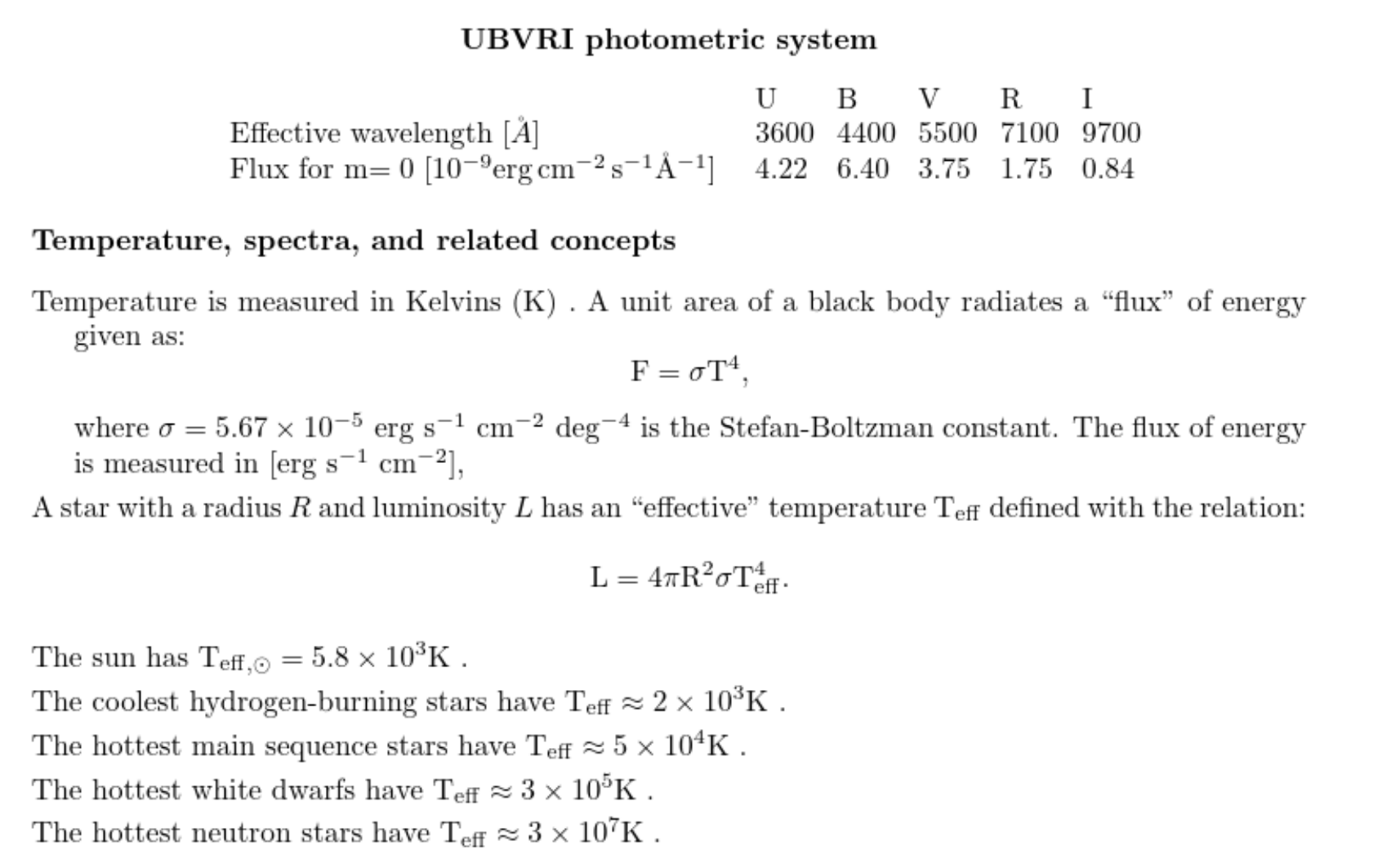}
 \caption{\label{fig:Astronomical terms}
  Astronomical terms and constants copied from www.astro.princeton.edu.}
 \end{figure}


%
%
%

\end{document}